\documentclass[9pt,journal]{IEEEtran}

\IEEEoverridecommandlockouts

\usepackage{color}

\usepackage{amsthm,amsmath,amssymb}
\usepackage{mathtools}
\usepackage{float}
\usepackage{verbatim}
\usepackage{url}

\DeclareFontFamily{U}{wncy}{}
\DeclareFontShape{U}{wncy}{m}{n}{<->wncyr10}{}
\DeclareSymbolFont{mcy}{U}{wncy}{m}{n}
\DeclareMathSymbol{\Sh}{\mathord}{mcy}{"58} 
\usepackage[pdftex]{graphicx}	
\usepackage{adjustbox}

\usepackage[switch]{lineno}
\usepackage[named]{algo}
\usepackage[noend]{algpseudocode}
\usepackage{algorithm}

\usepackage[inline]{enumitem}
\DeclareMathSymbol{\Chi}{\mathalpha}{operators}{"58}

\usepackage[flushleft]{threeparttable} 

\usepackage{upgreek}
\usepackage{bm}
\usepackage{nicefrac}

\usepackage{threeparttable}
\usepackage{multirow}

\ifCLASSOPTIONcompsoc
  \usepackage[caption=false,font=normalsize,labelfont=sf,textfont=sf]{subfig}
\else
  \usepackage[caption=false,font=footnotesize]{subfig}
\fi

\usepackage{cite}
\usepackage{balance}

\newtheorem{theorem}{Theorem}
\newtheorem{definition}{Definition}
\newtheorem{lemma}[theorem]{Lemma}
\newtheorem{proposition}[theorem]{Proposition}

\newtheorem{remark}{Remark}

\newcommand{\suml}{\sum_{\ell=0}^{L-1}}

\newcommand{\sumk}{\sum_{k=0}^{K-1}}
\newcommand{\sumq}{\sum_{q=0}^{Q-1}}

\newcommand{\bsym}{\boldsymbol}

\newcommand{\en}[1]{\left[#1\right]}
\newcommand{\alpharl}{[\bsym{{\alpha}}_r]_\ell}
\newcommand{\taurl}{[\bsym{\overline{\tau}}_r]_\ell}
\newcommand{\nurl}{[\bsym{\overline{\nu}}_r]_\ell}
\newcommand{\alphacq}{[\bsym{\alpha}_c]_q}
\newcommand{\taucq}{[\bsym{\overline{\tau}}_c]_q}
\newcommand{\nucq}{[\bsym{\overline{\nu}}_c]_q}
\DeclareMathOperator*{\minimize}{minimize }
\DeclareMathOperator*{\maximize}{maximize }

\usepackage{array}
\newcommand{\expec}[1]{\mathbb{E}\left[#1\right]}

\newcolumntype{P}[1]{>{\centering\arraybackslash}p{#1}}

\IEEEoverridecommandlockouts

\begin{document}

\title{Dual-Blind Deconvolution for Overlaid Radar-Communications Systems}

\author{Edwin Vargas~\IEEEmembership{Student Member,~IEEE}, Kumar Vijay Mishra~\IEEEmembership{Senior Member,~IEEE}, Roman Jacome~\IEEEmembership{Student Member,~IEEE}, Brian M. Sadler~\IEEEmembership{Life Fellow,~IEEE}, Henry Arguello~\IEEEmembership{Senior Member,~IEEE}
\thanks{E. V., R. J., and H. A. are with Universidad Industrial de Santander, Bucaramanga, Santander 680002 Colombia, e-mail: \{edwin.vargas4@correo., roman2162474@correo., henarfu@\}uis.edu.co.}
\thanks{K. V. M. and B. M. S. are with the United States CCDC Army Research Laboratory, Adelphi, MD 20783 USA, e-mail: kvm@ieee.org, brian.m.sadler6.civ@mail.mil.}
\thanks{This research was sponsored by the Army Research Office/Laboratory under Grant Number W911NF-21-1-0099, and the VIE project entitled ``Dual blind deconvolution for joint radar-communications processing''. K. V. M. acknowledges support from the National Academies of Sciences, Engineering, and Medicine via Army Research Laboratory Harry Diamond Distinguished Fellowship.}
\thanks{The conference precursor of this work was presented at the 2022 IEEE International Conference on Acoustics, Speech, and Signal Processing (ICASSP).}
}

\maketitle
\IEEEpeerreviewmaketitle

\begin{abstract}
The increasingly crowded spectrum has spurred the design of joint radar-communications systems that share hardware resources and efficiently use the radio frequency spectrum. We study a general spectral coexistence scenario, wherein the channels and transmit signals of both radar and communications systems are unknown at the receiver. In this \textit{dual-blind deconvolution} (DBD) problem, a common receiver admits a multi-carrier wireless communications signal that is overlaid with the radar signal reflected off multiple targets. The communications and radar channels are represented by \textit{continuous-valued} range-time and Doppler velocities of multiple transmission paths and multiple targets. We exploit the sparsity of both channels to solve the highly ill-posed DBD problem by casting it into a sum of multivariate atomic norms (SoMAN) minimization. We devise a semidefinite program to estimate the unknown target and communications parameters using the theories of positive-hyperoctant trigonometric polynomials (PhTP). Our theoretical analyses show that the minimum number of samples required for near-perfect recovery is dependent on the logarithm of the maximum of number of radar targets and communications paths rather than their sum. We show that our SoMAN method and PhTP formulations are also applicable to more general scenarios such as unsynchronized transmission, the presence of noise, and multiple emitters. Numerical experiments demonstrate great performance enhancements during parameter recovery under different scenarios.   
\end{abstract}

\section{Introduction}
\label{sec:intro}
With the advent of new wireless communications systems and novel radar technologies, the electromagnetic spectrum has become a contested resource \cite{mishra2019toward}. This has led to the development of various system engineering and signal processing approaches toward optimal, secure, and dynamic sharing of the spectrum \cite{paul2016survey}. In general, spectrum-sharing technologies follow three major approaches: co-design \cite{liu2020co}, cooperation \cite{bicua2018radar}, and coexistence \cite{wu2022resource}. While co-design requires developing new systems and waveforms to efficiently utilize the spectrum \cite{duggal2020doppler}, cooperation requires an exchange of additional information between radar and communications to enhance their respective performances \cite{paul2016survey}. In the coexistence approach, radar and communications independently access the spectrum and spectrum-sharing efforts are largely focused on receiver processing. Among all three approaches, separating the radar and communications signals at a coexistence receiver is highly challenging because the uncoordinated sharing of resources does not help in clearly distinguishing the two received signals. In this paper, we focus on the coexistence problem.

In general, coexistence systems employ different radar and communications waveforms and separate receivers, wherein the management of interference from different radio systems is key to retrieving useful information \cite{bicua2018radar}. When the received radar signal reflected off from the target is overlaid with communications messages occupying the same bandwidth, the knowledge of respective waveforms is useful in designing matched filters to extract the two signals \cite{paul2016survey}. Usually, the radar signal is known and the goal of the radar receiver is to extract unknown target parameters from the received signal. In a typical communications receiver, the channel is known (or estimated) and the unknown transmit message is of interest in the receive processing. However, these assumptions do not extend to a general scenario. For example, in a passive \cite{sedighi2021localization} or multi-static \cite{dokhanchi2019mmwave} radar system, the receiver may not have complete information about the radar transmit waveform. Similarly, in communications over dynamic channels such as millimeter-wave \cite{mishra2019toward} or Terahertz-band \cite{elbir2021terahertz}, the channel coherence times are very short. As a result, the channel state information changes rapidly and cannot be determined \textit{a priori}. Moreover, different transmitters are employed in the spectral coexistence scenario leading to a lack of synchronized arrivals of both signals at a common receiver. 

In this paper, we consider the aforementioned general spectral coexistence scenario, where both radar and communications channels and their respective transmit signals are unknown to the common receiver. We model the extraction of all four of these quantities as a \textit{dual-blind deconvolution} (DBD) problem. This formulation is a variant of \textit{blind deconvolution} (BD) --- a longstanding problem that occurs in a variety of engineering and scientific applications, such as
astronomy, image deblurring, system identification, and optics --- where two unknown signals are estimated from the observation of their convolution~\cite{jefferies1993restoration,ayers1988iterative,abed1997blind,ahmed2018leveraging}. This problem is ill-posed and, in general, structural constraints are imposed on signals to derive recovery algorithms with provable performance guarantees. The underlying ideas in these techniques are based on compressed sensing and low-rank matrix recovery, wherein signals lie in the low-dimensional random subspace and/or in high signal-to-noise ratio (SNR) regime \cite{lee2016blind,li2019rapid,ahmed2013blind,kuo2019geometry,ahmed2018leveraging}.

\subsection{Prior Art}
\color{black}
\begin{table*}[!t]
	\centering
	\caption{Comparison with Prior Art}
	\begin{adjustbox}{max width=\textwidth}

	\begin{threeparttable}%
    \begin{tabular}{l P{7.0cm} P{3cm} l}
		\hline \\
		\textbf{Problem} \vspace{0.5em}& 		\textbf{Measurements} \vspace{0.5em}& \textbf{Unknown parameters} &  \textbf{Algorithm} \vspace{0.5em}\\
		\hline 
		Multi-dimensional SR\tnote{a}\;\cite{xu2014precise} & $[\mathbf{y}]_{\widetilde{n}} = \sum\limits_{\ell=0}^{L-1} [\bsym{\alpha}]_\ell e^{-\mathrm{j}2\pi \mathbf{f}_\ell^T\mathbf{n}} $ 
		&$\bsym{\alpha},\mathbf{f}_\ell$ &Multi-dimensional ANM\vspace{1.5em}\\
		\hline
	    1-D BD\tnote{b}\; \cite{chi2016guaranteed} &\vspace{0.01em} $[\mathbf{y}]_n =  \suml [\bsym{\alpha}]_\ell e^{-\mathrm{j}2\pi n [\bsym\tau]_\ell}[\mathbf{g}]_n  $
		&$\mathbf{g}_n,\bsym{\alpha},\bsym\tau$&   
		ANM\vspace{1.5em}\\
		\hline
	    Blind 1-D SR\tnote{c} \cite{yang2016super} &\vspace{0.001em} $[\mathbf{y}]_n =  \suml [\bsym{\alpha}]_\ell e^{-\mathrm{j}2\pi n [\bsym\tau]_\ell}[\mathbf{g}_\ell]_n  $
	&$\mathbf{g}_\ell,\bsym{\alpha},\bsym\tau$& ANM\vspace{2.5em}\\
	\hline
		SR radar\tnote{d}\;\cite{heckel2016super} & \vspace{0.001em} $\mathbf{y} = \mathbf{G}\suml[\bsym{\alpha}]_\ell\mathbf{a}(\mathbf{r}_\ell)$
	&$\mathbf{r}_\ell,\bsym{\alpha}$	& ANM\vspace{2.5em}\\
		\hline
		Blind 2-D SR\tnote{e} \cite{suliman2022blind} &\vspace{0.001em}$[\mathbf{y}]_{\widetilde{n}}=\suml [\bsym{\alpha}]_\ell \mathbf{a}(\mathbf{r}_\ell)^H\mathbf{G}_{\widetilde{n}}\mathbf{g}_\ell$ 
		&$\bsym{\alpha},\mathbf{r}_\ell,\mathbf{g}_\ell$& ANM\vspace{2.5em}\\
		\hline
		
		 BdM\tnote{f}\;  \cite{ling2017blind} &{$\mathbf{y} =  \sum_{i=1}^{I} \operatorname{diag}(\mathbf{B}_i\mathbf{h}_i)\mathbf{A}_i\mathbf{x}_i$}
		&$\mathbf{h}_\ell,\mathbf{x}_\ell$& Nuclear norm minimization \vspace{2.5em}\\
		\hline
		
		 1-D SR-dM\tnote{g} \cite{li2019stable}\;   & $[\mathbf{y}]_{n}=\sum_{i=1}^{I} [\mathbf{g}]_{i, n} \cdot\left(\sum_{\ell=0}^{L_{i}-1} [\bsym{\alpha}_i]_{\ell} e^{-j 2 \pi n [\bsym\tau_{i}]_{\ell}}\right)$ 
		 
		&$\bsym\alpha_{i}, \bsym\tau_i$& SoAN minimization\vspace{2.5em}\\
		\hline
		
		This paper (DBD)\tnote{h} & \vspace{0.001em} $  [\mathbf{y}]_{\widetilde{n}} = \suml[\bsym{\alpha}_r]_\ell \mathbf{b}_n^H\mathbf{u}e  ^{-\mathrm{j}2\pi n\tau_e}e^{-\mathrm{j}2\pi(n[\bsym{\tau}_r]_\ell+p[\bsym{\nu}_r]_\ell  )}+\sumq [\bsym{\alpha}_c]_q \mathbf{d}_{\widetilde{n}}^H\mathbf{v}e^{-\mathrm{j}2\pi (n[\bsym{\tau}_c]_q+p[\bsym{\nu}_c]_q )}$
		&$\bsym{\alpha}_r,\mathbf{u},\bsym{\tau}_r,\bsym{\nu}_r,$ $ \bsym{\alpha}_c,\mathbf{v},\bsym{\tau}_{c},\bsym{\nu}_c$& SoMAN minimization \vspace{2.5em}\\	
		\hline
	\end{tabular}
	\begin{tablenotes}[para]
	\item[a]$\mathbf{f}_\ell$ is a d-dimensional vector of unknown off-the-grid poles, $[\bsym{\alpha}]_\ell$ is complex amplitude, $\widetilde{n}$ is a linear indexing sequence of the multidimensional measurement, and $L$ is the number of frequencies.\\ 
	\item[b]Vector $\mathbf{g}$ is the transmit signal, $[\bsym\tau]_\ell$ is the unknown off-the-grid pole.
	\\\item[c]This problem assumes a total of $L$ unknown signals of which the $\ell$-th signal and the $n$-th sample is $[\mathbf{g}_\ell]_n$.
	\\\item[d]$\mathbf{G}$ denotes the sensing Gabor matrix and $\mathbf{a}(\mathbf{r})$ are the atoms containing 2-D where $\mathbf{r}_\ell = [[\bsym\tau]_\ell,[\bsym\nu]_\ell]$, $[\bsym\tau]_\ell$ denote time-delay and $[\bsym\nu]_\ell$ Doppler-velocity.
	\\\item[e] $\mathbf{G}_n$ is the sensing Gabor matrix.  \\\item[f]Matrices $\mathbf{B}_\ell, \mathbf{A}_\ell$ model the unknown channels. The unknown signal vectors are $\mathbf{h}_\ell$ and $\mathbf{x}_\ell$. 
     \\\item[g]
     $L_i$ is the number of off-the-grid parameters of $i$-th signal $[\mathbf{g}]_{i,n}$. \\\item[h] See Section~\ref{sec:Signal_model} for details. Briefly, $\mathbf{b}_n$ and $\mathbf{d}_{\widetilde{n}}$ are columns of a representation basis; $\mathbf{u}$ and $\mathbf{v}$ are low-dimensional representations of the radar and communications signal, respectively. Note that, from \eqref{eq:tilden}, the index $\tilde{n}$ in the communications signal component $\mathbf{d}_{\widetilde{n}}$ depends on the pulse index $p$. The radar signal component $\mathbf{b}_n$ only depends on the discrete-time index $n$.
    \end{tablenotes}
    \end{threeparttable}
    \end{adjustbox}
	\label{tbl:1}\vspace{-0.5cm}
\end{table*}
 Unlike prior works on spectral coexistence (see, e.g., \cite{chiriyath2017radar,mishra2019toward} and references therein), we examine the overlaid radar-communications signal as the ill-posed DBD problem. Previously, \cite{farshchian2016dual} studied \textit{dual deconvolution} problem which assumed that the radar transmit signal and communications channel were known. Our approach toward a more challenging DBD problem is inspired by some recent works \cite{chi2016guaranteed,yang2016super} that have analyzed the basic BD for off-the-grid sparse scenarios. The radar and communications channels are usually sparse \cite{mishra2018sub}, more so at higher frequency bands, and their parameters are continuous-valued \cite{mishra2015spectral}. Hence, sparse reconstruction in off-the-grid or continuous parameter domain through techniques based on atomic norm minimization (ANM) \cite{off_the_grid} is appropriate for our application. In BD, sparsity and subspace constraints over the signals have shown great ability in reducing the search space to obtain unique solutions \cite{li2017identifiability,zhang2019structured}.

Our work has close connections with a rich heritage of research in several related signal processing problems; Table~\ref{tbl:1} compares our DBD problem with some important prior works. Our approach is based on the semidefinite program (SDP) derived for ANM-based spectral super-resolution (SR) of a high-dimensional signal in \cite{xu2014precise}. This formulation has previously been extended to bi-variate ANM for estimating delay and Doppler frequencies of targets from received radar signal samples \cite{heckel2016super}. The ANM-based recovery was also employed in the 1-D BD formulation of \cite{chi2016guaranteed}. In general, the absence of knowledge about the transmit signal requires one to define a structural assumption about the signal, e.g., a low-dimensional subspace representation based on incoherence and isotropy properties \cite{candes2011probabilistic}. This structural assumption was also used in \cite{yang2016unknown}, which considered blind SR with multiple modulating unknown waveforms. A 2-D ANM-based blind SR with unknown transmit signals was addressed in \cite{suliman2021mathematical} and extended to 3-D in \cite{suliman2019exact}. The ANM has also been useful for line spectrum denoising \cite{li2019atomic}. However, none of the aforementioned works deal with mixed radar-communications signal structures. 

A closely related problem is that of \textit{blind demixing} (BdM) where the sampled signal is a mixture of several BD problems \cite{ling2017blind,li2019stable}, all of which have a \textit{similar} structure. Per Table~\ref{tbl:1}, for $I=1$ and $2$, the BdM reduces to single and double BDs (with uniform structure in each convolution), respectively. However, our DBD problem has a different structure in each convolution operation (e.g., the communications messages change with each transmission). Further, the recovery guarantees previously derived for BdM do not consider special signal structures of overlaid radar-communications. The recovery algorithms in some BdM formulations \cite{ling2017blind} are based on nuclear norm minimization over a set of rank-one \textit{lifted} matrices. In \cite{dong2018blind}, the BdM recovery problem had multiple non-convex rank-one constraints to deal with multiple low-latency communications systems. These BdM studies do not employ an atomic decomposition over the transfer function and are not well-suited for the estimation of off-the-grid parameters.

\subsection{Our Contributions}
 Preliminary results of this work appeared in our conference publication \cite{vargas2022joint}, where only radar and communications channel parameters were estimated using SoMAN minimization. In this paper, we apply our methods also to estimate the communications messages and radar waveforms, provide detailed theoretical guarantees, include additional numerical validations, address multiple emitter scenarios, and consider unsynchronized and noisy signals. Our main contributions in this paper are:\\
\textbf{1) Solving spectral coexistence as a structured continuous-valued DBD.} Contrary to the above-mentioned BD, BdM, and SR problems, our DBD is based on practical radar-communications scenarios that lend a special structure to the signal as follows. Both communications and radar channels are sparse and their parameters are continuous-valued; the radar channel is a function of unknown target parameters such as range-time and Doppler velocities. A similar delay-Doppler channel is considered for communications paths. Whereas the radar transmits a train of pulses, the communications transmitter emits a multi-carrier signal. Our DBD formulation incorporates all these restrictions/structures.
\\
\textbf{2) Novel ANM-based recovery.} The continuous-valued nature of radar and communications channel parameters requires using ANM-based approaches. In particular, we cast the recovery problem as the minimization of the sum of multivariate atomic norms (SoMAN). Previously, the sum of 1-D atomic norms (SoAN) was minimized to solve a joint (non-blind) demixing (dM) and SR problem in \cite{li2019stable}. However, this 1-D SoAN formulation is non-blind and hence does not trivially extend to our multi-variable DBD. Further, the guarantees in the 1-D SoAN are provided for the same structure of multiple signals while, in our formulation, the structures of radar and communications signals are not identical. Moreover, in the sequel, we also generalize SoMAN to include synchronization errors for scenarios in which radar and communications systems have different clocks.\\
\textbf{3) Strong theoretical guarantees.} Our method is backed up by strong theoretical performance guarantees. We prove that the minimum number of samples required for perfect recovery scale logarithmically with the maximum of the radar targets and communications paths. This is counter-intuitive because the recovery in radar and communications receivers generally depends on the number of unknown scatterers and paths.\\
\textbf{4) Generalization to practical scenarios.} We consider several practical issues to generalize our approach (see Supplementary Material). We show that, in the presence of noise, our formulation is applicable by adding a regularization term to the dual problem. We address the unsynchronized transmission of radar and communications signals by multiplying the atomic set by a lag-dependent diagonal matrix. We also show that the n-tuple BD problem comprising multiple sources of radar and communications signals is addressed by minimizing the addition of multiple SoMANs. Finally, we handle the practical case of an unequal duration of radar pulse and communications symbol by formulating the SoMAN problem to compute multiple coefficient vectors for communications signals. Note that our recent follow-up work in \cite{jacome2022multid} also considers the multi-antenna DBD problem.\\
\textbf{5) Extensive numerical experiments:} We comprehensively validated our SoMAN approach for several scenarios, including randomly distributed targets, unequal numbers of radar targets and communications paths, and closely-spaced parameters. To validate the provided theoretical guarantees numerically, We provide statistical performance by varying system parameters and validate the theoretical result predicted by Theorem 2. Finally, we refer the reader to the Supplementary Material that compiles several numerical results for all the practical scenarios mentioned above. 

\color{black}

\subsection{Organization and Notations} 
The rest of the paper is organized as follows. In the next section, we present the signal model of overlaid radar and communications receiver. Section \ref{sec:formulation} casts the DBD into SoMAN minimization and presents the SDP of the dual problem. In Section \ref{sec:built_poly}, we derive the conditions for finding the dual certificate of support. We validate our proposed approach through extensive numerical experiments in Section \ref{sec:results} and conclude in Section \ref{sec:summ}. In the supplementary material, Appendix~\ref{app:generalization} generalizes our method to more complex scenarios.

Throughout this paper, we reserve boldface lowercase, boldface uppercase, and calligraphic letters for vectors, matrices, and index sets, respectively. The notation $\en{\mathbf{x}}_i$ indicates the $i$-th entry of the vector $\mathbf{x}$ and $\en{\mathbf{X}}_i$ the $i$-th row of $\mathbf{X}$. We denote the transpose, conjugate, and Hermitian by $(\cdot)^T$, $(\cdot)^*$, and $(\cdot)^H$, respectively. The identity matrix of size $N\times N$ is $\mathbf{I}_N$. $||\cdot||_p$ is the $\ell_p$ norm. The notation $\text{Tr}\left\lbrace \cdot \right\rbrace $ is the trace of the matrix, $|\cdot|$ is the cardinality of a set, $\operatorname{supp}(\dot)$ is the support set of its argument, $\mathbb{E}\left[ \cdot \right]$ is the statistical expectation function, and $\mathbb{P}$ denotes the probability. The functions $\text{max}$ and $\text{min}$ output the maximum and minimum values of their arguments, respectively. The sign function is defined as $\operatorname{sign}(c) = \frac{c}{|c|}$. The function $\text{diag}(\cdot)$ outputs a diagonal matrix with the input vector along its main diagonal. The block diagonal matrix with diagonal matrix elements $\mathbf{X}_1$, $\dots$, $\mathbf{X}_P$ is  $\mathbf{X} = \operatorname{blockdiag}[\mathbf{X}_1,\dots,\mathbf{X}_P]$.


\section{Signal Model}
\label{sec:Signal_model}
Consider a pulse Doppler radar that transmits a train of $P_r$ uniformly-spaced pulses $s(t)$ with a pulse repetition interval (PRI) $T_r$; its reciprocal $1/T_r$ is the pulse repetition frequency (PRF). The transmit signal of the radar is 
\begin{align}
x_r(t) = \sum_{p=0}^{P_r-1} s(t-pT_r),\;0\le t \le P_rT_r.
\end{align}
The entire duration of $P_r$ pulses is known as the coherent processing interval (CPI). The pulse $s(t)$ is a time-limited baseband function, whose continuous-time Fourier transform (CTFT) is $S(f)=\int_{-\infty}^{\infty} s(t) e^{-j 2\pi f t} \mathrm{d}t$. It is common to assume that most of the radar signal's energy lies within the frequencies $\pm B/2$, where $B$ denotes the effective signal bandwidth, i.e., $s(t) \approx \int_{-B/2}^{B/2} S(f) e^{j 2\pi f t} \mathrm{d}f$. 

The radar channel or \textit{target scene} consists of $L$ non-fluctuating point-targets, according to the {Swerling-I} target model \cite{skolnik2008radar}. The transmit signal is reflected back by the targets toward the radar receiver. The unknown target parameter vectors are ${\bsym{\alpha}}_r \in \mathbb{C}^L$, ${\overline{\bsym{\tau}}}_r \in \mathbb{R}^L$ and ${\overline{\bsym{\nu}}}_r \in \mathbb{R}^L$, where the $\ell$-th target is characterized by: time delay $\taurl$, which is linearly proportional to the target's range \textit{i.e.} $\taurl = 2d_\ell/c$ where $c$ is the speed of light and $d_\ell$ is the target range; Doppler frequency $\nurl$, proportional to the target's radial velocity \textit{i.e.} $\nurl = 4\pi f_c v_\ell/c $ where $f_c$ is the carrier frequency and $v_\ell$ is the radial velocity; and complex amplitude $\alpharl$ that models the path loss and target reflectivity. The target locations are defined with respect to the polar coordinate system of the radar and their range and Doppler are assumed to lie in the unambiguous time-frequency region, i.e., the time delays ($\tau$) are no longer than the PRI, and Doppler frequencies ($\nu$) are up to the PRF. The radar channel is the product of delta functions as
\begin{align}
    h_r(\nu, \tau) = \suml \en{\bsym{\alpha}_r}_\ell \delta(\nu-\en{{\overline{\bsym{\nu}}}_r}_\ell) \delta(\tau-\en{{\overline{\bsym{\tau}}}_r}_\ell).
\end{align}
 The delay-Doppler representation of the channel is obtained by taking the Fourier transform along the time-axis of the time-delay representation $h_r(t, \tau)$ as
\begin{align}
    h_r(t, \tau) = \suml \en{\bsym{\alpha}_r}_\ell \delta(\tau-\en{{\overline{\bsym{\tau}}}_r}_\ell)e^{-\mathrm{j}2\pi\en{{\overline{\bsym{\nu}}}_r}_\ell t}.\label{eq:radar_channel}
\end{align}

For communications, the transmit signal $x_c(t)$ comprises $P_c$ messages modeled as orthogonal frequency-division multiplexing (OFDM) signals with $K$ equi-bandwidth sub-carriers which together occupy a bandwidth of $B$ with a symbol duration of $T_c$ and have a frequency separation of $\Delta f$ \cite{cimini1985analysis}. Hence, the $m$-th message is $x_m(t) = \sumk [\mathbf{g}_m]_k e^{\mathrm{j}2\pi k\Delta f t}$, where $[\mathbf{g}_m]_k$ is $m$-th complex symbol modulated onto the $k$-th sub-carrier frequency $f_k=k\Delta f$. The communications transmitted signal is 
\begin{align}
x_c(t) = \sum_{m = 0}^{P_c-1} x_m(t-mT_c),\; 0\leq t \leq P_c T_c. \label{comm_transmitted}
\end{align}

The communications channel comprises $Q$ propagation paths characterized by their attenuation coefficients, delay, and frequency shifts, respectively, encapsulated in the parameter vectors $\bsym{\alpha}_c \in \mathbb{C}^Q$, $\overline{\bsym{\tau}}_c \in \mathbb{R}^Q$, and $\overline{\bsym{\nu}} \in \mathbb{R}^Q$. 
The delay-Doppler representation of the communications channel is
\begin{equation}
    h_c(\nu,\tau) = \sumq \alphacq \delta(\nu-\nucq) \delta (\tau-\taucq),
\end{equation}
from which we obtain the time-delay representation as 
\begin{equation}
    h_c(t,\tau) = \sumq \alphacq \delta (\tau-\taucq)e^{-\mathrm{j}2\pi\nucq t}. \label{eq:comms_channel}
\end{equation}
\begin{figure}[!t]
    \centering
    \includegraphics[width=1.0\columnwidth]{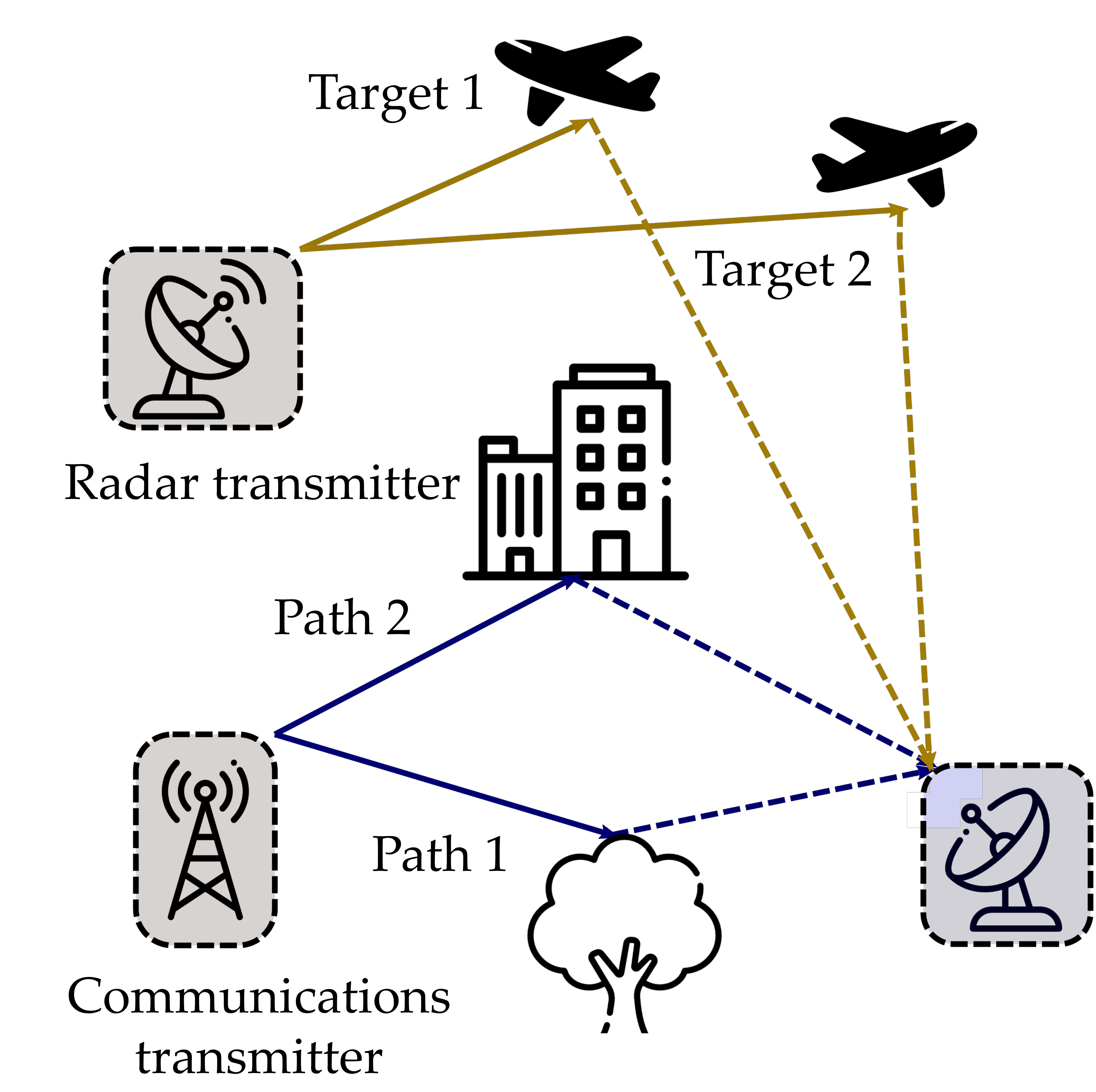}
    \caption{{In a spectral coexistence scenario, a joint receiver admits the overlaid radar and communications signals.}}
    \label{fig:coexistence}
\end{figure}
\subsection{Overlaid Radar-Communications Receiver}
\label{subsec:overlaid_rx}
In a spectral coexistence scenario, both radar and communications operate in the same spectrum (Fig.~\ref{fig:coexistence}). At the receiver, the baseband signal is a superposition of radar and communications signals as
\begin{align}
    y(t) &= x_r(t)*h_r(t)+x_c(t)*h_c(t),\nonumber\\
    &=\int_{-\infty}^{\infty} h_r(t,\tau)x_r(t-\tau)+ h_c(t,\tau)x_c(t-\tau)d\tau,
    \label{eq:received_signal_integral}
\end{align}
where $0\leq t \leq \textrm{max}(P_rT_r,P_cT_c)$.
Substituting the radar and communications channel models of, respectively, \eqref{eq:radar_channel} and \eqref{eq:comms_channel} in \eqref{eq:received_signal_integral} yields 
\begin{align}
    y(t) = &\suml \alpharl x_r(t-[\overline{\bsym{\tau}}_r]_\ell) e^{-\mathrm{j}2\pi \nurl t} {+} \sumq \alphacq x_c(t-\taucq) e^{-\mathrm{j}2\pi \nucq t}.\label{eq:rxsig_2}
\end{align}
Substituting expressions of $x_r(t)$ and $x_c(t)$ into \eqref{eq:rxsig_2} leads to
\begin{align}
    y(t) &= \sum_{p=0}^{P_r-1}\suml\alpharl s(t-pT_r-[\overline{\bsym{\tau}}_r]_\ell)e^{-\mathrm{j}2\pi\nurl t} \nonumber\\& {+} \sum_{m=0}^{P_c-1}\sumq \alphacq x_m(t-mT_c-\taucq)e^{-\mathrm{j}2\pi \nucq t}\\
 &\approx \sum_{p=0}^{P_r-1}\suml[\bsym{\alpha}_r]_\ell s(t-pT_r-[\overline{\bsym{\tau}}_r]_\ell)e^{-\mathrm{j}2\pi[\overline{\bsym{\nu}}_r]_\ell pT_r} \nonumber\\& {+}\sum_{m=0}^{P_c-1}\sumq [\bsym{\alpha}_c]_q x_m(t-mT_c-[\overline{\bsym{\tau}}_c]_q)e^{-\mathrm{j}2\pi[\overline{\bsym{\nu}}_c]_qmT_c},
    \label{model_1}
\end{align}
{where the last approximation results from the assumptions $[\overline{\bsym{\nu}}_r]_\ell T_r\ll1$ and $[\overline{\bsym{\nu}}_c]_q T_c \ll 1$ \cite{zheng2017super,muns2019beam} such that the phase rotation between each radar pulse or OFDM message in the communications is assumed constant during a CPI. Then, for every $P<t<P+1$, the values $[\bsym{\overline{\nu}}_r]_\ell t T$ and  $[\bsym{\overline{\nu}}_c]_\ell t T$ remain constant. The change occurs only during the next pulse. Thus, we replace $tT$ with the index $p$.}

\begin{figure}[!t]
    \centering
    \includegraphics[width=1.0\columnwidth]{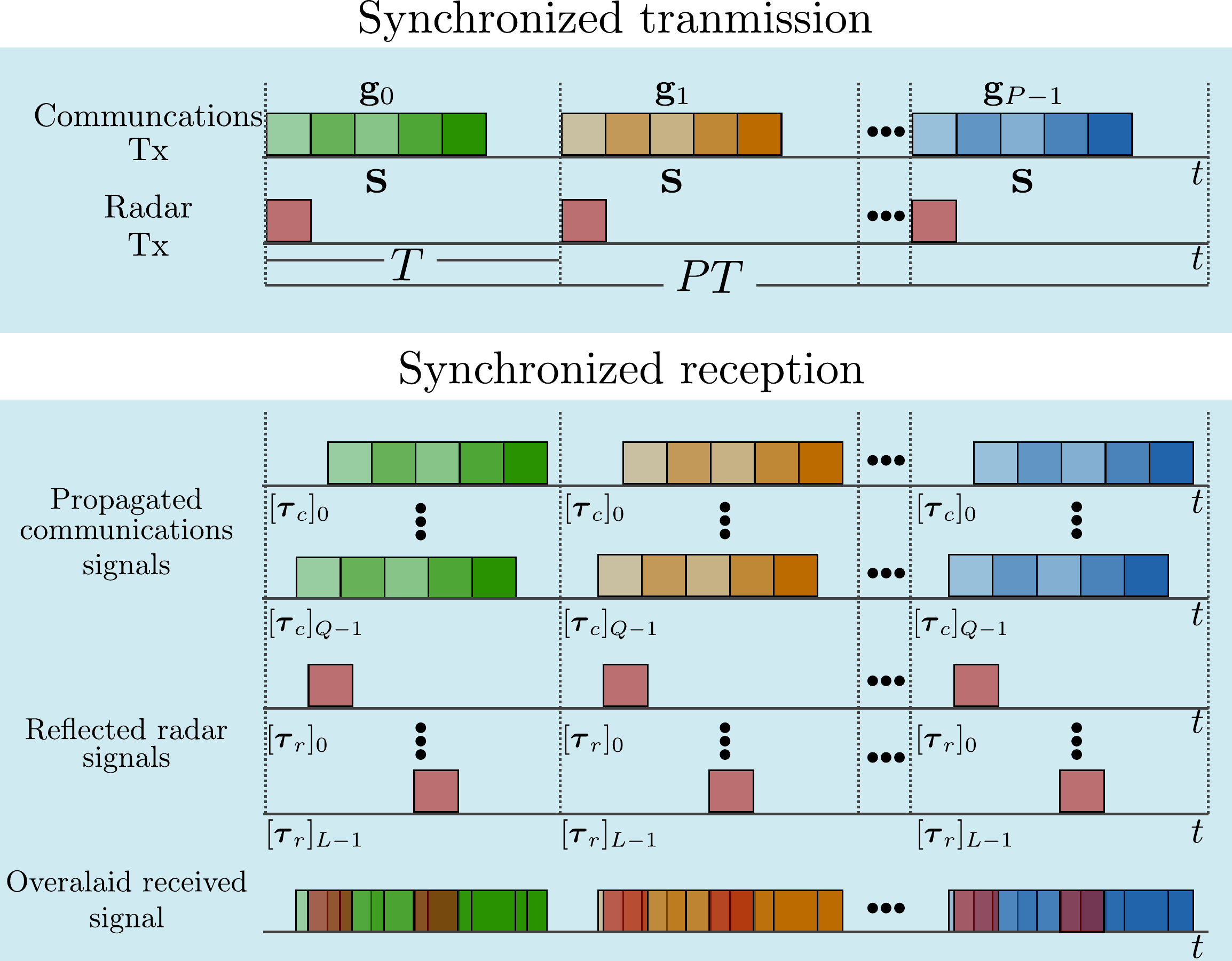}
    \caption{Sequence of transmission and reception of radar and communications signals. In synchronized transmission, the radar transmits each pulse simultaneously as the communications transmitter sends out messages. The common receiver of both systems simultaneously admits radar signals reflected off $L$ targets and communications messages through $Q$ paths. For the unsynchronized transmission, see Fig. 1 in the supplementary material.}
    \label{fig:sync_fig}
\end{figure}

For the sake of simplicity, assume that the number of pulses equals the number of messages, i.e., $P_r=P_c = P$ and the PRI $T_r$ is the same as the message duration $T_c$ \textit{i.e.} $T_r = T_c = T$ (Fig.~\ref{fig:sync_fig}). In the sequel (Section~\ref{subsec:generalziation_pulsewidth}), we also show that our formulation is applicable to the case when $T_r \neq T_c$. Following Fig.~\ref{fig:sync_fig}, the received signal in \eqref{model_1} becomes 
\begin{align}
y(t) = \sum_{p=0}^{P-1} \widetilde{y}_p(t)\label{yp},
\end{align}
where
\begin{align}
    \widetilde{y}_p(t) = &\suml[\bsym{\alpha}_r]_\ell s(t-pT-[\overline{\bsym{\tau}}_r]_\ell)e^{-\mathrm{j}2\pi\nurl pT}\nonumber\\& {+}\sumq [\bsym{\alpha}_c]_q x_p(t-pT-\taucq)e^{-\mathrm{j}2\pi\nucq pT},\;0\leq t \leq PT.
    \label{eq:x_p}
\end{align}
Express the measurements in terms of shifted signals $y_p(t) = \widetilde{y}_p(t+ pT)$, where the signals $\widetilde{y}_p(t+ pT)$ are time-aligned with $y_0(t)$. Then, the signal $y_0(t)$ and the shifted signals contain the same set of parameters. 

The CTFT of $y_p(t)$ produces 
\begin{align}
    Y_p(f) = &\int_{pT}^{pT+T} e^{-\mathrm{j}2\pi ft}\left(\suml\alpharl s(t-[\overline{\bsym{\tau}}_r]_\ell)e^{-\mathrm{j}2\pi\nurl pT} \right. \nonumber\\& \hspace{1cm} \left. {+}\sumq\alphacq x_p(t-\taucq)e^{-\mathrm{j}2\pi\nucq pT}\right)dt,
    \label{eq:ft_general}
\end{align}
which is a sum of the Fourier transforms of the $p$-th received symbol of the communications signal and the $p$-th received pulse of the radar system. In the integral above, computing the radar term and expanding $x_p(t)$ in the communications part yields 
\begin{align}
    Y_p(f) =& \suml[\bsym{\alpha}_r]_\ell S(f)e^{-\mathrm{j}2\pi f\taurl }e^{-\mathrm{j}2\pi\nurl pT} \nonumber\\&
    +\sumq [\bsym{\alpha}_c]_q e^{-\mathrm{j}2\pi f\taucq}e^{-\mathrm{j}2\pi\nucq pT} \sumk [\mathbf{g}_p]_k\times \nonumber\\
    &\hspace{2cm}\int_{pT}^{pT+T} e^{-\mathrm{j}2\pi ft}  e^{\mathrm{j}2\pi k\Delta f t}dt.
    \label{ft_1}
\end{align}

Sampling \eqref{ft_1} {uniformly at the frequencies} $f_n= \frac{Bn}{M}=n\Delta f$, with $n = -N,\dots,N$, $M = 2N+1$, and assuming that $M = K$, {i.e.,} sampling in the frequency domain at the OFDM separation frequency $\Delta f$ \cite{zheng2017super} gives
\begin{align}
    &Y_p(f_n) = \suml[\bsym{\alpha}_r]_\ell S(f_n)e^{-\mathrm{j}2\pi n \Delta f\taurl}e^{-\mathrm{j}2\pi\nurl pT}\nonumber\\
    &+\sumq [\bsym{\alpha}_c]_q e^{-\mathrm{j}2\pi n \Delta f\taucq }e^{-\mathrm{j}2\pi[\overline{\bsym{\nu}}_c]_{q}pT}\sumk [\mathbf{g}_p]_k\int_{pT}^{pT+T} e^{\mathrm{j}2\pi \Delta f (k-n) t}dt.
    \label{ft_2}
\end{align}
Using the property $\int_{0}^{T}e^{-\mathrm{j}2\pi \Delta f(k-n)t}dt=0$ with $k \neq n$ and concatenating the values $Y_p(f_n)$ in the vector $\mathbf{y}_p$, i.e., $\en{\mathbf{y}_p}_{n+N} = Y_p(f_n)$ for $n=-N,\cdots,N$, the discrete values of the Fourier transform \eqref{ft_1} are
\begin{align}
    [\mathbf{y}_p]_{n+N} = &\suml[\bsym{\alpha}_r]_\ell [\mathbf{s}]_{n+N} e^{-\mathrm{j}2\pi(n[{\bsym{\tau}}_r]_\ell+ p[{\bsym{\nu}}_r]_\ell )} \nonumber\\&{+}\sumq [\bsym{\alpha}_c]_q[\mathbf{g}_p]_{n+N} e^{-\mathrm{j}2\pi(n[{\bsym{\tau}}_c]_q+ p[{\bsym{\nu}}_c]_q )},
    \label{eq:y_p}
\end{align}
where $[{\bsym{\tau}}_r]_\ell = \frac{[\overline{\bsym{\tau}}_r]_\ell}{T} \in [0,1]$ and $[\bsym{\tau}_c]_\ell = \frac{[\overline{\bsym{\tau}}_c]_\ell}{T} \in [0,1]$ are the normalized delays; $[\bsym{\nu}_r]_\ell = \frac{[\overline{\bsym{\nu}}_r]_\ell}{\Delta f} \in [0,1]$ and 
$[\bsym{\nu}_c]_\ell = \frac{[\overline{\bsym{\nu}}_c]_\ell}{\Delta f} \in [0,1]$ are the normalized Doppler frequencies; the absolute values of  $[\bsym \alpha_r]_\ell$ and $[\bsym\alpha_c]_q$ are unit-norm after normalizing the  signal by its magnitude, i.e. $\vert[\bsym \alpha_r]_\ell\vert = \vert[\bsym \alpha_c]_q\vert=1$; and $\en{\mathbf{s}}_{n+N} = S(f_n)$. 

We collect all discrete Fourier transform samples in the measurement vector
\begin{align}
\mathbf{y} = \en{\mathbf{y}_0^T,\cdots,\mathbf{y}_{P-1}^T}^T. 
\end{align}
Then, all samples of $P$ pulses (and communications messages) are
\begin{align}
    [\mathbf{y}]_{\widetilde{n}} = &\suml[\bsym{\alpha}_r]_\ell [\mathbf{s}]_n e^{-\mathrm{j}2\pi(n[{\bsym{\tau}}_r]_\ell+ p[{\bsym{\nu}}_r]_\ell )} {+}\sumq [\bsym{\alpha}_c]_q[\mathbf{g}]_{\widetilde{n}} e^{-\mathrm{j}2\pi(n[{\bsym{\tau}}_c]_q+ p[{\bsym{\nu}}_c]_q )},
    \label{eq:y_v}
\end{align}
where \begin{align}
\label{eq:tilden}
    \widetilde{n}=n+N+Mp,
\end{align} 
is a linear indexing sequence of the multidimensional measurement with $n = -N,\dots,N$ and $p = 0,\dots,P-1$, and $\mathbf{g} = [\mathbf{g}_0^T,...,\mathbf{g}_{P-1}^T]$. Note that $\en{\mathbf{y}}_{\widetilde{n}} = \en{\mathbf{y}_p}_{n+N} $ and $\en{\mathbf{g}}_{\widetilde{n}} = \en{\mathbf{g}_p}_{n+N}$. 

\begin{remark}
    In general, a cyclic prefix (CP) of length $K_{\text{cp}}$ is added to each modulated OFDM symbol for mitigating inter-symbol interference (ISI). The OFDM symbol duration then becomes $T_{\text{cp}}+T_c$, where $T_{\text{cp}}$ denote the CP duration. Usually, $T_{\text{cp}}$ is chosen such that the maximum delay $[\bsym\tau_c]_{q_{\max}}< T_{\text{cp}}<T_c$. While we did not consider CP in the communications model above, our approach could be easily modified to include CP by changing the expression $x_m(t) = \sum_{k=0} ^{K-1}[\mathbf{g}_m]_ke^{\mathrm{j}2\pi k\Delta ft}$ to  $x_m(t) = \sum_{k=K-K_{\text{cp}}+1} ^{K-1}[\mathbf{g}_m]_ke^{\mathrm{j}2\pi k\Delta ft}$. The consequence of this change is that the communications messages $\mathbf{g}_m$ is of a larger dimension. This requires a different low-dimensional subspace representation described in the next subsection. This case also resembles the unequal PRI and symbol duration scenario explained in Section A.D of the Supplementary Material.
\end{remark}

\subsection{Representation in low-dimensional space }
\label{subsec:lifting}
Our goal is to estimate the {sets} of parameters $[\bsym{\alpha}_r]_\ell,[\bsym{\tau}_r]_\ell, [\bsym{\nu}_r]_\ell,[\bsym{\alpha}_c]_q,[\bsym{\tau}_c]_q$ and $[\bsym{\nu}_c]_q$, without knowing the radar pulse signal samples $\mathbf{s}$ and the communications symbols $\{\mathbf{g}_p\}_{p=0}^{P-1}$. This is an ill-posed problem because of a large number of unknown variables and fewer measurements. 
We now represent the unknown transmit radar and communications signals $\mathbf{s}$ and $\mathbf{g}_p$ in a low-dimensional subspace spanned by the columns of known random matrices $\mathbf{B}\in\mathbb{C}^{M\times J}$ and $\mathbf{D}_p\in\mathbb{C}^{M\times J}$, respectively. 
Therefore, $\mathbf{s} = \mathbf{B}\mathbf{u}$ and $\mathbf{g}_p = \mathbf{D}_p\mathbf{v}_p$, where $\mathbf{u} \in \mathbb{C}^J$ and  $\mathbf{v}_p \in \mathbb{C}^{J}$ are the unknown low-dimensional coefficient vector of the radar and communications signal respectively. This reduces the number of unknown variables because $J\ll M$.  The basis matrices 
are
\begin{subequations}
\begin{align}
    \mathbf{B} &= \en{\mathbf{b}_{-N},\dots, \mathbf{b}_{N}}^H\text{,}\;\mathbf{b}_n \in \mathbb{C}^J, \label{s_decomp_a} \\
    \mathbf{D}_p &= \en{\en{\mathbf{D}_p}_{-N},\cdots,\en{\mathbf{D}_p}_{N}}^H\text{, }\en{\mathbf{D}_p}_n \in \mathbb{C}^{J}. \label{s_decomp_b}
\end{align}
\end{subequations}
Rewrite the set of symbols $ \mathbf{g}$ compactly as 
\begin{align}
        \mathbf{g} = \mathbf{D}\mathbf{v}\text{,}\hspace{1em}&\mathbf{D} =\en{\mathbf{d}_{0},\cdots,\mathbf{d}_{MP-1}}^H =\text{blockdiag}\en{\mathbf{D}_0,\cdots,\mathbf{D}_{P-1}}.\label{g_decomp}
\end{align}
{The low-dimensional representation of $\mathbf{s}$ and $\mathbf{g}$ is inspired by radar and communications applications. For instance, this structure has been validated previously for a sensing-based SR of complex exponentials that are modulated with unknown waveforms \cite{yang2016super}. In the case of wireless communications, this assumption has been verified for multipath channels with unknown modulation \cite{ahmed2013blind} and multi-user communications \cite{luo2006low}.}
Using the basis decompositions \eqref{s_decomp_a}-\eqref{s_decomp_b} and \eqref{g_decomp}, the signal model in \eqref{eq:y_v} becomes
\begin{align}
    [\mathbf{y}]_{\widetilde{n}} = &\suml[\bsym{\alpha}_r]_\ell \mathbf{b}_n^H\mathbf{u}e^{-\mathrm{j}2\pi(n[\bsym{\tau}_r]_\ell+p[\bsym{\nu}_r]_\ell  )} {+}\sumq [\bsym{\alpha}_c]_q \mathbf{d}_{\widetilde{n}}^H\mathbf{v}e^{-\mathrm{j}2\pi (n[\bsym{\tau}_c]_q+p[\bsym{\nu}_c]_q )}.\label{signal}
\end{align}
Denote the channel vectors as 
\begin{align}
\mathbf{h}_r = \suml[\bsym{\alpha}_r]_\ell\mathbf{a}(\mathbf{r}_\ell),\label{eq:radar_ch_vec}
\end{align}
and
\begin{align}
     \mathbf{h}_c = \sumq[\bsym{\alpha}_c]_q\mathbf{a}(\mathbf{c}_q),\label{eq:comms_ch_vec}
\end{align}
 where $\mathbf{r}_\ell = [[\bsym{\tau}_r]_\ell,[\bsym{\nu}_r]_\ell]^T$ and $\mathbf{c}_q = [[\bsym{\tau}_c]_q,[\bsym{\nu}_c]_q]^T$ such that
\begin{align}
\label{eq:atom_r}
   &{\mathbf{a}(\mathbf{r}_\ell)}\nonumber\\ &= \mathbf{a}(\left[[\bsym{\tau}_r]_\ell,[\bsym{\nu}_r]_\ell\right]^T) \nonumber\\
   &= \big[e^{\mathrm{j}2\pi([\bsym{\tau}_r]_\ell (-N)+[\bsym{\nu}_r]_\ell (0))}, \dots, e^{\mathrm{j}2\pi([\bsym{\tau}_r]_\ell (N)+[\bsym{\nu}_r]_\ell(P-1))}\big]\in\mathbb{C}^{MP} ,
\end{align}
and
\begin{align}
\label{eq:atom_c}
   &{\mathbf{a}(\mathbf{c}_q)} \nonumber\\&= \mathbf{a}(\left[[\bsym{\tau}_c]_q,[\bsym{\nu}_c]_q\right]^T) \nonumber\\&= \big[e^{\mathrm{j}2\pi([\bsym{\tau}_c]_q (-N)+[\bsym{\nu}_c]_q] (0))}, \dots, e^{\mathrm{j}2\pi([\bsym{\tau}_c]_q (N)+[\bsym{\nu}_c]_q(P-1))}\big]\in\mathbb{C}^{MP} .
\end{align}
Then, the model in \eqref{signal} reduces to
\begin{align}
    [\mathbf{y}]_{\widetilde{n}} = \mathbf{h}_r^H \mathbf{e}_{\widetilde{n}} \mathbf{b}_n^H \mathbf{u} + \mathbf{h}_c^H\mathbf{e}_{\widetilde{n}}\mathbf{d}_{\widetilde{n}}^H\mathbf{v}\label{signal_2},
\end{align}
or equivalently
\begin{align}
    [\mathbf{y}]_{\widetilde{n}} &= \langle\mathbf{Z}_r,\mathbf{G}_{\widetilde{n}}^H\rangle + \langle\mathbf{Z}_c,\mathbf{A}_{\widetilde{n}}^H\rangle, \nonumber\\& = \text{Tr}(\mathbf{G}_{\widetilde{n}}\mathbf{Z}_r) +\text{Tr}(\mathbf{A}_{\widetilde{n}}\mathbf{Z}_c),
\end{align}
with the \textit{waveform-channel matrices} 
\begin{align}
    \mathbf{Z}_r &= \mathbf{u}\mathbf{h}_r^H \in \mathbb{C}^{J\times MP},\label{eq:radar_mat}\\
    \mathbf{Z}_c &=\mathbf{v}\mathbf{h}_c^H \in \mathbb{C}^{PJ\times MP},\label{eq:comms_mat}
\end{align}
 $\mathbf{G}_{\widetilde{n}} = \mathbf{e}_{\widetilde{n}}\mathbf{b}_n^H \in \mathbb{C}^{MP\times J}$ and $\mathbf{A}_{\widetilde{n}} = \mathbf{e}_{\widetilde{n}}\mathbf{d}_{\widetilde{n}}^H \in \mathbb{C}^{MP\times PJ}$, where $\mathbf{e}_{\widetilde{n}}$ is the $v$-th canonical vector of $\mathbb{R}^{MP}$. 

Define the linear operators $\aleph_r: \mathbb{C}^{J\times MP}\rightarrow \mathbb{C}^{MP}$ and $\aleph_c: \mathbb{C}^{PJ\times MP}\rightarrow \mathbb{C}^{MP}$  such that
\begin{align*}
    [\aleph_r(\mathbf{Z}_r)]_{\widetilde{n}} = \text{Tr}(\mathbf{G}_{\widetilde{n}}\mathbf{Z}_r),
\end{align*}
and
\begin{align*}
    [\aleph_c(\mathbf{Z}_c)]_{\widetilde{n}} = \text{Tr}(\mathbf{A}_{\widetilde{n}}\mathbf{Z}_c).
\end{align*}
The measurement vector now becomes
\begin{equation}
    \mathbf{y} = \aleph_r(\mathbf{Z}_r) + \aleph  _c(\mathbf{Z}_c),
    \label{eq:received_signal}
\end{equation}
where the matrices $\mathbf{Z}_r$ and $\mathbf{Z}_c$ encapsulate all the unknown variables to be estimated, i.e.,
channel vectors $\mathbf{h}_r$ and $\mathbf{h}_c$ along with the signal coefficients $\mathbf{u}$ and $\mathbf{g}$. This representation of the unknown variables in $1$-rank matrix is similar to the \textit{lifting} trick previously employed in other ill-posed problems such as BD \cite{ahmed2013blind,chi2016guaranteed} and phase retrieval \cite{candes2013phaselift}. Given that $\mathbf{Z}_r = \mathbf{u}\mathbf{h}_r^H$ ($\mathbf{Z}_c =\mathbf{v}\mathbf{h}_c^H $), one could estimate the vectors $\mathbf{u}$ and $\mathbf{h}_r$ $(\mathbf{v}$ and $\mathbf{h}_c$) up to a scaling factor as the left and right singular vectors of $\mathbf{Z}_r$ ($\mathbf{Z}_c$). 


\section{Parameter Recovery via SoMAN minimization}
\label{sec:formulation}
We model the radar and communications trails of the received signal in \eqref{eq:received_signal} as positive linear combinations of constituent \textit{atoms}, which are lifted rank-one matrices $\mathbf{u}\mathbf{a}(\mathbf{r})^H$ and $\mathbf{v}\mathbf{a}(\mathbf{c})^H$ belonging to the atomic sets
\begin{align}
    &\mathcal{A}_r = \Big\{\mathbf{u}\mathbf{a}(\mathbf{r})^H: \mathbf{r}\in[0,1)^2,||\mathbf{u}||_2 = 1   \Big\} \;\subset \mathbb{C}^{J\times MP},
    \label{eq:atomic_set_rad}
\end{align}
and
\begin{align}
    &\mathcal{A}_c = \Big\{\mathbf{v}\mathbf{a}(\mathbf{c})^H: \mathbf{c}\in[0,1)^2,||\mathbf{v}||_2 = 1   \Big\} \;\subset \mathbb{C}^{PJ\times MP}\text{,}
    \label{eq:atomic_set_com}
\end{align}
respectively. These atoms are basic units for synthesizing our signal with sparse channels. This definition leads to the following formulation of \textit{atomic norms} $||\mathbf{Z}_r||_{\mathcal{A}_r}$ and $||\mathbf{Z}_c||_{\mathcal{A}_c}$ - a sparsity-enforcing analog of $\ell_1$ norm for, respectively, general atomic sets $\mathcal{A}_r$ and $\mathcal{A}_c$:
\begin{align}
    &||\mathbf{Z}_r||_{{\mathcal{A}_r}} = \inf_{\stackrel{\alpharl \in \mathbb{C}, \boldsymbol{r}_\ell \in [0,1]^2}{||\mathbf{u}||_2 = 1}} \Bigg\{\sum_\ell |\alpharl| \Big| {\mathbf{Z}}_r = \sum_\ell \alpharl\mathbf{u}\mathbf{a}(\mathbf{r}_\ell)^H\Bigg\},
    \label{eq:atomic_norm_rad}
\end{align}
and
\begin{align}    
    &||\mathbf{Z}_c||_{\mathcal{A}_c} = \inf_{\stackrel{\alphacq \in \mathbb{C}, \boldsymbol{c}_q \in [0,1]^2}{||\mathbf{v}||_2 = 1}} \Bigg\{\sum_q |\alphacq| \Big| \mathbf{Z}_c = \sum_q \alphacq\mathbf{v}\mathbf{a}(\mathbf{c}_q)^H\Bigg\}.
    \label{eq:atomic_norm_com}
\end{align}

To estimate the unknown matrices $\mathbf{Z}_r$ and $\mathbf{Z}_c$, we minimize the SoMAN $||{\mathbf{Z}_r}||_{\mathcal{A}_r} + ||{\mathbf{Z}_c}||_{\mathcal{A}_c}$ among all possible matrices ${\mathbf{Z}_r}$ and ${\mathbf{Z}_c}$ leading to the same observed samples as $\mathbf{y}$. The \textit{primal} convex optimization problem for our DBD then becomes 
\begin{align}
    &\minimize_{{\mathbf{Z}}_r,{\mathbf{Z}}_c} ||{\mathbf{Z}}_r||_{\mathcal{A}_r} +||{\mathbf{Z}}_c||_{\mathcal{A}_c} 
    \;\text{subject to }     \mathbf{y} = \aleph_r({{\mathbf{Z}}_r}) + \aleph_c({\mathbf{{Z}}_c}).
    \label{eq:primal_problem}
\end{align}
A semidefinite characterization of minimization of SoMAN in \eqref{eq:primal_problem} does not directly result from the primal problem as is the case with the standard atomic norm \cite{mishra2015spectral,xu2014precise}. Since the primal problem has only equality constraints, Slater's condition is satisfied, and strong duality holds. This implies, solving the dual problem also yields an exact solution to the primal problem. Therefore, we now analyze the \textit{dual} problem of \eqref{eq:primal_problem} and derive a new semidefinite program for SoMAN minimization using theories of positive-hyperoctant trigonometric polynomials (PhTP). 

\subsection{Dual problem} 
\label{subsec:dual_problem}
The dual problem of \eqref{eq:primal_problem} is obtained via standard Lagrangian analysis. The Lagrangian function of problem \eqref{eq:primal_problem} is 
\begin{align}
    \mathcal{L}(\mathbf{Z}_r,\mathbf{Z}_c,\mathbf{q})&=||\mathbf{Z}_r||_{\mathcal{A}_r} +||\mathbf{Z}_c||_{\mathcal{A}_c}+  \langle\mathbf{q,y}-\aleph_r(\mathbf{Z}_r) - \aleph_c(\mathbf{Z}_c)\rangle,
\end{align}
where $\mathbf{q}$ is the dual variable. Define the dual function $g(\mathbf{q})$ as 
\begin{align}
    g(\mathbf{q})=&\inf_{\mathbf{Z}_r,\mathbf{Z}_c}\mathcal{L}(\mathbf{Z}_r,\mathbf{Z}_c,\mathbf{q})\nonumber\\
    =&\langle\mathbf{q,y}\rangle + \inf_{\mathbf{Z}_r,\mathbf{Z}_c} (||\mathbf{Z}_r||_{\mathcal{A}_r} +||\mathbf{Z}_c||_{\mathcal{A}_c}) 
    -\langle\mathbf{q},\aleph_r(\mathbf{Z}_r)\rangle-\langle\mathbf{q}, \aleph_c(\mathbf{Z}_c)\rangle) \nonumber \\
    =&\langle\mathbf{q,y}\rangle - \sup_{\mathbf{Z}_r} \left(\langle\aleph^*_r(\mathbf{q}),\mathbf{Z}_r \rangle-||\mathbf{Z}_r||_{\mathcal{A}_r}\right)- \sup_{\mathbf{Z}_c} \left(\langle\aleph^*_c(\mathbf{q}),\mathbf{Z}_c \rangle-||\mathbf{Z}_c||_{\mathcal{A}_c}\right) , 
    \label{eq:dual}
\end{align}
where $\aleph_r^*: \mathbb{C}^{MP}\rightarrow\mathbb{C}^{J \times MP}$ and $\aleph_c^*: \mathbb{C}^{MP}\rightarrow\mathbb{C}^{PJ \times MP}$  are the adjoint operators of $\aleph_r$ and $\aleph_c$, respectively, such that $\aleph_r^*(\boldsymbol{q}) = \sum_{p=0}^{P-1}\sum_{n=-N}^{N}[\mathbf{q}]_{\widetilde{n}}\mathbf{G}_{\widetilde{n}}^H$ and $\aleph_c^*(\mathbf{q}) = \sum_{p=0}^{P-1}\sum_{n=-N}^{N}[\mathbf{q}]_{\widetilde{n}}\mathbf{A}_{\widetilde{n}}^H $. The supremum values in \eqref{eq:dual} correspond to the convex conjugate function of the atomic norms $||\mathbf{Z}_r||_{\mathcal{A}_r}$ and $||\mathbf{Z}_c||_{\mathcal{A}_c}$. 

Using the expansion of the dual function $g(\mathbf{q})$ in \eqref{eq:dual}, the dual problem of \eqref{eq:primal_problem} is 
\begin{align}
    &\maximize_{\mathbf{q}}\langle\mathbf{q,y}\rangle_{\mathbb{R}}\nonumber\\&\text{subject to } \Vert\aleph_r^\star(\mathbf{q})\Vert^\star_{\mathcal{A}_r}\leq1 \nonumber\\&\hphantom {\text {subject to } } \Vert \aleph_c^\star(\mathbf{q})\Vert^\star_{\mathcal{A}_c}\leq1, \label{dual_problem}
\end{align}
where $\|\cdot\|^*$ represents the dual norm. This dual norm is defined as 
\begin{align} 
||\mathbf{Z}||_{\mathcal{A}}^* = \sup_{||\mathbf{U}||_{\mathcal{A}}\leq 1} \langle \mathbf{U},\mathbf{Z} \rangle, \label{dual_norm}
\end{align}
where the indicator function
\[
  h^*(\mathbf{Z}) =
  \begin{cases}
                                   0 & \text{if } ||\mathbf{Z}||_{\mathcal{A}}^*\leq 1,\\
                                   \infty & \text{otherwise}, \\
  \end{cases}
\]
is the conjugate function of the dual norm.

We identify that the constraints in \eqref{dual_problem} are positive-hyperoctant trigonometric polynomials (PhTP) in $\mathbf{r}\in [0,1]^2$ and $\mathbf{c} \in [0,1]^2$.  Using the definition of dual-norm \eqref{dual_norm}, rewrite these constraints in \eqref{dual_problem} as
\begin{align*}
    \Vert \aleph_r^*(\mathbf{q})\Vert^*_{\mathcal{A}_r} = \sup_{\mathbf{r}\in[0,1]^2, \Vert \mathbf{u}\Vert=1} \left\vert\langle\mathbf{u}, \aleph_r^*(\mathbf{q})^H\mathbf{a}(\mathbf{r})\rangle\right\vert=  \left\Vert \aleph_r^*(\mathbf{q})^H\mathbf{a}(\mathbf{r})\right\Vert_2, \nonumber\\
    \Vert \aleph_c^*(\mathbf{q})\Vert^*_{\mathcal{A}_c} = \sup_{\mathbf{c}\in[0,1]^2, \Vert \mathbf{v}\Vert=1} \left\vert\langle\mathbf{v}, \aleph_c^*(\mathbf{q})^H\mathbf{a}(\mathbf{c})\rangle\right\vert =  \left\Vert \aleph_c^*(\mathbf{q})^H\mathbf{a}(\mathbf{c})\right\Vert_2.
\end{align*}
In the expressions above, define the bi-variate (delay and Doppler) vector 
polynomials in $\mathbf{r}$ and $\mathbf{c}$ as
\begin{align}
    &\mathbf{f}_r(\mathbf{r}) = \aleph_r^*(\mathbf{q})^H\mathbf{a}(\mathbf{r})= \sum_{p=0}^{P-1}\sum_{n=-N}^{N}[\mathbf{q}]_{\widetilde{n}} \mathbf{G}_{\widetilde{n}}^H \mathbf{a}(\mathbf{r}) \in \mathbb{C}^{J},\label{eq:poly_r}\\
    &\mathbf{f}_c(\mathbf{c})=\aleph_c^*(\mathbf{q})^H\mathbf{a}(\mathbf{c})=\sum_{p=0}^{P-1}\sum_{n=-N}^{N}[\mathbf{q}]_{\widetilde{n}}\mathbf{A}_{\widetilde{n}}^H \mathbf{a}(\mathbf{c}) \in \mathbb{C}^{JP}.\label{eq:poly_c}
\end{align}
Evidently, the inequality constraints in \eqref{dual_problem} are equivalent to the Euclidean norms of the polynomials $\mathbf{f}_r(\mathbf{r})$ and $\mathbf{f}_c(\mathbf{c})$ being upper bounded by unity. 
It follows from the Bounded Real Lemma (BRL) \cite{dumitrescu2007positive} that the bound on the magnitude of a multivariate trigonometric polynomial is characterized by a linear matrix inequality (LMI). More precisely, given a bi-variate PhTP $R(\bsym{\lambda}) = \sum_{\mathbf{t}} k_\mathbf{t} e^{-\mathrm{j}2\pi\bsym{\lambda}^T\mathbf{t}}$, where $\bsym{\lambda}\in [0,1]^2, \mathbf{t} = [t_1,t_2], 0\leq t_1\leq L_1, 0\leq t_2\leq L_2$, with integers $L_1$ and $L_2$. The inequality $| R(\bsym{\lambda})|<1$ implies that there exists a positive semidefinite \textit{Gram} matrix $\mathbf{K}$ such that \cite[p. 135]{dumitrescu2007positive}
\begin{align}
    \delta_\mathbf{n} = \text{Tr}\left(\bsym{\Theta}_\mathbf{n}\mathbf{K}\right),  \begin{bmatrix}
        \mathbf{K} & {\mathbf{k}}^H \\
        {\mathbf{k}} & \mathbf{I}_J 
        \end{bmatrix}
    \succeq0,
 \end{align}
where $\mathbf{n} = [n_1,n_2]$, $0<n_1<m_1$, $-m_2<n_2<m_2$, $\mathbf{k}$ is the vector containing all the polynomial coefficients, 
\begin{align}
\delta_\mathbf{n} = 
            \begin{dcases}
                1, \;n_1=n_2=0,\\ 
                0,\;\textrm{otherwise}, 
            \end{dcases}
\end{align}
and $\bsym{\Theta}_\mathbf{n}= \bsym{\Theta}_{n_2}\otimes \bsym{\Theta}_{n_1}$, where  $\bsym{\Theta}_{n_1}$, $\bsym{\Theta}_{n_2}$ are two elementary Toeplitz matrices with ones in the $n_1$-th and $n_2$-th diagonal respectively and zero elsewhere. 

In the case of uni-variate trigonometric polynomials where the size of the \textit{Gram} matrix is fixed. However, for the multi-variate case, the size of $\mathbf{K}$ depends on the sum-of-squares relaxation degree vector $\mathbf{m}=[m_1,m_2]$, where $m_1\geq L_1$ and $m_2\geq L_2$. A higher relaxation degree leads to a better approximation but it also leads to a higher computational complexity. In practice, this degree may be chosen to be the minimum possible value, i.e., $m_1 = L_1$ and $m_2 = L_2$ and still yield the  optimal result \cite{xu2014precise,heckel2016super}. Furthermore, it is obvious that when such a positive semidefinite matrix $\mathbf{K}$ exists, $|R(\bsym{\lambda})|<1$. Following the SDP characterization of PhTP shown heretofore, we convert the constraints of the dual optimization problem \eqref{dual_problem} to the following LMIs
\begin{align*}
    \begin{bmatrix}
        \mathbf{K} & \widehat{\mathbf{K}}_r^H \\
        \widehat{\mathbf{K}}_r & \mathbf{I}_J 
        \end{bmatrix}
    \succeq0,
    \begin{bmatrix}
        \mathbf{K} & \widehat{\mathbf{K}}_c^H \\
        \widehat{\mathbf{K}}_c & \mathbf{I}_{JP} 
        \end{bmatrix}\succeq 0, \mathbf{K}\succeq 0,
 \end{align*}
 where $\widehat{\mathbf{K}}_r =  \sum_{p=0}^{P-1}\sum_{n=-N}^{N}[\mathbf{q}]_{\widetilde{n}}\mathbf{G}_{\widetilde{n}}^H \in \mathbb{C}^{MP\times J}$ and $\widehat{\mathbf{K}}_c = \sum_{p=0}^{P-1}\sum_{n=-N}^{N} [\mathbf{q}]_{\widetilde{n}} \mathbf{A}_{\widetilde{n}}^H \in \mathbb{C}^{MP\times PJ}$ are the coefficients of the PhTPs. This leads to the SDP of \eqref{dual_problem} as
 \begin{align}
    &\maximize_{\mathbf{q,K}}\quad \langle\mathbf{q,y}\rangle_{\mathbb{R}}\nonumber\\
    &\text{subject to }\mathbf{K}\succeq 0\nonumber\\&\hphantom{\text{subject to }}  
    \begin{bmatrix}
        \mathbf{K} & \widehat{\mathbf{K}}_r^H \\
        \widehat{\mathbf{K}}_r & \mathbf{I}_J 
        \end{bmatrix}
    \succeq0\nonumber\\&\hphantom{\text{subject to }} 
    \begin{bmatrix}
        \mathbf{K} & \widehat{\mathbf{K}}_c^H \\
        \widehat{\mathbf{K}}_c & \mathbf{I}_{JP} 
        \end{bmatrix}\succeq 0
    \nonumber\\&\hphantom{\text{subject to }}
    \text{Tr}(\boldsymbol{\Theta}_\mathbf{n}\mathbf{K}) = \delta_{\mathbf{n}}.\label{dual_opt}
\end{align}
 This SDP is solved using an off-the-shelf solver such as CVX \cite{grant2009cvx}. 
 {\begin{remark}
 We highlight some key differences in our SDP with respect to some related prior works that apply ANM to other (non-BD) applications. For instance, the SDP in \cite{xu2014precise}, the coefficient of the polynomial is just the dual variable. In the non-blind SR problem of \cite{heckel2016super}, the coefficient of the polynomial in the LMI is defined as a product of the inverse Fourier matrix basis, the Gabor sensing matrix, and the dual variable. For blind 2-D SR of \cite{suliman2021mathematical}, the coefficients are defined in terms of a low-dimensional subspace representation. Furthermore, these ANM approaches have only one LMI while SoMAN has as many LMIs as the number of atomic norms. A multitude of LMIs occur in the SDP for ANM applied to super-resolution problem in \cite{mishra2015spectral} but it employed only one PhTP that was also univariate.
  \end{remark}}
 After the dual variable is obtained by solving the SDP in \eqref{dual_opt}, we build the PhTPs $\mathbf{f}_r\left(\mathbf{r}\right)$, $\mathbf{f}_c\left(\mathbf{c}\right)$. The set of delay and Doppler parameters for radar $\widehat{\mathcal{R}} = \left\{\widehat{\mathbf{r}}_\ell \right\}_{\ell=0}^{L-1}$ and communications $\widehat{\mathcal{C}} = \left\{\widehat{\mathbf{r}}_\ell \right\}_{\ell=0}^{Q-1}$ are localized in the respective delay-Doppler planes when, respectively, the PhTPs $\Vert\mathbf{f}_r(\widehat{\mathbf{r}}_\ell)\Vert$ and $\Vert\mathbf{f}_c(\widehat{\mathbf{r}}_q)\Vert1$ assume a maximum modulus of unity. This procedure yields channel information for both systems. Thereafter, to extract the radar and communications transmit waveform, we solve a system of linear equations in Section~\ref{subsec:tx_est}.

\color{black}
\subsection{Dual Certificate}
The primal problem \eqref{eq:primal_problem} has only equality constraints. Hence, Slater's condition is satisfied and strong duality holds. This allows us to state the following Proposition~\ref{prop:dual_certificate}, which presents the dual-certificate of support for the optimizer of \eqref{eq:primal_problem}.
\begin{proposition}
\label{prop:dual_certificate}
Consider the signal $\mathbf{y}$ in \eqref{eq:received_signal}. The atomic sets $\mathcal{A}_r$ and $\mathcal{A}_c$ are as defined in \eqref{eq:atomic_set_rad} and \eqref{eq:atomic_set_com}, respectively. Denote $\mathcal{R} = \{\mathbf{r}_\ell\}_{\ell=0}^{L-1}$ and $\mathcal{C} = \{\mathbf{c}_q\}_{q=0}^{Q-1}$. The pair $\{{\mathbf{Z}}_r^\star,\mathbf{Z}_c^\star\}$ is the unique solution of \eqref{eq:primal_problem} if there exist two bi-variate PhTPs defined as in \eqref{eq:poly_r}-\eqref{eq:poly_c}, one each in $\{\boldsymbol{\tau}_r$, $\bsym{\nu}_r\}$ and $\{\boldsymbol{\tau}_c$, $\bsym{\nu}_c\}$, with complex coefficients $\mathbf{q}$ such that
\begin{subequations}
\begin{align}
    \mathbf{f}_r(\mathbf{r}_\ell) &= \mathrm{sign}( [\bsym{\alpha}_r]_\ell) \mathbf{u},\;   \forall \mathbf{r}_\ell \in \mathcal{R}, \label{eq:cert_1}\\
    \mathbf{f}_c(\mathbf{c}_q) &= \mathrm{sign}( [\bsym{\alpha}_c]_q) \mathbf{v},\;  \forall\mathbf{c}_q \in \mathcal{C}, 
    \label{eq:cert_2}\\
    \Vert \mathbf{f}_r(\mathbf{r}) \Vert_2^2 &< 1,\; \forall \mathbf{r} \in [0,1]^2 \setminus \mathcal{R},  \label{eq:cert_3}\\
    \Vert \mathbf{f}_c(\mathbf{c}) \Vert_2^2  &< 1,\;  \forall \mathbf{c} \in [0,1]^2 \setminus \mathcal{C}.\label{eq:cert_4}
\end{align}
\end{subequations}
\end{proposition}

\begin{IEEEproof}
The variable $\mathbf{q}$ is dual feasible. This gives \par\noindent
\begin{flalign}
    &\langle\mathbf{q,y}\rangle_{\mathbb{R}} = \langle\mathbf{q,\aleph_r(\mathbf{Z}_r)}\rangle_{\mathbb{R}}+ \langle\aleph_c(\mathbf{Z}_c)\rangle_{\mathbb{R}}\nonumber\\
    &= \langle\aleph_r^*(\mathbf{q}),\mathbf{Z}_r\rangle_{\mathbb{R}}+ \langle\aleph_c^*(\mathbf{q}),\mathbf{Z}_c\rangle_{\mathbb{R}}\nonumber\\
    & =\langle\aleph_r^*(\mathbf{q}),\sum_{\ell=0}^{L-1} [\bsym{\alpha}_r]_\ell\mathbf{u}\mathbf{a}(\mathbf{r}_\ell)^H\rangle_{\mathbb{R}} + \langle\aleph_c^*(\mathbf{q}),\sum_{q=0}^{Q-1} [\bsym{\alpha}_c]_q\mathbf{v}\mathbf{a}(\mathbf{c}_q)^H\rangle_{\mathbb{R}} \nonumber \\
    & = \sum_{\ell=0}^{L-1} [\bsym{\alpha}_r]_\ell^* \langle\aleph_r^*(\mathbf{q}),\mathbf{u}\mathbf{a}(\mathbf{r}_\ell)^H\rangle_{\mathbb{R}} + \sum_{q=0}^{Q-1} [\bsym{\alpha}_c]_q^* \langle\aleph_c^*(\mathbf{q}),\mathbf{v}\mathbf{a}(\mathbf{c}_q)^H\rangle_{\mathbb{R}}\nonumber \\
    &= \sum_{\ell=0}^{L-1} [\bsym{\alpha}_r]_\ell^* \langle\mathbf{f}_r(\mathbf{r}_\ell),\mathbf{u}\rangle_{\mathbb{R}} + \sum_{q=0}^{Q-1} [\bsym{\alpha}_c]_q^* \langle\mathbf{f}_c(\mathbf{c}_q),\mathbf{v}\rangle_{\mathbb{R}}\nonumber\\
    & =\sum_{\ell=0}^{L-1}[\bsym{\alpha}_r]_\ell^*\textrm{sign}( [\bsym{\alpha}_r]_\ell) + \sum_{q=0}^{Q-1} [\bsym{\alpha}_c]_q^*\textrm{sign}( [\bsym{\alpha}_c]_q)\nonumber\\
    &= \sum_{\ell=0}^{L-1} |[\bsym{\alpha}_r]_\ell| + \sum_{q=0}^{Q-1} |[\bsym{\alpha}_c]_q|\nonumber \\
    & \geq ||\mathbf{Z}_r||_{\mathcal{A}_r} + ||\mathbf{Z}_c||_{\mathcal{A}_c}\label{eq:lower_bound_dual}.
\end{flalign}
On the other hand, based on the H\"{o}lder inequality we have
\begin{align}
\langle\mathbf{q,y}\rangle_{\mathbb{R}}&=\langle\aleph_r^*(\mathbf{q}),\mathbf{Z}_r\rangle_{\mathbb{R}}+ \langle\aleph_c^*(\mathbf{q}),\mathbf{Z}_c\rangle_{\mathbb{R}}\nonumber\\
&\leq ||\aleph_r^*(\mathbf{q})||_{\mathcal{A}_r}^*||\mathbf{Z}_r||_{\mathcal{A}_r} +   ||\aleph_c^*(\mathbf{q})||_{\mathcal{A}_c}^*||\mathbf{Z}_c||_{\mathcal{A}_c}\nonumber\\
&\leq ||\mathbf{Z}_r||_{\mathcal{A}_r}+||\mathbf{Z}_c||_{\mathcal{A}_c},
\label{eq:upper_bound_dual}
\end{align}
where the first inequality follows from the Cauchy-Schwartz inequality and the last follows from \eqref{eq:cert_1}-\eqref{eq:cert_4}. From \eqref{eq:lower_bound_dual} and \eqref{eq:upper_bound_dual}, $\langle\mathbf{q,y}\rangle_{\mathbb{R}}=||\mathbf{Z}_r||_{\mathcal{A}_r}+||\mathbf{Z}_c||_{\mathcal{A}_c}$ so that the pair $\{\mathbf{Z}_r,\mathbf{Z}_c\}$ is primal optimal and $\mathbf{q}$ is dual optimal based on strong duality.
\end{IEEEproof}

\subsection{Performance guarantees}
\label{subsec:perf_huar}
Recall the following useful properties. 
\begin{definition}[Isotropy \cite{candes2011probabilistic}] The distribution $\mathcal{F}$ is isotropic if  given a random vector $\mathbf{x} \in \mathbb{C}^N \sim \mathcal{F}$ satisfies the following condition: 
    \begin{equation}
    \label{eq:isotropy}
        \mathbb{E}[\mathbf{x}\mathbf{x}^H] =\mathbf{I}_N.
    \end{equation}
   which means that the columns have unit variance and are uncorrelated.
\end{definition}
\begin{definition}[Incoherence \cite{candes2011probabilistic}] The distribution $\mathcal{F}$ is incoherent if given a vector $\mathbf{x}=[x_0,\dots,x_{N-1}] \sim \mathcal{F}$ and given coherence parameter $\mu$ it satisfies that:
\begin{equation}
        {\max_{0<i<N-1}|x_i|^2\leq\mu.}
    \end{equation}
\end{definition}

Following these definitions, we now state our main result.
\begin{theorem}
\label{th:main}
Assume the columns of $\mathbf{B}$ and $\mathbf{D}$ in, respectively, \eqref{s_decomp_a} and \eqref{g_decomp} are drawn from a distribution that satisfies the isotropy and incoherence properties. Further, if $\Vert\mathbf{v}\Vert_2^2=\Vert\mathbf{u}\Vert_2^2=1$ and $\vert\mathbf{\alpha_i}\vert=1$, then there exists a numerical constant $C$ such that
\begin{align}
    MP &\geq  C\mu \operatorname{max}(L,Q)J\log^2\left(\frac{MP J}{\delta}\right)\nonumber\\
    & \times\operatorname{max}\left\{\log\left(\frac{MPQJ}{\delta}\right),\log\left(\frac{MPLJ}{\delta}\right)\right\},  \label{eq:th_main} 
\end{align}
is sufficient to guarantee that, with probability at least $1-\delta$, the pair $\{\mathbf{Z}_r,\mathbf{Z}_c\}$ is perfectly recovered from $MP$ measurements of the observation vector $\mathbf{y}$ in \eqref{eq:received_signal}.
\end{theorem}
\begin{IEEEproof}
See Section~\ref{sec:built_poly}.
\end{IEEEproof}

\subsection{Estimating the communications messages and radar waveform}\label{subsec:tx_est}
Solving \eqref{eq:primal_problem} yields matrices $\mathbf{Z}_r$ and $\mathbf{Z}_c$, wherein the information about the channel vectors $\mathbf{h}_r$ and $\mathbf{h}_c$ is obtained by following the procedure explained at the end of Section ~\ref{subsec:dual_problem}. It follows from \eqref{eq:radar_mat} and \eqref{eq:comms_mat}, that the only unknowns left to estimate are now the radar and communications waveform vectors $\mathbf{u}$ and $\mathbf{v}$. To this end, define the matrices 
\begin{equation}
\mathbf{W}_r = 
\left[\begin{array}{ccc}
\mathbf{a}\left(\widehat{\mathbf{r}}_{0}\right)^{H} \mathbf{A}_{0} & \ldots & \mathbf{a}\left(\widehat{\mathbf{r}}_{L-1}\right)^{H}  \mathbf{A}_{0} \\
\vdots & \ddots & \vdots \\
\mathbf{a}\left(\widehat{\mathbf{r}}_{0}\right)^{H} \mathbf{A}_{MP-1} & \ldots & \mathbf{a}\left(\widehat{\mathbf{r}}_{L-1}\right)^{H}  \mathbf{A}_{MP-1}
\end{array}\right],
\end{equation}
where $\mathbf{W}_r \in \mathbb{C}^{MP\times LJ}$ and
\begin{equation}
\mathbf{W}_c = 
\left[\begin{array}{ccc}
\mathbf{a}\left(\widehat{\mathbf{c}}_{0}\right)^{H} \mathbf{G}_{0} & \ldots & \mathbf{a}\left(\widehat{\mathbf{c}}_{Q-1}\right)^{H}  \mathbf{G}_{0} \\
\vdots & \ddots & \vdots \\
\mathbf{a}\left(\widehat{\mathbf{c}}_{0}\right)^{H} \mathbf{G}_{MP-1} & \ldots & \mathbf{a}\left(\widehat{\mathbf{c}}_{Q-1}\right)^{H}  \mathbf{G}_{MP-1}
\end{array}\right],
\end{equation}
where $\mathbf{W}_c \in \mathbb{C}^{MP\times PQJ}$. Concatenate these matrices as $\mathbf{W_0} = [\mathbf{W}r, \mathbf{W}c] \in \mathbb{C}^{MP\times J(L+PQ)}$ and the coefficient vectors as
$$\mathbf{z} =[ \left[\bsym{\alpha}_r\right]_{0}\mathbf{u}^T,\cdot,\left[\bsym{\alpha}_r\right]_{L-1}\mathbf{u}^T,\left[\bsym{\alpha}_c\right]_{0}\mathbf{v}^T,\cdots,\left[\bsym{\alpha}_c\right]_{Q-1}\mathbf{v}^T]^T,$$
where $\mathbf{z}  \in \mathbb{C}^{J(L+PQ)}$. We recover this vector $\mathbf{z}$ through a least-squares solution of the following problem 
\begin{equation}
    \minimize_{\widehat{\mathbf{z}}} \Vert \mathbf{y} - \mathbf{W}_0 \mathbf{\widehat{z}} \Vert_2.\label{recover_gs}\end{equation}
This only requires that the columns of the matrix $[\mathbf{W}_r\; \mathbf{W}_c]$ are linearly independent because the matrices depend on the values of steering vectors $\mathbf{a}(\widehat{\mathbf{r}}_0),\dots,\mathbf{a}(\widehat{\mathbf{r}}_0)$,  $\mathbf{a}(\widehat{\mathbf{c}}_0),\dots,\mathbf{a}(\widehat{\mathbf{c}}_0)$. The radar waveform $\mathbf{s}$ and the communications messages $\mathbf{g}$ are then estimated following the definitions in \eqref{s_decomp_a}-\eqref{s_decomp_b} and \eqref{g_decomp} as $\widehat{\mathbf{s}} = \mathbf{B}\widehat{\mathbf{u}}$ and $\widehat{\mathbf{g}} =\mathbf{D} \widehat{\mathbf{v}}$. Here, note that the communications messages change at every pulse. Therefore, $\mathbf{g}$ is a concatenation of all messages over $P$ pulses.

\subsection{Computational Complexity}
The present work is the first investigation into the spectral coexistence DBD problem. Here, our focus is perfectly recovering the unknown continuous-valued channel and signal parameters. In this context, ANM-based SDP excels and yields strong theoretical guarantees. The complexity of the proposed algorithm is $\mathcal{O}\left(M^{3.5}P^{3.5}\right)$, which corresponds to the complexity of the SDP solver SDPT3 \cite{zhang2019efficient,krishnan2005interior}. With non-ANM non-SDP heuristics, the exact recovery of continuous-valued parameters severely degrades. However, other approaches may be employed to reduce the computational complexity of the SDP. These techniques include reweighting the continuous dictionary \cite{mingjiu2022gridless}, accelerating proximal gradient \cite{wang2018ivdst}, and localizing the solution space using prior knowledge and applying block iterative $\ell_1$ minimization (BL1M) \cite{cho2015block}. A promising alternative is also mentioned in our recent follow-up work \cite{monsalve2022beurling}, where we have formulated a simple delay-only DBD problem and solved it via a low-rank Hankel matrix recovery approach. However, this method has not been evaluated for our general DBD problem.

\section{Proof of Theorem \ref{th:main}}
\label{sec:built_poly}
To prove Theorem ~\ref{th:main}, we first construct the radar and communications dual polynomials and then demonstrate that they satisfy the required conditions \eqref{eq:cert_1}-\eqref{eq:cert_4} mentioned in the Proposition \ref{prop:dual_certificate}. The existence of dual  polynomials
guarantees the optimal solution $\{\mathbf{Z}_r^\star, \mathbf{Z}_c^\star\}$ of the primal problem \eqref{eq:primal_problem}. 

\subsection{Construction of the dual polynomials} 
Using the notation $\mathbf{f}^{(m',n')}(\mathbf{r}) = \frac{\partial ^{m'}}{\partial \tau^{m'}}\frac{\partial ^{n'}}{\partial \nu^{n'}} \mathbf{f}(\mathbf{r})$, the following conditions ensure that the dual polynomials $\mathbf{f}_r(\mathbf{r})$ and $\mathbf{f}_c(\mathbf{c})$ achieve the maxima at true parameter values
\begin{subequations}
\label{eq:cert_opt_1_to_6}
\begin{align}
&\mathbf{f}_r(\mathbf{r}_\ell) = \mathrm{sign}( [\bsym{\alpha}_r]_\ell) \mathbf{u}, &\forall \mathbf{r}_\ell \in \mathcal{R},\\
& -\mathbf{f}_r^{(1,0)}(\mathbf{r}_\ell) = \mathbf{0}_{J\times 1}, &\forall \mathbf{r}_\ell \in \mathcal{R},\\
& -\mathbf{f}_r^{(0,1)}(\mathbf{r}_\ell) = \mathbf{0}_{J\times 1}, &\forall \mathbf{r}_\ell \in \mathcal{R},\\
& \hspace{1em}\mathbf{f}_c(\mathbf{c}_q) = \mathrm{sign}( [\bsym{\alpha}_c]_q) \mathbf{v},  & \forall \mathbf{c}_q \in \mathcal{C},\\
& -\mathbf{f}_c^{(1,0)}(\mathbf{c}_q) = \mathbf{0}_{PJ\times 1}, &\forall \mathbf{c}_q \in \mathcal{C}, \\
& -\mathbf{f}_c^{(0,1)}(\mathbf{c}_q) = \mathbf{0}_{PJ\times 1}, &\forall \mathbf{c}_q \in \mathcal{C}. 
\end{align}
\end{subequations}

In order to build the polynomials $\mathbf{f}_r(\mathbf{r})$ and $\mathbf{f}_c(\mathbf{c})$ that satisfy the conditions in \eqref{eq:cert_opt_1_to_6}, we now introduce random kernels that are obtained after approximating the dual variable $\mathbf{q}$ in \eqref{eq:poly_r} and \eqref{eq:poly_c}. Unlike previous related works \cite{heckel2016super,candes_superresolution,chi2016guaranteed,off_the_grid}, an additional complication arises in the DBD problem because the dual variable $\mathbf{q}$ must be the same for both radar and communications. Our strategy is to make a coarse approximation of the polynomials via an initial estimate ${\mathbf{q}_0} \neq \mathbf{q}$ of the dual variable. To this end, we solve the following weighted constrained minimization problem with weights $\omega_0$, $\cdots$, $\omega_{MP-1}$ (to be determined later)
\begin{align}
\minimize_{{\mathbf{q}_0}} & \hspace{1em}\| \mathbf{W} {\mathbf{q}_0} \|_{2}^{2} \nonumber\\
\text { subject to } & \hspace{1em} \mathbf{f}_r(\mathbf{r}_\ell) = \mathrm{sign}( [\bsym{\alpha}_r]_\ell) \mathbf{u} &\forall \mathbf{r}_\ell \in \mathcal{R} \nonumber\\
& -\mathbf{f}_r^{(1,0)}(\mathbf{r}_\ell) = \mathbf{0}_{J\times 1} &\forall \mathbf{r}_\ell \in \mathcal{R}\nonumber\\
& -\mathbf{f}_r^{(0,1)}(\mathbf{r}_\ell) = \mathbf{0}_{J\times 1} &\forall \mathbf{r}_\ell \in \mathcal{R}\nonumber\\
& \hspace{1em}\mathbf{f}_c(\mathbf{c}_q) = \mathrm{sign}( [\bsym{\alpha}_c]_q) \mathbf{v}  & \forall \mathbf{c}_q \in \mathcal{C}\nonumber\\
& -\mathbf{f}_c^{(1,0)}(\mathbf{c}_q) = \mathbf{0}_{PJ\times 1} &\forall \mathbf{c}_q \in \mathcal{C}\nonumber\\
& -\mathbf{f}_c^{(0,1)}(\mathbf{c}_q) = \mathbf{0}_{PJ\times 1} &\forall \mathbf{c}_q \in \mathcal{C},\nonumber\\
\label{eq:dual_first}
\end{align}
where 
\begin{equation}
\mathbf{W}=\text{diag}([\omega_0,\cdots,\omega_{MP-1}]^T) \label{eq:weights_1}
\end{equation}
is a diagonal matrix. Note that {the} first and fourth constraints are the same as $\eqref{eq:cert_1}$ and $\eqref{eq:cert_2}$, while the remaining constraints are conditions necessary to satisfy $\eqref{eq:cert_3}$ {and} $\eqref{eq:cert_4}$. Using the expressions in \eqref{s_decomp_a}, \eqref{eq:atom_r}, and \eqref{eq:atom_c}, define the matrix $\mathbf{F} \in \mathbb{C}^{(3LJ + 3QPJ ) \times MP}$ as
\[
\mathbf{F}=\left[\begin{smallmatrix}
\mathbf{b}_{-N}\mathbf{e}_{-NP}^{H} \mathbf{a}\left(\mathbf{r}_{0}\right) & \cdots & \mathbf{b}_{N}\mathbf{e}_{NP}^{H} \mathbf{a}\left(\mathbf{r}_{0}\right) \\
\vdots & \ddots & \vdots \\
\mathbf{b}_{-N}\mathbf{e}_{-NP}^{H} \mathbf{a}\left(\mathbf{r}_{L-1}\right)  & \hdots & \mathbf{b}_{N}\mathbf{e}_{NP}^{H} \mathbf{a}\left(\mathbf{r}_{L-1}\right)  \\
-\mathrm{j}2\pi(-N)\mathbf{b}_{-N}\mathbf{e}_{-NP}^{H} \mathbf{a}\left(\mathbf{r}_{0}\right)& \hdots &-\mathrm{j}2\pi(N)\mathbf{b}_{N}\mathbf{e}_{NP}^{H} \mathbf{a}\left(\mathbf{r}_{0}\right)\\
\vdots & \ddots & \vdots \\
-\mathrm{j}2\pi(-N)\mathbf{b}_{-N}\mathbf{e}_{-NP}^{H} \mathbf{a}\left(\mathbf{r}_{L-1}\right)& \hdots &-\mathrm{j}2\pi(N)\mathbf{b}_{N}\mathbf{e}_{NP}^{H} \mathbf{a}\left(\mathbf{r}_{L-1}\right)\\
-\mathrm{j}2\pi(-P)\mathbf{b}_{-N}\mathbf{e}_{-NP}^{H} \mathbf{a}\left(\mathbf{r}_{0}\right)& \hdots &-\mathrm{j}2\pi(P)\mathbf{b}_{N}\mathbf{e}_{NP}^{H} \mathbf{a}\left(\mathbf{r}_{0}\right)\\
\vdots & \ddots & \vdots \\
-\mathrm{j}2\pi(-P)\mathbf{b}_{-N}\mathbf{e}_{-NP}^{H} \mathbf{a}\left(\mathbf{r}_{L-1}\right)& \hdots &-\mathrm{j}2\pi(P)\mathbf{b}_{N}\mathbf{e}_{NP}^{H} \mathbf{a}\left(\mathbf{r}_{L-1}\right)\\
\mathbf{d}_{-NP}\mathbf{e}_{-NP}^{H} \mathbf{a}\left(\mathbf{c}_{0}\right) & \cdots & \mathbf{d}_{NP}\mathbf{e}_{NP}^{H} \mathbf{a}\left(\mathbf{c}_{0}\right) \\
\vdots & \ddots & \vdots \\
\mathbf{d}_{-NP}\mathbf{e}_{-NP}^{H} \mathbf{a}\left(\mathbf{c}_{Q-1}\right) & \cdots & \mathbf{d}_{NP}\mathbf{e}_{NP}^{H} \mathbf{a}\left(\mathbf{c}_{Q-1}\right)  \\
-\mathrm{j}2\pi(-N)\mathbf{d}_{-NP}\mathbf{e}_{-NP}^{H} \mathbf{a}\left(\mathbf{c}_{0}\right)& \hdots &-\mathrm{j}2\pi(N)\mathbf{d}_{NP}\mathbf{e}_{NP}^{H} \mathbf{a}\left(\mathbf{c}_{0}\right)\\
\vdots & \ddots & \vdots \\
-\mathrm{j}2\pi(-N)\mathbf{d}_{-NP}\mathbf{e}_{-NP}^{H} \mathbf{a}\left(\mathbf{c}_{Q-1}\right)& \hdots &-\mathrm{j}2\pi(N)\mathbf{d}_{NP}\mathbf{e}_{NP}^{H} \mathbf{a}\left(\mathbf{c}_{Q-1}\right)\\
-\mathrm{j}2\pi(-P)\mathbf{d}_{-NP}\mathbf{e}_{-NP}^{H} \mathbf{a}\left(\mathbf{c}_{0}\right)& \hdots &-\mathrm{j}2\pi(P)\mathbf{d}_{NP}\mathbf{e}_{NP}^{H} \mathbf{a}\left(\mathbf{c}_{0}\right)\\
\vdots & \ddots & \vdots \\
-\mathrm{j}2\pi(-P)\mathbf{d}_{-NP}\mathbf{e}_{-NP}^{H} \mathbf{a}\left(\mathbf{c}_{Q-1}\right)& \hdots &-\mathrm{j}2\pi(P)\mathbf{d}_{NP}\mathbf{e}_{NP}^{H} \mathbf{a}\left(\mathbf{c}_{Q-1}\right)
\end{smallmatrix}\right]
\label{eq:matrix_dual}
\]
and the vector {$\mathbf{t}= [\mathbf{t}_r^T, \mathbf{t}_c^T]^T \in \mathbb{C}^{(3LJ + 3QPJ)\times 1}$}, where
\begin{align}
    \mathbf{t}_r &= [ \textrm{sign}(\en{\bsym{\alpha}_r}_0)\mathbf{u}^T,\cdots, \textrm{sign}(\en{\bsym{\alpha}_r}_{L-1})\mathbf{u}^T,\mathbf{0}_{J\times 1}^T,\cdots,\mathbf{0}_{J\times 1}^T]^T, \nonumber \\
    \mathbf{t}_c &= [ \textrm{sign}(\en{\bsym{\alpha}_c}_0)\mathbf{v}^T,\cdots, \textrm{sign}(\en{\bsym{\alpha}_c}_{Q-1})\mathbf{v}^T,\mathbf{0}_{PJ\times 1}^T,\cdots,\mathbf{0}_{PJ\times 1}^T]^T. \nonumber\\   
    \label{eq:RHS_dual_certificate}
\end{align}
Then, the optimization problem in \eqref{eq:dual_first} becomes
\begin{align}
\minimize_{{\mathbf{q}_0}} & \hspace{1em}\| \mathbf{W}{\mathbf{q}_0} \|_{2}^{2} \nonumber\\
\text{subject to } & \hspace{1em}\mathbf{F} {\mathbf{q}_0}=\mathbf{t}.
\label{eq:dual_first_v2}
\end{align}
The least-squares solution of \eqref{eq:dual_first_v2} is ${\mathbf{q}_0}= \left(\mathbf{W}^H\mathbf{W}\right)^{-1}\mathbf{F}^H \bsym{\lambda}$ with 
\begin{align}
    &\bsym{\lambda} = [\bsym{\beta}^T,\bsym{\gamma}^T,\bsym{\zeta}^T,\bsym{\eta}^T,\bsym{\theta}^T,\bsym{\xi}^T]^T,
\label{eq:coefficients_kernels}
\end{align}
where
    $\bsym{\beta} = [\bsym{\beta}_0^T,\cdots,\bsym{\beta}_{L-1}^T]^T \in \mathbb{C}^{JL\times1}$, 
    $\bsym{\gamma} = [\bsym{\gamma}_0^T,\cdots,\bsym{\gamma}_{L-1}^T]^T \in \mathbb{C}^{JL\times1}$, 
    $\bsym{\zeta} = [\bsym{\zeta}_0^T,\cdots,\bsym{\zeta}_{L-1}^T]^T \in \mathbb{C}^{JL\times1}$, $\bsym{\eta} = [\bsym{\eta}_0^T,\cdots,\bsym{\eta}_{Q-1}^T]^T \in \mathbb{C}^{QPJ\times1}$, 
    $\bsym{\theta} = [\bsym{\theta}_0^T,\cdots,\bsym{\theta}_{Q-1}^T]^T \in \mathbb{C}^{QPJ\times1}$, and
    $\bsym{\xi} = [\bsym{\xi}_0^T,\cdots,\bsym{\xi}_{Q-1}^T]^T \in \mathbb{C}^{QPJ\times1}$.
    
Using the expressions of $\mathbf{F}$ and $\bsym{\lambda}$, we rewrite the product $\mathbf{b}=\mathbf{F}^H \bsym{\lambda}$ in the expression of ${\mathbf{q}_0}$ as
\begin{align}
\mathbf{b} &=\suml\left[\begin{smallmatrix}
\mathbf{a} \left(\mathbf{r}_{\ell}\right)^{H} \mathbf{e}_{-NP}\mathbf{b}_{-N}^{H}  \\
\vdots \\
\mathbf{a}\left(\mathbf{r}_{\ell}\right)^{H}\mathbf{e}_{NP}\mathbf{b}_{N}^H
\end{smallmatrix}\right]\bsym{\beta}_\ell +\left[\begin{smallmatrix}
\mathrm{j}2\pi(-N)\mathbf{a}\left(\mathbf{r}_{\ell}\right)\mathbf{b}_{-N}\mathbf{e}_{-NP}^{H}  \\
\vdots \\
\mathrm{j}2\pi(N)\mathbf{a}\left(\mathbf{r}_{\ell}\right)\mathbf{b}_{N}\mathbf{e}_{NP}^{H} 
\end{smallmatrix}\right]\bsym{\gamma}_\ell \nonumber\\
&\hspace{3.8cm}+ \left[\begin{smallmatrix}
\mathrm{j}2\pi(-P)\mathbf{a}\left(\mathbf{r}_{\ell}\right)\mathbf{b}_{-N}\mathbf{e}_{-NP}^{H}   \\
\vdots \\
\mathrm{j}2\pi(P)\mathbf{a}\left(\mathbf{r}_{\ell}\right)\mathbf{b}_{N}\mathbf{e}_{NP}^{H} 
\end{smallmatrix}\right]\bsym{\zeta}_\ell \nonumber\\ &+
\sumq\left[\begin{smallmatrix}
\mathbf{a} \left(\mathbf{c}_{q}\right)^{H} \mathbf{e}_{-NP}\mathbf{d}_{-NP}^{H}  \\
\vdots \\
\mathbf{a}\left(\mathbf{c}_{q}\right)^{H}\mathbf{e}_{NP}\mathbf{d}_{NP}^H
\end{smallmatrix}\right]\bsym{\eta}_q + 
\left[\begin{smallmatrix}
\mathrm{j}2\pi(-N)\mathbf{a}\left(\mathbf{c}_{q}\right)\mathbf{d}_{-NP}\mathbf{e}_{-NP}^{H}   \\
\vdots \\
\mathrm{j}2\pi(N)\mathbf{a}\left(\mathbf{c}_{q}\right)\mathbf{d}_{NP}\mathbf{e}_{NP}^{H} 
\end{smallmatrix}\right]\bsym{\theta}_q \nonumber\\
&\hspace{3.8cm}+ \left[\begin{smallmatrix}
\mathrm{j}2\pi(-P)\mathbf{a}\left(\mathbf{c}_{q}\right)\mathbf{d}_{-NP}\mathbf{e}_{-NP}^{H}   \\
\vdots \\
\mathrm{j}2\pi(P)\mathbf{a}\left(\mathbf{c}_{q}\right)\mathbf{d}_{NP}\mathbf{e}_{NP}^{H} 
\end{smallmatrix}\right]\bsym{\xi}_q. 
\label{eq:first_dual_v3}
\end{align}
Substituting $\mathbf{q}_0$ for the approximation of $\mathbf{q}$  in \eqref{eq:poly_r} yields 
\begin{align}
    &\mathbf{f}_r(\mathbf{r})\nonumber\\
    &=\sum_{\ell=0}^{L-1}\left[\left(\sum_{\widetilde{n}}\frac{1}{\omega_{\widetilde{n}}^2} \mathbf{b}_{n}\mathbf{e}_{\widetilde{n}}^{H} \mathbf{a}(\mathbf{r})^H \mathbf{a}(\mathbf{r}_\ell) \mathbf{e}_{\widetilde{n}}\mathbf{b}_{n}^H\right) \bsym{\beta}_{\ell} \right. \nonumber\\
& \hspace{1cm} \left. +\left(\sum_{\widetilde{n}}\frac{1}{\omega_{\widetilde{n}}^2} \mathrm{j}2\pi(n)\mathbf{b}_{n}\mathbf{e}_{\widetilde{n}}^{H} \mathbf{a}(\mathbf{r})^H \mathbf{a}\left(\mathbf{r}_{\ell}\right)\mathbf{e}_{\widetilde{n}}\mathbf{b}_{n}^H\right)\bsym{\gamma}_\ell  \right. \nonumber\\
& \hspace{1cm} \left.+\left(\sum_{\widetilde{n}}\frac{1}{\omega_{\widetilde{n}}^2} \mathrm{j}2\pi(p)\mathbf{b}_{n}\mathbf{e}_{\widetilde{n}}^{H} \mathbf{a}(\mathbf{r})^H \mathbf{a}\left(\mathbf{r}_{\ell}\right)\mathbf{e}_{\widetilde{n}}\mathbf{b}_{n}^H\right) \bsym{\zeta}_\ell\right] \nonumber\\
& + \sum_{q=0}^{Q-1}\left[\left(\sum_{\widetilde{n}}\frac{1}{\omega_{\widetilde{n}}^2} \mathbf{b}_{n}\mathbf{e}_{\widetilde{n}}^{H} \mathbf{a}(\mathbf{c})^H \mathbf{a}(\mathbf{c}_q) \mathbf{e}_{\widetilde{n}}\mathbf{d}_{\widetilde{n}}^H\right) \bsym{\eta}_{q}\right. \nonumber\\
& \hspace{1cm} \left.+\left(\sum_{\widetilde{n}}\frac{1}{\omega_{\widetilde{n}}^2} \mathrm{j}2\pi(n)\mathbf{b}_{n}\mathbf{e}_{\widetilde{n}}^{H} \mathbf{a}(\mathbf{r})^H \mathbf{a}\left(\mathbf{r}_{\ell}\right)\mathbf{e}_{\widetilde{n}}\mathbf{d}_{\widetilde{n}}^H\right)\bsym{\theta}_\ell \right. \nonumber\\
& \hspace{1cm} \left. + \left(\sum_{\widetilde{n}}\frac{1}{\omega_{\widetilde{n}}^2} \mathrm{j}2\pi(p)\mathbf{b}_{n}\mathbf{e}_{\widetilde{n}}^{H} \mathbf{a}(\mathbf{r})^H \mathbf{a}\left(\mathbf{r}_{\ell}\right)\mathbf{e}_{\widetilde{n}}\mathbf{d}_{\widetilde{n}}^H\right) \bsym{\xi}_q\right].\label{eq:expr_fr_1}
\end{align}
Using
\begin{align}
  \mathbf{b}_n\mathbf{e}_{\widetilde{n}}^H\mathbf{a}(\mathbf{r})&=e^{\mathrm{j}2\pi(n\tau + p \nu)}\mathbf{b}_n, \\
 \mathbf{a}(\mathbf{r})^H   \mathbf{e}_{\widetilde{n}}\mathbf{b}_n^H&=e^{-\mathrm{j}2\pi(n\tau + p \nu)}\mathbf{b}_n^H,
\end{align}
the expression in \eqref{eq:expr_fr_1} becomes\par\noindent\small
\begin{align}
    &\hspace{-3mm}\mathbf{f}_r(\mathbf{r}) \nonumber\\
    &\hspace{-3mm}=\sum_{\ell=0}^{L-1}\sum_{\widetilde{n}}\frac{1}{\omega_{\widetilde{n}}^2}\left[     e^{\mathrm{j}2\pi n(\tau -\tau_\ell)}e^{\mathrm{j}2\pi p(\nu - \nu_\ell)}\mathbf{b}_{n}\mathbf{b}_{n}^H \bsym{\beta}_{\ell} \right. \nonumber\\
    & \hspace{1.5cm}\left.+ \mathrm{j}2\pi(n)e^{\mathrm{j}2\pi n(\tau-\tau_\ell)}e^{\mathrm{j}2\pi p(\nu - \nu_\ell)}\mathbf{b}_{n}\mathbf{b}_{n}^H\bsym{\gamma}_\ell  \right. \nonumber\\
    & \hspace{1.5cm}\left. +\mathrm{j}2\pi(p) e^{\mathrm{j}2\pi n(\tau-\tau_\ell)}e^{\mathrm{j}2\pi p(\nu - \nu_\ell)}\mathbf{b}_{n}\mathbf{b}_{n}^H \bsym{\zeta}_\ell\right]\nonumber\\
    &+\sum_{q=0}^{Q-1}\sum_{\widetilde{n}}\frac{1}{\omega_{\widetilde{n}}^2}\left[e^{\mathrm{j}2\pi n(\tau-\tau_q)}e^{\mathrm{j}2\pi p(\nu - \nu_q)} \mathbf{b}_{n}\mathbf{d}_{\widetilde{n}}^H \bsym{\eta}_{q}\right. \nonumber\\
    & \hspace{1.5cm}\left. + \mathrm{j}2\pi(n)e^{\mathrm{j}2\pi n(\tau-\tau_q)}e^{\mathrm{j}2\pi p(\nu - \nu_q)}\mathbf{b}_{n}\mathbf{d}_{\widetilde{n}}^H\bsym{\theta}_\ell  \right. \nonumber\\
    & \hspace{1.5cm}\left. + \mathrm{j}2\pi(p)e^{\mathrm{j}2\pi n(\tau-\tau_q)}e^{\mathrm{j}2\pi p(\nu - \nu_q)}\mathbf{b}_{n}\mathbf{d}_{\widetilde{n}}^H \bsym{\xi}_q\right].\label{eq:expr_fr_2}
\end{align}\normalsize
The expression above involves the random matrices 
\begin{align}
    \mathbf{M}\left(\mathbf{r}\right)
    =&\sum_{\widetilde{n}} \frac{1}{\omega_{\widetilde{n}}^2}e^{\mathrm{j}2\pi n\tau}e^{\mathrm{j}2\pi p\nu}    \mathbf{b}_{n}\mathbf{b}_{n}^H \in \mathbb{C}^{J\times J},
\end{align}
and
\begin{align}
    \mathbf{N}\left(\mathbf{r}\right)=&\sum_{\widetilde{n}} \frac{1}{\omega_{\widetilde{n}}^2}e^{\mathrm{j}2\pi n\tau}e^{\mathrm{j}2\pi p\nu}    \mathbf{b}_{n}\mathbf{d}_{\widetilde{n}}^H \in \mathbb{C}^{J\times PJ}.
\end{align}
Using the notation $\mathbf{X}^{(m',n')}(\mathbf{r}) = \frac{\partial ^{m'}}{\partial \tau^{m'}}\frac{\partial ^{n'}}{\partial \nu^{m'}} \mathbf{X}(\mathbf{r})$, we rewrite the polynomial in \eqref{eq:expr_fr_2} as
\begin{align}
   \mathbf{f}_r (\mathbf{r}) &= \sum_{\ell=0}^{L-1} \mathbf{M}(\mathbf{r-r}_\ell)\bsym{\beta}_\ell + \mathbf{M}^{(1,0)}(\mathbf{r-r}_\ell)\bsym{\gamma}_\ell + \mathbf{M}^{(0,1)}(\mathbf{r-r}_\ell)\bsym{\zeta},_\ell \nonumber\\
   &+\sum_{q=0}^{Q-1} {\mathbf{N}}(\mathbf{r- c}_q)\bsym{\eta}_q + {\mathbf{N}}^{(1,0)}(\mathbf{r-c}_q)\bsym{\theta}_q + 
    {\mathbf{N}}^{(0,1)}(\mathbf{r-c}_q)\bsym{\xi}_q.
    \label{eq:polyr_random_v1}
\end{align}

Recall from \eqref{eq:tilden} that $\tilde{n}=N+n + Mp$,  $n=-N,\cdots,N$, $p=0,\cdots,P-1$, and $M=2N+1$. We choose the weights in \eqref{eq:weights_1} as
\begin{equation}
\omega_{\tilde{n}} = \sqrt{\frac{N}{g_N(n)}} \sqrt{\frac{P}{g_P(p)}}, \label{eq:weights_2}
\end{equation}
such that $\mathbf{M}$ and $\mathbf{N}$ are generated from the $2$-D squared Fej\'er kernel

\begin{align}
    \varphi(\mathbf{r}) = \varphi_N(\tau)\varphi_P(\nu),
    \label{eq:kernel_2D}
\end{align}
where
\begin{align}
    \varphi_N(\tau) &= \left(\frac{\text{sin}(T\pi \tau)}{T\text{sin}(\pi \tau)}\right)^4, \hspace{1em} T=\frac{N}{2}+1,\nonumber\\
   \varphi_N(\tau) &= \sum_{n=-N}^{N} g_N(n) \textrm{e}^{\textrm{j}2\pi \tau n},
\end{align}
with
\begin{align}
    g_N(n) = \frac{1}{N}\sum_{k=\text{max}\{ n-N,-N\}}^{\text{min}\{ n+N,N\}} \left(1-\frac{\vert k\vert}{M}\right)\left(1-\frac{\vert n-k\vert}{M}\right).\nonumber 
\end{align}
The  Fej\'er kernel has the maximum modulus of unity and then it rapidly decays to zero. It has been successfully applied in the construction of polynomials that guarantee optimality for SR and BD problems \cite{off_the_grid,candes_superresolution,heckel2016super}.
 {\begin{remark}
 Unlike prior ANM-based BD approaches \cite{chi2016guaranteed} that construct the polynomial with one random kernel, the radar polynomial \eqref{eq:polyr_random_v1} in our DBD problem is constructed using two distinct random kernels and their partial derivatives; as mentioned next, the same is true for the communications polynomial. Note that the selection of the random kernels ensures that the dual radar polynomial in \eqref{eq:polyr_random_v1} has the same form as \eqref{eq:poly_r}. However, the basis vectors $\mathbf{b}_n$ and $\mathbf{d}_{\widetilde{n}}$ are statistically independent. Therefore, the contribution in expectation of the random kernel $\mathbf{N}$ to interpolate the sign pattern in the dual certificate is zero. 
\end{remark}}
The dual communications polynomial $\mathbf{f}_c$ is obtained \textit{mutatis mutandis} as \par\noindent\small
\begin{align}
    \mathbf{f}_c(\mathbf{c}) =& \sum_{\ell=0}^{L-1} \mathbf{N}(\mathbf{c-r}_\ell)^H\bsym{\beta}_\ell + \mathbf{N}^{(1,0)}(\mathbf{c-r}_\ell)^H\bsym{\gamma}_\ell + \mathbf{N}^{(0,1)}(\mathbf{c-r}_\ell)^H\bsym{\zeta}_\ell \nonumber\\&+\sum_{q=0}^{Q-1} {\mathbf{P}}^{(0,0)}(\mathbf{c-c}_q)\bsym{\eta}_q + {\mathbf{P}}^{(1,0)}(\mathbf{c-c}_q)\bsym{\theta}_q + 
    {\mathbf{P}}^{(0,1)}(\mathbf{c-c}_q)\bsym{\xi}_q,
    \label{eq:polyc_random_final}
\end{align}\normalsize
where 
\begin{align}
    \mathbf{P}\left(\mathbf{r}\right)=&\frac{1}{MP}\sum_{\widetilde{n}} g_N(n) g_P(p) e^{\mathrm{j}2\pi n\tau}e^{\mathrm{j}2\pi p\nu}    \mathbf{d}_{\widetilde{n}}\mathbf{d}_{\widetilde{n}}^H.
\end{align}
Note that the kernel $\mathbf{N}(\mathbf{r})$ is common between the expressions of radar and communications polynomials.

Next, we select the vectors $\bsym{\beta}$, $\bsym{\gamma}$, $\bsym{\zeta}$, $\bsym{\eta}$, $\bsym{\theta}$, and $\bsym{\xi}$  to satisfy the conditions \eqref{eq:cert_1} and \eqref{eq:cert_2} for the radar and communications polynomials, respectively.
\subsection{To satisfy conditions \eqref{eq:cert_1} and \eqref{eq:cert_2} }
 Take the derivatives of kernels 
\begin{flalign}
    &\mathbf{M}^{(m',n')}(\mathbf{r}) =\sum_{\widetilde{n}} \frac{1}{\omega_{\widetilde{n}}^2}(\mathrm{j}2\pi n)^{n'}(\mathrm{j}2\pi p)^{m'}e^{\mathrm{j}2\pi n \tau}e^{\mathrm{j}2\pi n \nu }\mathbf{b}_n\mathbf{b}_n^H\nonumber\\
    &\mathbf{N}^{(m',n')}(\mathbf{r}) =\left.\sum_{\widetilde{n}} \frac{1}{\omega_{\widetilde{n}}^2}(\mathrm{j}2\pi n)^{n'}(\mathrm{j}2\pi p)^{m'}e^{\mathrm{j}2\pi n \tau }e^{\mathrm{j}2\pi n \nu}\mathbf{b}_n\mathbf{d}_{\widetilde{n}}^H\right. \nonumber\\
    &\mathbf{P}^{(m',n')}(\mathbf{c}) =\left.\sum_{\widetilde{n}} \frac{1}{\omega_{\widetilde{n}}^2}(\mathrm{j}2\pi n)^{n'}(\mathrm{j}2\pi p)^{m'}e^{\mathrm{j}2\pi n \tau}e^{\mathrm{j}2\pi n \nu }\mathbf{d}_{\widetilde{n}}\mathbf{d}_{\widetilde{n}}^H.\right.   \nonumber   
\end{flalign}
Then, rewrite the derivatives of radar and communications polynomials as
\begin{align}
    \mathbf{f}_r^{(m',n')}(\mathbf{r})&= \sum_{\ell=0}^{L-1} \mathbf{M}^{(m',n')}(\mathbf{r-r}_\ell)\bsym{\beta}_\ell  + \mathbf{M}^{(m'+1,n')}(\mathbf{r-r}_\ell)\bsym{\gamma}_\ell \nonumber\\
& \hspace{3.8cm} + \mathbf{M}^{(m',n'+1)}(\mathbf{r-r}_\ell)\bsym{\zeta}_\ell \nonumber\\
    &+\sum_{q=0}^{Q-1} {\mathbf{N}}^{(m',n')}(\mathbf{r-c}_q)\bsym{\eta}_q + {\mathbf{N}}^{(m'+1,n')}(\mathbf{r-c}_q)\bsym{\theta}_q \nonumber\\
& \hspace{3.8cm}+ 
    {\mathbf{N}}^{(m',n'+1)}(\mathbf{r-c}_q)\bsym{\xi}_q,
\end{align}
and,
\begin{align}
    \mathbf{f}_c^{(m',n')}(\mathbf{c})
    &= \sum_{\ell=0}^{L-1} \mathbf{T}^{(m',n')}(\mathbf{c-r}_\ell)\bsym{\beta}_\ell + \mathbf{T}^{(m'+1,n')}(\mathbf{c-r}_\ell)\bsym{\gamma}_\ell \nonumber\\
& \hspace{3.8cm}+ \mathbf{T}^{(m',n'+1)}(\mathbf{c-r}_\ell)\bsym{\zeta}_\ell \nonumber\\
    &+\sum_{q=0}^{Q-1} {\mathbf{P}}^{(m',n')}(\mathbf{c-c}_q)\bsym{\eta}_q + {\mathbf{P}}^{(m'+1,n')}(\mathbf{c-c}_q)\bsym{\theta}_q
    \nonumber\\
& \hspace{3.8cm}+ 
    {\mathbf{P}}^{(m',n'+1)}(\mathbf{c-c}_q)\bsym{\xi}_q.
    \label{eq:polyc_derivative}
\end{align}

Next, define the vectors
\begin{align}
    \mathbf{j}_r = [\bsym{\beta}^T,{\kappa}\bsym{\gamma}^T,\kappa \bsym{\zeta}^T]^T,\hspace{1em} \mathbf{j}_c = [\bsym{\eta}^T,{\kappa}\bsym{\theta}^T,\kappa \bsym{\xi}^T]^T,
\end{align}
and matrices $\mathbf{H}_1 \in \mathbb{C}^{3LJ \times 3LJ}$, $\mathbf{H}_2 \in \mathbb{C}^{3LJ \times 3PQJ}$, and $\mathbf{H}_3 \in \mathbb{C}^{3PQJ \times 3PQJ}$ such that
$$
\mathbf{H}_i=\left[\begin{array}{ccc}
{\mathbf{E}_i^{(0,0)}}& \frac{1}{\kappa} {\mathbf{E}}_i^{(1,0)} & \frac{1}{\kappa} {\mathbf{E}}_i^{(0,1)}\\
-\frac{1}{\kappa} {\mathbf{E}}_i^{(1,0)} & -\frac{1}{\kappa^{2}} {\mathbf{E}}_i^{(2,0)} & -\frac{1}{\kappa^{2}} {\mathbf{E}}_i^{(1,1)} \\
-\frac{1}{\kappa} {\mathbf{E}}_i^{(0,1)} & -\frac{1}{\kappa^{2}} {\mathbf{E}}_i^{(1,1)} & -\frac{1}{\kappa^{2}} {\mathbf{E}}_i^{(0,2)}
\end{array}\right],\;i=1,2,3
$$
with $\kappa = \sqrt{\varphi''(0)}$. The matrix ${\mathbf{E}_1}^{(m',n')} \in \mathbb{C}^{LJ \times LJ}$ comprises a total of $L^2$ block matrices, each of size $J\times J$, as

$$
{\mathbf{E}_1}^{\left(m^{\prime}, n^{\prime}\right)}=\left[\begin{array}{ccc}
\mathbf{M}^{\left(m^{\prime}, n^{\prime}\right)} \left(\mathbf{r}_{1}-\mathbf{r}_{1}\right) & \ldots & \mathbf{M}^{\left(m^{\prime}, n^{\prime}\right)}\left(\mathbf{r}_{1}- \mathbf{r}_{L}\right) \\
\vdots & \ddots & \vdots \\
\mathbf{M}^{\left(m^{\prime},n^{\prime}\right)}\left(\mathbf{r}_{L}-\mathbf{r}_{1}\right) & \ldots & \mathbf{M}^{\left(m^{\prime},n^{\prime}\right)}\left(\mathbf{r}_{L}-\mathbf{r}_{L}\right)
\end{array}\right].
$$
Similarly, the {matrices} ${\mathbf{E}_2}^{(m',n')} \in \mathbb{C}^{LJ \times QPJ}$ and ${\mathbf{E}_3}^{(m',n')} \in \mathbb{C}^{QPJ \times QPJ}$ {consist} of $L Q$ block matrices of size $J\times PJ$ {and} $Q^2$ block matrices of size $PJ\times PJ$ {respectively}
$$
{\mathbf{E}_2}^{\left(m^{\prime}, n^{\prime}\right)}=\left[\begin{array}{ccc}
\mathbf{N}^{\left(m^{\prime}, n^{\prime}\right)} \left(\mathbf{r}_{1}- \mathbf{c}_{1}\right) & \ldots & \mathbf{N}^{\left(m^{\prime}, n^{\prime}\right)}\left(\mathbf{r}_{1}- \mathbf{c}_{Q}\right) \\
\vdots & \ddots & \vdots \\
\mathbf{N}^{\left(m^{\prime}, n^{\prime}\right)}\left(\mathbf{r}_{L}-\mathbf{c}_{1}\right) & \ldots & \mathbf{N}^{\left(m^{\prime}, n^{\prime}\right)}\left(\mathbf{r}_{L}-\mathbf{c}_{Q}\right)
\end{array}\right],
$$

$$
{\mathbf{E}_3}^{\left(m^{\prime}, n^{\prime}\right)}=\left[\begin{array}{ccc}
\mathbf{P}^{\left(m^{\prime}, n^{\prime}\right)} \left(\mathbf{c}_{1}-\mathbf{c}_{1}\right) & \ldots & \mathbf{P}^{\left(m^{\prime}, n^{\prime}\right)}\left(\mathbf{c}_{1}-\mathbf{c}_{Q}\right) \\
\vdots & \ddots & \vdots \\
\mathbf{P}^{\left(m^{\prime}, n^{\prime}\right)}\left(\mathbf{c}_{Q}-\mathbf{c}_{1}\right) & \ldots & \mathbf{P}^{\left(m^{\prime}, n^{\prime}\right)}\left(\mathbf{c}_{Q}-\mathbf{c}_{Q}\right)
\end{array}\right].
$$

Consequently, we rewrite the conditions in \eqref{eq:cert_opt_1_to_6} as the following system of linear equations
\begin{align}
\underbrace{\left[\begin{array}{ccc}
{\mathbf{H}_1} & {\mathbf{H}_2}   \\
 {\mathbf{H}_2^H} & {\mathbf{H}_3}
\end{array}\right]}_{ = \mathbf{H} \in \mathbb{C}^{3(QPJ+LJ) \times 3(QPJ+LJ)}  }\left[\begin{array}{c}
\mathbf{j}_r \\
\mathbf{j}_c 
\end{array}\right]=\left[\begin{array}{c}
\mathbf{t}_r \\
\mathbf{t}_c
\end{array}\right],
\label{eq:system_coefficients}
\end{align}
where $\mathbf{t}_r$ and $\mathbf{t}_c$ {have been} defined in \eqref{eq:RHS_dual_certificate}. Clearly, if the matrix $\mathbf{H}$ is invertible, the coefficients $\mathbf{j}_r$ ($\mathbf{j}_c$) for the radar (communications) polynomials could be obtained. In order to establish the invertibility of $\mathbf{H}$, we show that $\expec{\mathbf{H}}$ is invertible and then prove that $\mathbf{H}$ is close to $\expec{\mathbf{H}}$.

Using the following notation of the vectors
\begin{align*}
   \bsym{\upsilon}_{\widetilde{n}} = \left[\begin{array}{c} 1\\\frac{-\mathrm{j} 2 \pi n}{\kappa}\\\frac{-\mathrm{j} 2 \pi p}{\kappa}
    \end{array}\right] \otimes \left[\begin{array}{c} e^{-\mathrm{j} 2 \pi(n[\tau_r]_0+p[\nu_r]_0)}\\\vdots\\e^{-\mathrm{j} 2 \pi(n[\tau_r]_{L-1}+p[\nu_r]_{L-1})}
    \end{array}\right]\in\mathbb{C}^{3L},
\end{align*}
and
\begin{align*}
    \bsym{\varrho}_{\widetilde{n}} = \left[\begin{array}{c} 1\\\frac{-\mathrm{j} 2 \pi n}{\kappa}\\\frac{-\mathrm{j} 2 \pi p}{\kappa}
    \end{array}\right] \otimes \left[\begin{array}{c} e^{-\mathrm{j} 2 \pi(n[\tau_c]_0+p[\nu_c]_0)}\\\vdots\\e^{-\mathrm{j} 2 \pi(n[\tau_c]{L-1}+p[\nu_c]_{L-1})}
    \end{array}\right]\in\mathbb{C}^{3Q},
\end{align*}
the constituent matrices of $\mathbf{H}$ become 
\begin{align}
\mathbf{H}_1 &=\frac{1}{MP} \sum_{\widetilde{n}} g_N(n)g_P(p) \left(\bsym{\upsilon}_{\widetilde{n}} \otimes  \mathbf{b}_{n}\right)\left(\bsym{\upsilon}_{\widetilde{n}} \otimes  \bsym{\varrho}_{\widetilde{n}}\right)^{H} \nonumber\\
&=\frac{1}{MP} \sum_{\widetilde{n}} g_N(n) g_P(p) \left(\bsym{\upsilon}_{\widetilde{n}} \bsym{\upsilon}_{\widetilde{n}}^{H}\right) \otimes\left(\mathbf{b}_{n} \mathbf{b}_{n}^{H}\right),\label{H1}\end{align}\begin{align}
\mathbf{H}_2&=\frac{1}{MP} \sum_{\widetilde{n}} g_N(n) g_P(p) \left(\bsym{\upsilon}_{\widetilde{n}}  \bsym{\varrho}_{\widetilde{n}}^{H}\right) \otimes\left(\mathbf{b}_{n} \mathbf{d}_{\widetilde{n}}^{H}\right),\label{H2}\end{align}\begin{align}
\mathbf{H}_3&=\frac{1}{MP} \sum_{\widetilde{n}} g_N(n) g_P(p) \left(\bsym{\varrho}_{\widetilde{n}} \bsym{\varrho}_{\widetilde{n}}^{H}\right) \otimes\left(\mathbf{d}_{\widetilde{n}} \mathbf{d}_{\widetilde{n}}^{H}\right),\label{H3}
\end{align}

The expected values of these matrices follow from the isotropy property in \eqref{eq:isotropy} as 
$\overline{\mathbf{H}}_1=\mathbb{E}\left[\mathbf{H}_1\right]= \frac{1}{MP} \sum_{\widetilde{n}} g_N(n) g_P(p) \left(\bsym{\upsilon}_{\widetilde{n}} \bsym{\upsilon}_{\widetilde{n}}^{H}\right) \otimes \mathbf{I}_J$, $\overline{\mathbf{H}}_3=\mathbb{E}\left[\mathbf{H}_3\right]=\frac{1}{MP} \sum_{\widetilde{n}} g_N(n) g_P(p) \left(\bsym{\varrho}_{\widetilde{n}} \bsym{\varrho}_{\widetilde{n}}^{H}\right) \otimes \mathbf{I}_{PJ}$ and $\overline{\mathbf{H}}_2 = \mathbb{E}\left[\mathbf{H}_2\right]= \mathbf{0}_{3QPJ\times3LJ}$.

The expectation of the matrix $\mathbf{H}$ is 
\begin{align}
\overline{\mathbf{H}}=\mathbb{E}\left[\mathbf{H}\right]=\left[\begin{array}{ccc}
{\overline{\mathbf{H}}_1} & \overline{\mathbf{H}}_2    \\
 \overline{\mathbf{H}}_2^H   & \overline{{\mathbf{H}}}_3
\end{array}\right].
\label{eq:system_coefficients_expectation}
\end{align}
The matrix $\overline{\mathbf{H}}$ in \eqref{eq:system_coefficients_expectation} is a blockdiagonal matrix with blocks $\overline{\mathbf{H}}_1$ and $\overline{\mathbf{H}}_3$. Thus, $\overline{\mathbf{H}}$ is invertible if and only if  $\overline{\mathbf{H}}_1$ and $\overline{\mathbf{H}}_3$ are invertible. Note that $\overline{\mathbf{H}}_1$ and  $\overline{\mathbf{H}}_3$ are built upon the 2-D Fej\'er kernel. This allows us to apply the result in \cite[Lemma C.2]{candes_superresolution} that directly implies that $\overline{\mathbf{H}}_1$ and $\overline{\mathbf{H}}_3$ are invertible. The following Lemma \ref{lemma:distance H-E[H]} shows that $\mathbf{H}$ is very close to $\mathbb{E}\left[\mathbf{H}\right]$ and hence invertible with high probability. 

\begin{lemma}
\label{lemma:distance H-E[H]}
Consider the event 
$$
\mathcal{E}_{\epsilon_1}=\{   \|  \mathbf{H} - \mathbb{E}\left[\mathbf{H}\right] \| \leq \epsilon_1 \},
$$
for every real $\epsilon_1 \in (0,0.8)$. Then the event $\mathcal{E}_{\epsilon_1}$ occurs with probability $1-4\delta$ for every $\delta>0$ provided that 
$$
MP> \frac{90 \mu J}{\epsilon_1^2}\max\left(L,Q\right)\textnormal{log}\left(\max\left(\frac{6LJ}{\delta},\frac{6PQJ}{\delta}\right)\right).
$$
\label{lemma:concentration}
\end{lemma}

\begin{IEEEproof}
See Appendix~\ref{app:lemma_concentration}.
\end{IEEEproof}
From Lemma \ref{lemma:distance H-E[H]}, and the following inequalities \cite{heckel2016super} $$\|\mathbf{I}-\overline{\mathbf{H}}_1\|\leq 0.1908,$$ 
and $$\|\mathbf{I}-\overline{\mathbf{H}}_3\|\leq 0.1908,$$
 it follows that $\mathbf{H}$ is invertible. Thus, one could obtain  $\bsym{\beta}_\ell$, $\bsym{\gamma}_\ell$, $\bsym{\zeta}_\ell$, $\bsym{\eta}_q$, $\bsym{\theta}_q$, and $\bsym{\xi}_q$, $\ell=0,\cdots,L-1$, $q=0,\cdots,Q-1$ from the solution of \eqref{eq:system_coefficients}. By the construction above, the conditions \eqref{eq:cert_1} and \eqref{eq:cert_2} are \textit{ipso facto} satisfied. Next, we prove that the polynomials $\mathbf{f}_r$ and $\mathbf{f}_c$ satisfy the conditions \eqref{eq:cert_3} and \eqref{eq:cert_4}. 

\subsection{To satisfy conditions \eqref{eq:cert_3} and \eqref{eq:cert_4}}
To prove  \eqref{eq:cert_3} and \eqref{eq:cert_4} for, respectively, $\mathbf{f}_r(\mathbf{r})$ and $\mathbf{f}_c(\mathbf{c})$, we utilize the strategy initially proposed in \cite{off_the_grid}. Note again that, in our case, we build each polynomial, i.e. $\mathbf{f}_r(\mathbf{r})$ and $\mathbf{f}_c(\mathbf{c})$,  with two different random kernels. Moreover, one of the kernels $\mathbf{N}(\mathbf{r})$ is common between the two polynomials. This makes the proof in this subsection different from \cite{off_the_grid}. We show that the random dual polynomials $\mathbf{f}_r(\mathbf{r})$ and $\mathbf{f}_c(\mathbf{c})$ concentrate around, respectively, $J$ and $PJ$-dimensional form of the scalar-valued dual polynomials constructed in \cite{candes_superresolution} on a discrete set $\Omega_{\textrm{grid}} \in [0,1)^2$. Thereafter, we consider an extension to the continuous set $[0,1)^2$. Here, we show this explicitly for $\mathbf{f}_r(\mathbf{r})$. The proof for $\mathbf{f}_c(\mathbf{c})$ is similar.

Consider the polynomial
\begin{align}
\boldsymbol{\Psi}_{r}^{(m^\prime,n^\prime)}(\mathbf{r})=& \frac{1}{\kappa^{m^\prime+n^\prime}}\left[\begin{array}{c}
\mathbf{M}^{(m^\prime,n^\prime)}\left(\mathbf{r}-\mathbf{r}_{0}\right)^{H} \\
\vdots \\
\mathbf{M}^{(m^\prime,n^\prime)}\left(\mathbf{r}-\mathbf{r}_{L-1}\right)^{H} \\ \frac{1}{\kappa}\mathbf{M}^{(m^{\prime}+1,n^\prime)}\left(\mathbf{r}-\mathbf{r}_{0}\right)^{H} \\
\vdots \\
\frac{1}{\kappa} \mathbf{M}^{(m^{\prime}+1,n^\prime)}\left(\mathbf{r}-\mathbf{r}_{L-1}\right)^{H}\\
\frac{1}{\kappa} \mathbf{M}^{(m^{\prime},n^{\prime}+1)}\left(\mathbf{r}-\mathbf{r}_{0}\right)^{H}\\
\vdots\\
\frac{1}{\kappa} \mathbf{M}^{(m^{\prime},n^{\prime}+1)}\left(\mathbf{r}-\mathbf{r}_{L-1}\right)^{H}\\
\end{array}\right]   \in \mathbb{C}^{3JL \times J}
\end{align}
with statistical expectation
\begin{align}
\mathbb{E} &[\boldsymbol{\Psi}_{r}^{(m^\prime,n^\prime)}(\mathbf{r})] \\
&= \frac{1}{MP} \sum_{\widetilde{n}}  g_{M}(n)g_{P}(p)\left(\frac{-\mathrm{j} 2 \pi n}{\kappa}\right)^{m'}\left(\frac{-\mathrm{j} 2 \pi p}{\kappa}\right)^{n'} \nonumber\\
& \hspace{3.8cm} \times e^{-\mathrm{j} 2 \pi( n \tau+p\nu)} \bsym{\upsilon}_{\widetilde{n}} \otimes \mathbf{I}_{J}\nonumber\\
&=\frac{1}{\kappa^{m^\prime+n^\prime}}\underbrace{\left[\begin{smallmatrix}
\varphi_{M}^{m^\prime}\left(\tau-[\bsym\tau_r]_{0}\right)^{*}\varphi_{P}^{n^\prime}\left(\nu-[\bsym\nu_r]_{0}\right)^{*} \\\vdots\\\varphi_{M}^{m^\prime}\left(\tau-[\bsym\tau_r]_{L-1}\right)^{*}\varphi_{P}^{n^\prime}\left(\nu-[\bsym\nu_r]_{Q-1}\right)^{*} \\
\frac{1}{\kappa} \varphi_{M}^{m^\prime+1}\left(\tau-[\bsym\tau]_{0}\right)^{*}\varphi_{P}^{n^\prime}\left(\nu-[\bsym\nu]_{0}\right)^{*}  \\
\vdots \\\frac{1}{\kappa} \varphi_{M}^{m^\prime+1}\left(\tau-[\bsym\tau]_{L-1}\right)^{*}\varphi_{P}^{n^\prime}\left(\nu-[\bsym\nu]_{Q-1}\right)^{*} \\
\frac{1}{\kappa} \varphi_{M}^{m^\prime}\left(\tau-[\bsym\tau]_{0}\right)^{*}\varphi_{P}^{n^\prime+1}\left(\nu-[\bsym\nu]_{0}\right)^{*}\\\vdots\\\frac{1}{\kappa} \varphi_{M}^{m^\prime}\left(\tau-[\bsym\tau]_{L-1}\right)^{*}\varphi_{P}^{n^\prime+1}\left(\nu-[\bsym\nu]_{Q-1}\right)^{*}\\
\end{smallmatrix}\right] }_{\bsym{\psi}_r^{(m^\prime,n^\prime)}(\mathbf{r})}\otimes \mathbf{I}_{J}\nonumber \\
&\triangleq \frac{1}{\kappa^{{m^\prime+n^\prime}}}\bsym{\psi}_r^{(m^\prime,n^\prime)}(\mathbf{r})\otimes\mathbf{I}_{J}. \nonumber
\end{align}
The analogous polynomial for communications is
\begin{align}
\label{eq:Psi_c}
\boldsymbol{\Psi}_{c}^{(m^\prime,n^\prime)}(\mathbf{r}) &= \frac{1}{\kappa^{m^\prime+n^\prime}}\left[\begin{array}{c}
\mathbf{N}^{(m^\prime,n^\prime)}\left(\mathbf{r}-\mathbf{c}_{0}\right)^{H} \\
\vdots \\
\mathbf{N}^{(m^\prime,n^\prime)}\left(\mathbf{r}-\mathbf{c}_{Q-1}\right)^{H} \\ \frac{1}{\kappa}\mathbf{N}^{m^{\prime}+1,n^\prime}\left(\mathbf{r}-\mathbf{c}_{0}\right)^{H} \\
\vdots \\
\frac{1}{\kappa} \mathbf{N}^{m^{\prime}+1,n^\prime}\left(\mathbf{r}-\mathbf{c}_{Q-1}\right)^{H}\\
\frac{1}{\kappa} \mathbf{N}^{m^{\prime},n^{\prime}+1}\left(\mathbf{r}-\mathbf{c}_{0}\right)^{H}\\
\vdots\\
\frac{1}{\kappa} \mathbf{N}^{m^{\prime},n^{\prime}+1}\left(\mathbf{r}-\mathbf{c}_{Q-1}\right)^{H}\\
\end{array}\right] \nonumber\\
=& \frac{1}{MP} \sum_{\widetilde{n}} g_{M}(n)g_{P}(p)\left(\frac{-\mathrm{j} 2 \pi n}{\kappa}\right)^{m'}\left(\frac{-\mathrm{j} 2 \pi p}{\kappa}\right)^{n'} \nonumber\\
& \hspace{1.5cm} \times e^{-\mathrm{j} 2 \pi (n \tau+p\nu)}  \bsym{\varrho}_{\widetilde{n}} \otimes \mathbf{d}_{\widetilde{n}} \mathbf{b}_{n}^{H} \in \mathbb{C}^{3QPJ \times J}.
\end{align}
Using $\mathbb{E}[\mathbf{d}_{\widetilde{n}}\mathbf{b}_n]=\mathbf{0}_{JP\times J}$, we have
\begin{align}
\mathbb{E}[\bsym{\Psi}_c^{(m^\prime,n^\prime)}(\mathbf{r})]=\frac{1}{\kappa^{m^\prime+n^\prime}}\bsym{\psi}_r^{(m^\prime,n^\prime)}\otimes \mathbf{0}_{PJ\times J} = \mathbf{0}_{3JPL\times J}
\end{align}

In order to simplify the bound computation we follow the procedure in \cite{off_the_grid} and we express $\mathbf{H}^{-1}$ in terms of the matrices $\mathbf{L}_1 \in \mathbb{C}^{3LJ\times LJ}, \mathbf{R}_1 \in \mathbb{C}^{3LJ\times 2JL}\widehat{\mathbf{L}}_1 \in \mathbb{C}^{3QPJ\times LJ},\widehat{\mathbf{R}}_1 \in \mathbb{C}^{3LJ\times 2LJ}, \mathbf{L}_2 \in \mathbb{C}^{3LJ\times PJQ}, \mathbf{R}_2 \in \mathbb{C}^{3LJ\times 2PJQ}, \widehat{\mathbf{L}}_2 \in \mathbb{C}^{3QPJ\times PJQ},\widehat{\mathbf{R}}_2 \in \mathbb{C}^{3QPJ\times 2PJQ},\widehat{\mathbf{L}}_2 \in \mathbb{C}^{3QPJ\times PQJ},\widehat{\mathbf{R}}_2 \in \mathbb{C}^{3LJ\times 2PQJ}$ as follows 
$$\mathbf{H}^{-1} \triangleq \left[\begin{array}{cccc}
\mathbf{L}_1 & \mathbf{R_1} & \mathbf{L}_2 & \mathbf{R_2}\\\widehat{\mathbf{L}}_1 & \widehat{\mathbf{R}}_1 & \widehat{\mathbf{L}}_2 & \widehat{\mathbf{R}}_2
\end{array}\right].
$$
and
$$
\overline{\mathbf{H}}^{-1} = \left[\begin{array}{cccc}
\mathbf{L}_1^\prime \otimes \mathbf{I}_J& \mathbf{R}_1^\prime \otimes \mathbf{I}_J& \mathbf{0}_{3LJ\times PJQ} & \mathbf{0}_{3LJ\times 2PJQ} \\\mathbf{0}_{3QPJ\times LJ} & \mathbf{0}_{3QPJ\times 2LJ} & \widehat{\mathbf{L}}_2^\prime\otimes \mathbf{I}_{PJ}& \widehat{\mathbf{R}}_2^\prime\otimes \mathbf{I}_{PJ}
\end{array}\right].
$$ 
Then, the solutions of \eqref{eq:system_coefficients} are
\begin{align*}
\mathbf{j}_r=\left[\begin{array}{c}
\bsym{\beta} \\
{\kappa}\bsym{\gamma}\\
{\kappa}\bsym{\zeta}
\end{array}\right]=\mathbf{L}_1\widetilde{\mathbf{u}}+\mathbf{L}_2\widetilde{\mathbf{v}},
\end{align*}
and
\begin{align*}
\mathbf{j}_c=\left[\begin{array}{c}
\bsym{\eta} \\
{\kappa}\bsym{\theta}\\
{\kappa}\bsym{\xi}
\end{array}\right]=\widehat{\mathbf{L}}_1\widetilde{\mathbf{u}}+\widehat{\mathbf{L}}_2\widetilde{\mathbf{v}},
\end{align*}
where $\widetilde{\mathbf{u}}=\textrm{sign}(\bsym{\alpha}_r)\otimes\mathbf{u}$ and $\widetilde{\mathbf{v}}=\textrm{sign}(\bsym{\alpha}_c)\otimes\mathbf{v}$.

Using the expressions above, the radar polynomial becomes
\begin{align}
&\frac{1}{\kappa^{{m^\prime+n^\prime}}} \mathbf{f}^{(m^\prime,n^\prime)}_r(\mathbf{r}) \nonumber\\
& = \left[\begin{array}{cc}
     \bsym{\Psi}_r^{(m^\prime,n^\prime)}(\mathbf{r})^H  & \bsym{\Psi}_c^{(m^\prime,n^\prime)}(\mathbf{r})^H
\end{array}\right]\left[\begin{array}{c}   \mathbf{j}_r \nonumber\\
     \mathbf{j}_c 
\end{array}\right]\\
&= \underbrace{\bsym{\Psi}_r^{(m^\prime,n^\prime)}\mathbf{r})^H \mathbf{L}_1\widetilde{\mathbf{u}}}_{=\bsym\Theta_r^{(m^\prime,n^\prime)}(\mathbf{r})}+\underbrace{ \bsym{\Psi}_r^{(m^\prime,n^\prime)}\mathbf{L}_2\widetilde{\mathbf{v}}}_{=\widetilde{\boldsymbol{\Theta}}_r^{(m^\prime,n^\prime)}(\mathbf{r})} \nonumber\\
& +\underbrace{\bsym{\Psi}_c^{(m^\prime,n^\prime)}(\mathbf{r})^H  \widehat{\mathbf{L}}_1\widetilde{\mathbf{u}}}_{=\boldsymbol{\Theta}_c^{(m^\prime,n^\prime)}(\mathbf{r})}+\underbrace{\bsym{\Psi}_c^{(m^\prime,n^\prime)}(\mathbf{r})^H\widehat{\mathbf{L}}_2\widetilde{\mathbf{v}}}_{= \widetilde{\boldsymbol{\Theta}}_c^{(m^\prime,n^\prime)}(\mathbf{r})}.\label{eq:fr_non_decom}
\end{align}

Expanding the matrix $\bsym{\Theta}^{(m^\prime,n^\prime)}_r(\mathbf{r})$, we obtain
\begin{align}
&\boldsymbol{\Theta}_r^{(m^\prime,n^\prime)}(\mathbf{r}) \nonumber\\
&=\left(\bsym{\Psi}_r^{(m^\prime,n^\prime)}(\mathbf{r})-\mathbb{E} \left[\bsym{\Psi}_r^{(m^\prime,n^\prime)}(\mathbf{r})\right]\mathbb{E}\left[ \bsym{\Psi}_r^{(m^\prime,n^\prime)}(\mathbf{r})\right]\right)^{H} \nonumber\\
& \times \left(\mathbf{L}_1-\mathbf{L}_1^{\prime} \otimes \mathbf{I}_{J}+\mathbf{L}_1^{\prime} \otimes \mathbf{I}_{J}\right) \widetilde{\mathbf{u}} \nonumber\\
&=\mathbb{E}\left[ \bsym{\Psi}_r^{(m^\prime,n^\prime)}(\mathbf{r})\right]^{H}\left(\mathbf{L}_1^{\prime} \otimes \mathbf{I}_{J}\right) \widetilde{\mathbf{u}} \nonumber\\
& +\underbrace{\left(\bsym{\Psi}_r^{(m^\prime,n^\prime)}(\mathbf{r})-\mathbb{E} \left[\bsym{\Psi}_r^{(m^\prime,n^\prime)}(\mathbf{r})\right]\right)^{H} \mathbf{L}_1 \widetilde{\mathbf{u}}}_{=\mathbf{J}_{1}^{(m^\prime,n^\prime)}(\mathbf{r})} \nonumber\\&+\hphantom{=}\underbrace{\mathbb{E}\left[ \bsym{\Psi}_r^{(m^\prime,n^\prime)}(\mathbf{r})\right]^{H}\left(\mathbf{L}_1-\mathbf{L}_1^{\prime} \otimes \mathbf{I}_{J}\right) \widetilde{\mathbf{u}}}_{=\mathbf{J}_{2}^{(m^\prime,n^\prime)}(\mathbf{r})},\label{eq:decomposition_theta_r}
\end{align}
where the first term evaluates to
$$
\begin{aligned}
&\mathbb{E}\left[ \bsym{\Psi}_r^{(m^\prime,n^\prime)}(\mathbf{r})\right]^{H}\left(\mathbf{L}_1^{\prime} \otimes \mathbf{I}_{J}\right) \widetilde{\mathbf{u}}\\
&= \left(\bsym{\psi}_r^{(m^\prime,n^\prime)}(\mathbf{r}) \otimes \mathbf{I}_{J}\right)^{H}\left(\mathbf{L}_1^{\prime} \otimes \mathbf{I}_{J}\right) (\textrm{sign}(\bsym{\alpha}_r)\otimes{\mathbf{u}})\\
&=\left(\bsym{\psi}_r^{(m^\prime,n^\prime)}(\mathbf{r}) \otimes \mathbf{I}_{J}\right)^{H}\left(\mathbf{L}_1^{\prime} \textrm{sign}(\bsym{\alpha}_r) \otimes \mathbf{u}\right)\\
&=\left(\bsym{\psi}_r^{(m^\prime,n^\prime)}(\mathbf{r})^{H}\mathbf{L}_1^{\prime} \textrm{sign}(\bsym{\alpha}_r)\right)\mathbf{u}\triangleq \frac{1}{\kappa^{(m'+n')}}f^{(m',n')}(\mathbf{r})\mathbf{u},
\end{aligned}
$$
where $
 \mathbf{L}_1^{\prime} \textrm{sign}(\bsym{\alpha}_r) \triangleq  \left[\begin{array}{c}
\overline{\bsym{\beta}} \;
{\kappa}\overline{\bsym{\gamma}} \;
{\kappa}\overline{\bsym{\zeta}}
\end{array}\right]^T \in \mathbb{R}^{3L}
$ and the scalar-valued polynomial 
\begin{align}
    f^{(m^\prime, n^\prime)} &= \sum_{\ell=0}^{L-1} \varphi^{(m^\prime, n^\prime)}(\mathbf{r-r}_\ell)\overline{\bsym{\beta}}_\ell + \varphi^{(m^\prime+1, n^\prime)}(\mathbf{r-r}_\ell)\overline{\bsym{\gamma}}_\ell \nonumber\\
& \hspace{1cm} +\varphi^{(m^\prime, n^\prime+1)}(\mathbf{r-r}_\ell)\overline{\bsym{\zeta}}_\ell,
\end{align}
where $\varphi$ is the Fej\'er kernel defined in \eqref{eq:kernel_2D}.

Similarly,  
$$
\begin{aligned}
\widetilde{\boldsymbol{\Theta}}_r^{(m^\prime,n^\prime)}(\mathbf{r})=&\underbrace{\left(\bsym{\Psi}_r^{(m^\prime,n^\prime)}(\mathbf{r})^H-\mathbb{E} \bsym{\Psi}_r^{(m^\prime,n^\prime)}(\mathbf{r})^{H}\right) \mathbf{L}_2 \widetilde{\mathbf{v}}}_{\mathbf{J}_{3}^{(m^\prime,n^\prime)}(\mathbf{r})} \nonumber\\
& + \underbrace{\mathbb{E}\left[ \bsym{\Psi}_r^{(m^\prime,n^\prime)}(\mathbf{r})\right]^{H} \left(\mathbf{L}_2 \right) \widetilde{\mathbf{v}}(\mathbf{r})}_{\mathbf{J}_{4}^{(m^\prime,n^\prime)}(\mathbf{r})},
\end{aligned} 
$$

Substituting various expressions above, \eqref{eq:fr_non_decom}  becomes
\begin{align}
      \frac{1}{\kappa^{(m'+n')}}\mathbf{f}_r^{(m^\prime, n^\prime)}(\mathbf{r}) &= \frac{1}{\kappa^{m'+n'}}   f^{(m^\prime, n^\prime)}(\mathbf{r})\mathbf{u} \nonumber\\
    & +  \mathbf{J}_1^{(m^\prime,n^\prime)}(\mathbf{r}) +\mathbf{J}_2^{(m^\prime,n^\prime)}(\mathbf{r}) \nonumber\\
    &  + \mathbf{J}_3^{(m^\prime,n^\prime)}(\mathbf{r}) +\mathbf{J}_4^{(m^\prime,n^\prime)}(\mathbf{r}) \nonumber\\&+ \boldsymbol{\Theta}_c^{(m^\prime,n^\prime)}(\mathbf{r}) + \widetilde{\boldsymbol{\Theta}}_c^{(m^\prime,n^\prime)}(\mathbf{r}).
      \label{decomp_f}
\end{align}

To demonstrate that  $ \frac{1}{\kappa}\mathbf{f}_r^{(m^\prime, n^\prime)}(\mathbf{r}) $ is close to $\frac{1}{\kappa^{m'+n'}}   f^{(m^\prime, n^\prime)}(\mathbf{r})\mathbf{u}$, we prove that the spectral norms of ~$\mathbf{J}_1^{(m^\prime,n^\prime)}(\mathbf{r})$, $\mathbf{J}_2^{(m^\prime,n^\prime)}(\mathbf{r})$, $\mathbf{J}_3^{(m^\prime,n^\prime)}(\mathbf{r}),\mathbf{J}_4^{(m^\prime,n^\prime)}(\mathbf{r})$, $ \boldsymbol{\Theta}_c^{(m^\prime,n^\prime)}(\mathbf{r})$, and  ~$\widetilde{\boldsymbol{\Theta}}_c^{(m^\prime,n^\prime)}(\mathbf{r})$ are small with high probability on the set of grid points $\Omega_{\textrm{grid}}$. Compared with the single-BD approaches such as \cite{chi2016guaranteed,yang2016super} where bounding the dual polynomial requires bounding $\mathbf{J}_1^{(m^\prime,n^\prime)}(\mathbf{r})$ and $\mathbf{J}_2^{(m^\prime,n^\prime)}(\mathbf{r})$, the DBD problem requires the same on the additional terms $\mathbf{J}_3^{(m^\prime,n^\prime)}(\mathbf{r}),\mathbf{J}_4^{(m^\prime,n^\prime)}(\mathbf{r})$, $ \boldsymbol{\Theta}_c^{(m^\prime,n^\prime)}(\mathbf{r})$, and  ~$\widetilde{\boldsymbol{\Theta}}_c^{(m^\prime,n^\prime)}(\mathbf{r})$. 
Then, we show, with high probability, that $ \frac{1}{\kappa^{m'+n'}}\mathbf{f}_r^{(m^\prime, n^\prime)}(\mathbf{r}) $ is close to $\frac{1}{\kappa^{m'+n'}}   f^{(m^\prime, n^\prime)}(\mathbf{r})\mathbf{u}$ on $\Omega_{\textrm{grid}}$. Finally, we show that the polynomial constructed as above satisfies the expected condition $\|\mathbf{f}_r(\mathbf{r})\|_2<1$, $\mathbf{r}\setminus \mathcal{R}$. With a proper grid size, we extend this result for the continuous domain $[0,1]^2$.

\subsubsection{Bounds {for} $\Vert\mathbf{J}_i^{(m^\prime,n^\prime)}(\mathbf{r})\Vert_2$, where $i=1,2,3,4$, $ \Vert\boldsymbol{\Theta}_c^{(m^\prime,n^\prime)}(\mathbf{r})\Vert_2$, and  ~$\Vert\widetilde{\boldsymbol{\Theta}}_c^{(m^\prime,n^\prime)}(\mathbf{r})\Vert_2$} 
\label{section:bound_J1} 
To bound these matrices, we use the fact that they can be decomposed as a sum of random zero-mean matrices and, consequently, matrix Bernstein inequality is applicable. See similar argument in \cite{yang2016super,off_the_grid}. The following Lemmata \ref{lemma:J1_last} and \ref{lemma:J2_last}  show that $\Vert\mathbf{J}_1^{(m^\prime,n^\prime)}(\mathbf{r})\Vert_2$ and $\Vert\mathbf{J}_2^{(m^\prime,n^\prime)}(\mathbf{r})\Vert_2$ are bounded on $\Omega_{\textrm{grid}}$.

\begin{lemma}\label{lemma:J1_last}
{Assume} that $\mathbf{u} \in \mathbb{C}^{J}$ {is} sampled {on} the complex unit sphere. Let $0 < \delta < 1 $ the failure probability, and $\epsilon_2>0$. There exists a numerical constant $C$ such that if 
\begin{align*}
    MP &\geq C\mu \operatorname{max}(L,Q)J  \operatorname{max}\left\{\frac{1}{\epsilon_2^2}\log\left(\frac{\vert\Omega_{\textnormal{\textrm{grid}}}\vert JL}{\delta}\right)\log^2\left(\frac{\vert\Omega_{\textnormal{\textrm{grid}}}\vert J}{\delta}\right), \right. \nonumber\\
& \hspace{4cm} \left. \log\left(\frac{LJ}{\delta}\right),\log\left(\frac{PQJ}{\delta}\right)\right\},
\end{align*}
the following inequality is satisfied 
\begin{align*}
    \mathbb{P}\left\{\underset{\widetilde{\mathbf{r}}_d\in\Omega_{\textnormal{\text{grid}}}}{sup}\left\Vert\mathbf{J}_1^{(m^\prime,n^\prime)}(\widetilde{\mathbf{r}})\right\Vert_2\geq   \epsilon_2, m^\prime,n^\prime=0,1,2,3 \right\}\leq 48\delta.
\end{align*}
\end{lemma}
\begin{IEEEproof}
See Appendix~\ref{app:bound_J1}.
\end{IEEEproof}

\begin{lemma}\label{lemma:J2_last}
{Assume} that $\mathbf{u} \in \mathbb{C}^{J}$ {is sampled on} the complex unit sphere. Let $0 < \delta < 1 $ the failure probability, and $\epsilon_3>0$. There exists a numerical constant $C$ such that if
\begin{align*}
    MP &\geq \frac{C\mu J\operatorname{max}(L,Q)}{\epsilon_3^2} \log^2\left(\frac{\vert\Omega_{\textnormal{\textrm{grid}}}\vert J}{\delta}\right) \nonumber\\
& \times \operatorname{max}\left\{\log\left(\frac{JL}{\delta}\right),\log\left(\frac{PQJ}{\delta}\right)\right\},
\end{align*}
{the following inequality is satisfied}
\begin{align*}
    \mathbb{P}\left\{\underset{\widetilde{\mathbf{r}}_d\in\Omega_{\text{grid}}}{sup}\left\Vert\mathbf{J}_2^{(m^\prime,n^\prime)}(\widetilde{\mathbf{r}})\right\Vert_2\geq   \epsilon_3, m^\prime, n^\prime=0,1,2,3\right\}\leq 32\delta.
\end{align*}
\end{lemma}
\begin{IEEEproof}
See Appendix~\ref{app:J2_last}.
\end{IEEEproof}
Similarly, the following Lemmata \ref{lemma:J3_last} and \ref{lemma:J4_last} show that $\Vert\mathbf{J}_3^{(m^\prime,n^\prime)}(\mathbf{r})\Vert_2$, and  $\Vert\mathbf{J}_4^{(m^\prime,n^\prime)}(\mathbf{r})\Vert_2$ are bounded on $\Omega_{\textrm{grid}}$.
\begin{lemma}\label{lemma:J3_last}
{Assume} that $\mathbf{v} \in \mathbb{C}^{J}$ {is sampled on} the complex unit sphere. Let $0 < \delta < 1 $ the failure probability and $\epsilon_4>0$. There exists a numerical constant $C$ such that if
\begin{align*}
    MP &\geq C\mu \operatorname{max}(L,Q)J\operatorname{max}\left\{\frac{1}{\epsilon_4^2}\log\left(\frac{\vert\Omega_{\textnormal{\textrm{grid}}}\vert JL}{\delta}\right)\log^2\left(\frac{\vert\Omega_{\textnormal{\textrm{grid}}}\vert J}{\delta}\right), \right. \nonumber\\
& \hspace{4cm} \left.  \log\left(\frac{JL}{\delta}\right),\log\left(\frac{PQJ}{\delta}\right)\right\},
\end{align*}
{the following inequality is satisfied}
\begin{align*}
    \mathbb{P}\left\{\underset{\widetilde{\mathbf{r}}_d\in\Omega_{\textnormal{\text{grid}}}}{sup}\left\Vert\mathbf{J}_3^{(m^\prime,n^\prime)}(\widetilde{\mathbf{r}})\right\Vert_2\geq   \epsilon_4, m^\prime,n^\prime=0,1,2,3 \right\}\leq 48\delta.
\end{align*}
\end{lemma}
\begin{IEEEproof}
The proof follows \textit{mutatis mutandis} from the proof of Lemma \ref{lemma:J1_last} using the fact that $\mathbf{J}_1^{(m^\prime,n^\prime)}$ is the sum of zero-mean random matrices. 
\end{IEEEproof}

\begin{lemma}\label{lemma:J4_last}
{Assume} that $\mathbf{v} \in \mathbb{C}^{J}$ {is sampled on} the complex unit sphere. Let $0<\delta<1$ be the failure probability and $\epsilon_5>0$ a constant. There exists a numerical constant $C$ such that if
\begin{align*}
    MP \geq& C\mu \operatorname{max}(L,Q)J\operatorname{max}\left\{\log^2\left(\frac{\vert\Omega_{\textnormal{\textrm{grid}}}\vert J}{\delta}\right),\log\left(\frac{JL}{\delta}\right),\log\left(\frac{PQJ}{\delta}\right)\right\},
\end{align*}
{the following inequality is satisfied}
\begin{align*}
    \mathbb{P}\left\{\underset{\widetilde{\mathbf{r}}_d\in\Omega_{\textnormal{\textrm{grid}}}}{sup}\left\Vert\mathbf{J}_4^{(m^\prime,n^\prime)}(\widetilde{\mathbf{r}})\right\Vert_2\geq   \epsilon_5, m^\prime, n^\prime=0,1,2,3 \right\}\leq 32\delta.
\end{align*} 
\end{lemma}
\begin{IEEEproof}
The proof follows \textit{mutatis mutandis} from the proof of Lemma \ref{lemma:J2_last} using the fact that $\mathbf{J}_2^{(m^\prime,n^\prime)}$ is the sum of zero-mean random matrices. 
\end{IEEEproof}

Finally, Lemmata \ref{lemma:theta_last} and \ref{lemma:theta_tilde_last} show that $ \Vert\boldsymbol{\Theta}_c^{(m^\prime,n^\prime)}(\mathbf{r})\Vert_2$, and  ~$\Vert\widetilde{\boldsymbol{\Theta}}_c^{(m^\prime,n^\prime)}(\mathbf{r})\Vert_2$ are bounded on $\Omega_{\textrm{grid}}$.
\begin{lemma}\label{lemma:theta_last}
{Assume} that $\mathbf{v} \in \mathbb{C}^{PJ}$ {is sampled on} the complex unit sphere. Let $0<\delta<1$ be the failure probability and $\epsilon_6>0$ a constant. There exists a numerical constant $C$ such that if
\begin{align*}
    MP \geq & C\mu \operatorname{max}(L,Q)J\operatorname{max}\left\{\frac{1}{\epsilon_6^2}\log\left(\frac{\vert\Omega_{\textnormal{\textrm{grid}}}\vert PQJ}{\delta}\right)\log^2\left(\frac{\vert\Omega_{\textnormal{\textrm{grid}}}\vert J}{\delta}\right), \right. \nonumber\\
& \hspace{4cm} \left. \log\left(\frac{LJ}{\delta}\right),\log\left(\frac{PQJ}{\delta}\right)\right\},
\end{align*}
{then, the following inequality is satisfied}
\begin{align*}
    \mathbb{P}\left\{\underset{\widetilde{\mathbf{r}}_d\in\Omega_{\textnormal{\textrm{grid}}}}{sup}\left\Vert{\bsym{\Theta}}_c^{(m^\prime,n^\prime)}(\widetilde{\mathbf{r}})\right\Vert_2\geq   \epsilon_6, m^\prime, n^\prime=0,1,2,3 \right\}\leq 48\delta.
\end{align*}
\end{lemma}
\begin{IEEEproof}
See Appendix~\ref{app:bound_theta}.
\end{IEEEproof}

\begin{lemma}\label{lemma:theta_tilde_last}
{Assume} that $\mathbf{v} \in \mathbb{C}^{PJ}$ {is sampled on} the complex unit sphere. Let $0<\delta<1$ be the failure probability and $\epsilon_6>0$ a constant. There exists a numerical constant $C$ such that if
\begin{align*}
    MP \geq& C\mu \operatorname{max}(L,Q)J\operatorname{max}\left\{\log^2\left(\frac{\vert\Omega_{\textnormal{\textrm{grid}}}\vert J}{\delta}\right),\log\left(\frac{LJ}{\delta}\right),\log\left(\frac{PQJ}{\delta}\right)\right\},
\end{align*}
{the following inequality is satisfied}
\begin{align*}
    \mathbb{P}\left\{\underset{\widetilde{\mathbf{r}}_d\in\Omega_{\textnormal{\textrm{grid}}}}{sup}\left\Vert\widetilde{\bsym{\Theta}}_r^{(m^\prime,n^\prime)}(\widetilde{\mathbf{r}})\right\Vert_2\geq   \epsilon_7, m^\prime, n^\prime=0,1,2,3 \right\}\leq 32\delta.
\end{align*}
\end{lemma}
\begin{IEEEproof}
The proof is obtained, \textit{mutatis mutandis}, from that of Lemma~\ref{lemma:theta_last}.
\end{IEEEproof}

\subsubsection{$ \frac{1}{\kappa}\mathbf{f}_r^{(m^\prime, n^\prime)} $ is close to $\frac{1}{\kappa^{m'+n'}}   f^{(m^\prime, n^\prime)}\mathbf{u}$ on $\Omega_{\textnormal{\textrm{grid}}}$}
We obtained this as a consequence of the following Lemma~\ref{lemma:bound_f_f_tilde_discrete}.
\begin{lemma}\label{lemma:bound_f_f_tilde_discrete}
Assume that $\Omega_{\textrm{grid}} \in [0,1)^2$ is a finite set of points. Let $0<\delta<1$ be the failure probability and $\epsilon>0$ a constant. Then, there exists a numerical constant $C$ such that if\par\noindent\small
\begin{align*}
    MP &\geq  C\mu \operatorname{max}(L,Q)J\operatorname{max}\left\{\frac{1}{\epsilon^2}\log\left(\frac{\vert\Omega_{\textnormal{\textrm{grid}}}\vert PQJ}{\delta}\right), \right. \nonumber\\
& \hspace{3cm} \left.   \frac{1}{\epsilon^2}\log\left(\frac{\vert\Omega_{\textnormal{\textrm{grid}}}\vert LJ}{\delta}\right), \log^2\left(\frac{\vert\Omega_{\textnormal{\textrm{grid}}}\vert J}{\delta}\right), \right. \nonumber\\
& \hspace{3cm} \left. \log\left(\frac{LJ}{\delta}\right),  \log\left(\frac{PQJ}{\delta}\right)\right\},
\end{align*}\normalsize
then 
$$
\mathbb{P}[\mathcal{E}_{\epsilon}]\geq 1 - \delta,
$$
where
\begin{align*}
 \mathcal{E}_{\epsilon}&=\left\{\underset{\widetilde{\mathbf{r}}_d\in\Omega_{\textnormal{\textrm{grid}}}}{\sup}\left\|\frac{1}{\kappa^{m'+n'}}\mathbf{f}_r^{(m^\prime, n^\prime)} (\widetilde{\mathbf{r}})-\frac{1}{\kappa^{m'+n'}}   f^{(m^\prime, n^\prime)}(\widetilde{\mathbf{r}})\mathbf{u}\right\|_2\leq \frac{\epsilon}{3}, \right. \nonumber\\
& \hspace{6cm} \left. 
m^\prime,n'=0,1,2,3\right\},
\end{align*}
\end{lemma}
\begin{IEEEproof}
Using Lemmata \ref{lemma:J1_last}-\ref{lemma:theta_tilde_last}, it directly follows that the event $\mathcal{E}_{\epsilon}$ holds with a high probability. This completes the proof. 
\end{IEEEproof}

\subsubsection{$\|\mathbf{f}_r(\mathbf{r})\|_2<1$ for $\mathbf{r} \in [0,1]^2 \setminus \mathcal{R}$}
Define the minimum separation condition for the off-the-grid delay and Doppler parameters as, respectively,
\begin{align}
    \max\left( \left\vert [\bsym{\tau
    }_r]_{\ell} - [\bsym\tau]_r]_{\ell^\prime}\right\vert,\left\vert\bsym{\tau
    }_c]_{q} - [\bsym\tau]_c]_{q^\prime}\right\vert\right) \geq \Delta_\tau,  \label{delta_m}
\end{align}
and
\begin{equation}
    \max\left( \left\vert [\bsym{\nu
    }_r]_{\ell} - [\bsym\nu]_r]_{\ell^\prime}\right\vert,\left\vert[\bsym\nu_c]_{q} - [\bsym\nu_c]_{q^\prime}\right\vert\right)\geq \Delta_\nu, \label{delta_p}
\end{equation}
where $\ell \neq\ell^\prime, q \neq q^{\prime}$, and $\vert a-b\vert$ is the wrap-around metric on [0,1), e.g. $|0.2-0.1| = 0.1$ but $|0.8-0.1|=0.3$. The following lemma shows that $ \frac{1}{\kappa}\mathbf{f}_r^{(m^\prime, n^\prime)} $ is close to $\frac{1}{\kappa^{m'+n'}}   f^{(m^\prime, n^\prime)}\mathbf{u}$ everywhere on $[0,1)^2$

\begin{lemma}\label{lemma:certificate_less_1_everywhere}
Assume that $\Delta_\tau \leq 1/M$ and $\Delta_\nu \leq 1/P$. Let $0<\delta<1$ the failure probability and $\epsilon>0$. Then, for all $\mathbf{r}=[\tau,\nu]$ and $m^\prime,n'=0,1,2,3$, there exists a numerical constant C such that if 
\begin{align*}
    &MP \geq  C\mu \operatorname{max}(L,Q)J\operatorname{max}\left\{\frac{1}{\epsilon^2}\log\left(\frac{MP QJ}{\delta}\right),\right.\\&\frac{1}{\epsilon^2}\log\left(\frac{MP LJ}{\delta}\right),\log^2\left(\frac{MP J}{\delta}\right),\left.\log\left(\frac{LJ}{\delta}\right),\log\left(\frac{PQJ}{\delta}\right)\right\},
\end{align*}
the following inequality is satisfied
$$
\left\|\frac{1}{\kappa^{m'+n'}}\mathbf{f}_r^{(m^\prime, n^\prime)}(\mathbf{r}) -\frac{1}{\kappa^{m'+n'}}   f^{(m^\prime, n^\prime)}(\mathbf{r})\mathbf{u}\right\|_2\leq \frac{\epsilon}{3}$$
\end{lemma}
\begin{IEEEproof}
See Appendix~\ref{app:certificate_less_1_everywhere}.
\end{IEEEproof}

The following lemma states that $\|\mathbf{f}_r(\mathbf{r})\|_2<1$ with a high probability.
\begin{lemma}\label{lemma:certificate_less_1}
There exists a constant $C$ such that if 
\begin{align}
\label{eq:sample_complexity}
    &MP \geq  C\mu \operatorname{max}(L,Q)J\log^2\left(\frac{MP J}{\delta}\right)\nonumber\\
    &\operatorname{max}\left\{\log\left(\frac{MPQJ}{\delta}\right),\log\left(\frac{MPLJ}{\delta}\right)\right\},
\end{align}
then with probability $1-\delta$, where $0<\delta<1$, $\|\mathbf{f}_r(\mathbf{r})\|_2<1$, for all $\mathbf{r} \in [0,1]^2 \setminus \mathcal{R}$.
\end{lemma}
\begin{IEEEproof}
The proof requires showing $\|\mathbf{f}_r(\mathbf{r})\|_2<1$ in the set of points near and far from $\mathbf{r}_j \in \mathcal{R}$. We refer the reader to \cite[Lemma 14]{suliman2021mathematical} for details, which proved an analogous result for single-BD with multi-dimensional bivariate dual polynomials. In our DBD problem, the dual polynomials are also multi-dimensional bi-variate but contain cross-terms (see \eqref{decomp_f}). Following the procedure in \cite[Lemma 14]{suliman2021mathematical} but instead using Lemma~\ref{lemma:certificate_less_1_everywhere} for our specific polynomial with $\epsilon=0.0005$ concludes the proof. 
\end{IEEEproof}

\subsection{Proof of the theorem}
Using the aforementioned results, a dual polynomial can be constructed to satisfy \eqref{eq:cert_1}, \eqref{eq:cert_2}, \eqref{eq:cert_3}, and \eqref{eq:cert_4} as long as the condition on $MP$ as in 
\eqref{eq:sample_complexity} holds. In particular, Lemma \ref{lemma:distance H-E[H]} proves $\eqref{eq:cert_1}$ and $\eqref{eq:cert_2}$ by showing that the linear system of equations in \eqref{eq:system_coefficients} is invertible with high probability. Conditioned on this event, Lemmata \ref{lemma:J1_last}-\ref{lemma:theta_tilde_last} lead to Lemma \ref{lemma:bound_f_f_tilde_discrete} showing a measure concentration of $\mathbf{f}_r(\mathbf{r})$ over $ f^{(m^\prime, n^\prime)}\mathbf{u}$ on $\Omega_{\textrm{grid}}$. Finally, this result is extended in Lemma \ref{lemma:certificate_less_1} for the continuous domain $[0,1]^2$ thereby proving that the polynomial satisfies \eqref{eq:cert_3}. The proof for satisfying the condition \eqref{eq:cert_4} for the communications polynomial $\mathbf{f}_c(\mathbf{c})$ follows a similar procedure as above for the radar polynomial. This completes the proof of Theorem \ref{th:main}.

\section{Numerical Experiments}
\label{sec:results}
We evaluated our model and methods through numerical experiments using the CVX SDP3 \cite{grant2009cvx} solver for the problem in \eqref{dual_opt}. Throughout all experiments, we generated the columns of the transformation matrices $\mathbf{B}$ and $\mathbf{D}_p$ as $\mathbf{b}_n= [1, e^{\mathrm{j}2\pi \sigma_n}, \dots, e^{\mathrm{j}2\pi(J-1) \sigma_n}]$, where $\sigma_n \sim \mathcal{N}(0,1)$. The parameters $\bsym{\alpha}_r$ and $\bsym{\alpha}_c$ were drawn from a normal distribution with $|[\bsym{\alpha}_r]_\ell|=|[\bsym{\alpha}_c]_q|=1$, $\forall q$ and 
$\ell$. The real and imaginary components of the coefficients vectors $\mathbf{u}$ and $\mathbf{v}$ followed a uniform distribution over the interval $[0,1]$. 
\begin{figure*}[!t]
    \centering
    \includegraphics[width=0.95\textwidth]{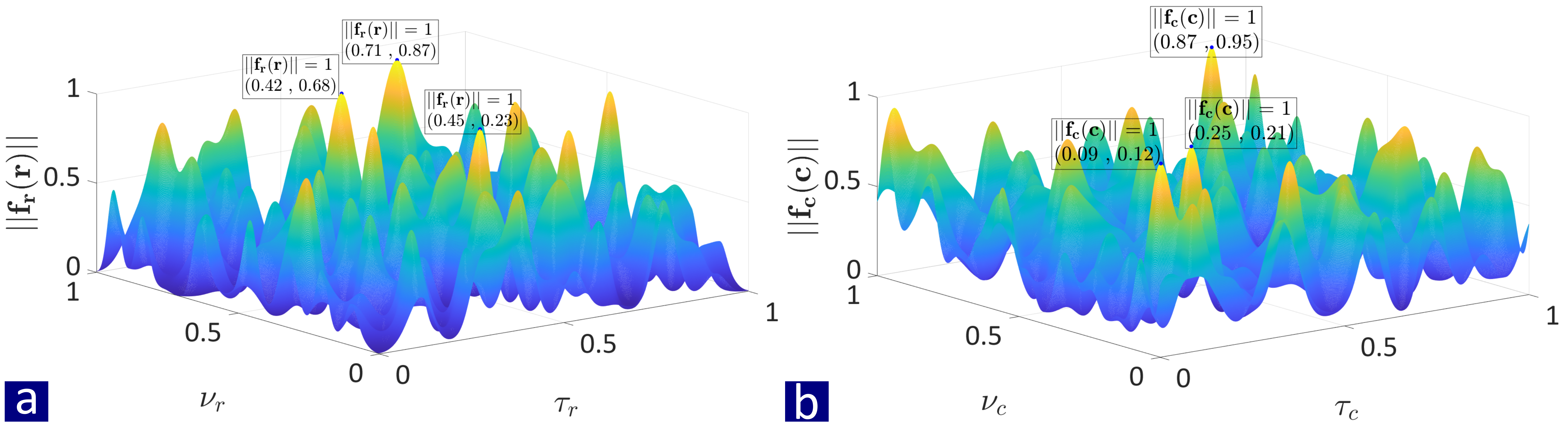}
    \caption{Noiseless channel parameter localization for (a) radar with $L=3$ targets and (b) communications with $Q=3$ paths. Here, $M = 13$, $P=9$, and $J=3$. The parameter estimates are localized in the delay-Doppler plane when the norm of the polynomials assumes a value of unity. 
    }
    \label{fig:results_dual}
\end{figure*}
\begin{figure*}[!t]
    \centering
    \includegraphics[width=0.95\textwidth]{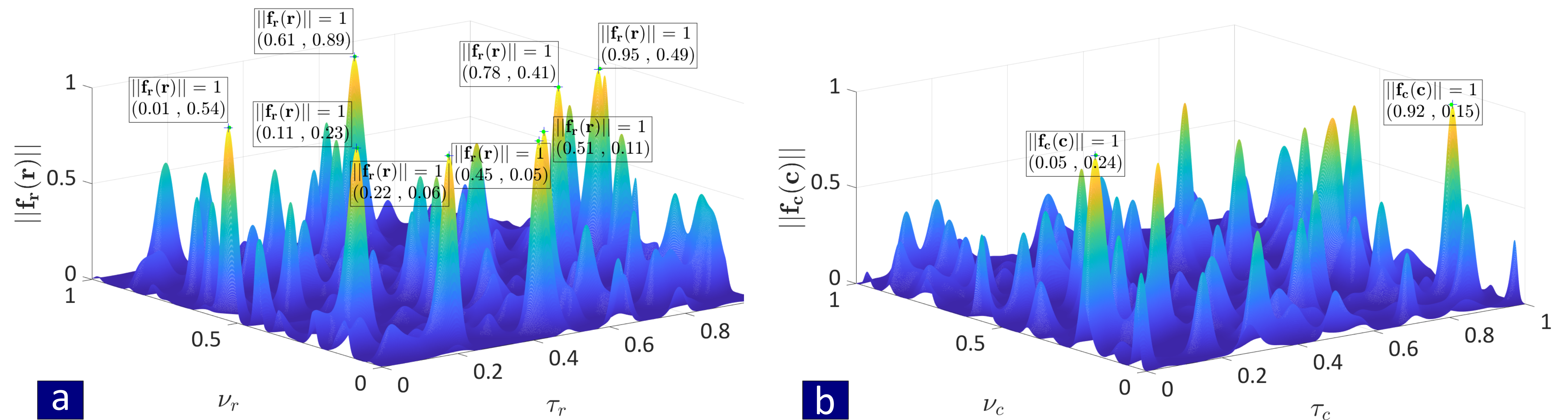}
    \caption{As in Fig.~\ref{fig:results_dual} but for (a) $L=8$ targets and (b) $Q=2$ paths.
    } 
    \label{fig:q_l}
\end{figure*}
\begin{figure*}[t]
    \centering
    \includegraphics[width=0.95\textwidth]{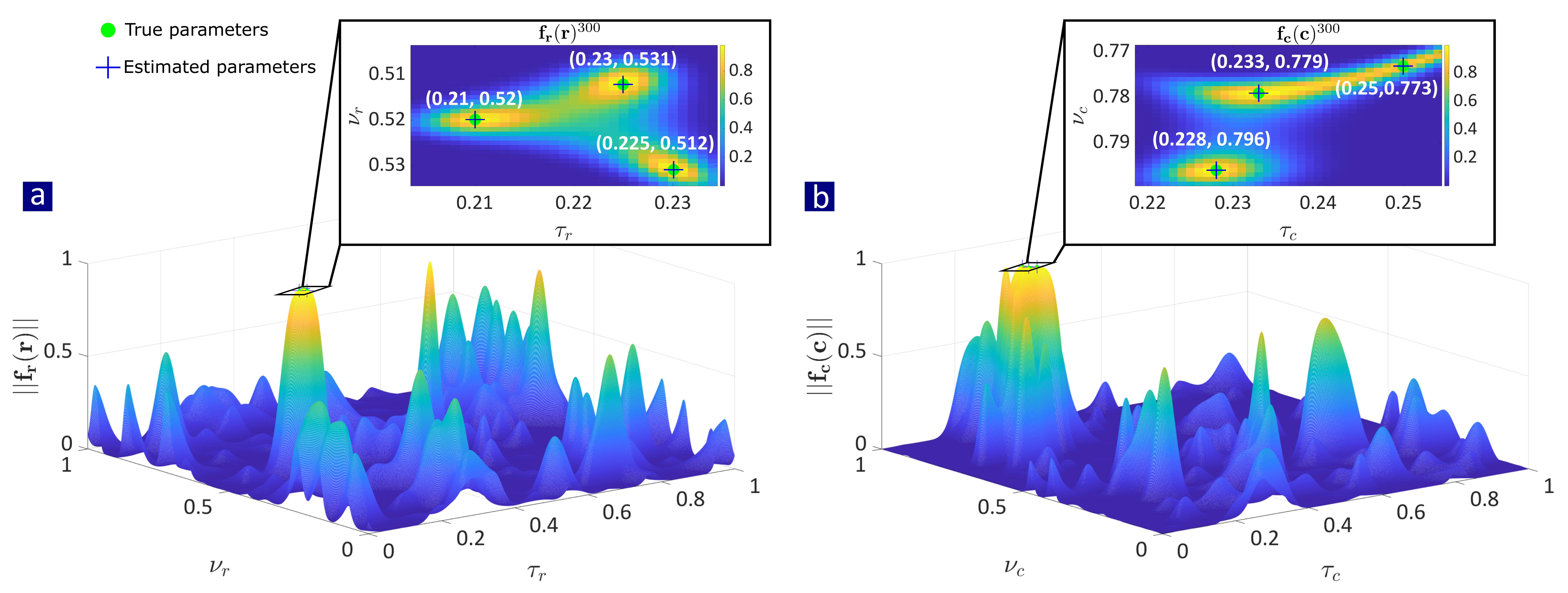}
    \caption{As in Fig.~\ref{fig:results_dual} but for closely spaced targets. The inset plots show the cross-section of PhTPs in the delay-Doppler plane.}
    \label{fig:close}
\end{figure*}

\subsection{Channel parameter estimation}
We first evaluated our approach for specific realizations of various channel conditions. 
\\
\noindent\textbf{Randomly spaced targets}: Fig.~\ref{fig:results_dual} illustrates a simple recovery scenario with number of samples $M = 13$, pulses/messages $P=9$, radar targets $L=3$, communications paths $Q=3$, and size of the low-dimensional subspace $J=3$. The target delays and Dopplers were drawn from the interval $[0,1]$ uniformly at random (without replacement). In particular, we used $\bsym{\tau}_r = [0.23, 0.68,0.87]$, $\bsym{\nu}_r = [0.45, 0.42,0.71]$, $\bsym{\tau}_c = [0.12, 0.21,0.95]$, and $\bsym{\nu}_c = [0.09, 0.25,0.87] $. Fig. ~\ref{fig:results_dual} shows the resulting dual polynomials $\mathbf{f}_r(\mathbf{r})$ and $\mathbf{f}_c(\mathbf{c})$ after solving \eqref{dual_opt}. The estimates of the channel parameters of radar and communications are located at $\Vert\mathbf{f}_r(\mathbf{r})\Vert = 1$ and $\Vert\mathbf{f}_c(\mathbf{c})\Vert = 1$. The SDP in \eqref{dual_opt} perfectly recovers the delay-Doppler pairs of both channels. \\
\noindent\textbf{Unequal number of targets and paths}: Fig.~\ref{fig:q_l} shows the perfect recovery of all parameters for a scenario with $L\neq Q$. In particular, we set $L=8$, $Q=2$, $M=17$ and $P=11$. The delay-Doppler parameters were drawn uniformly at random as before so that $\bsym{\tau}_r = [0.54,0.11,0.49,0.89,0.05, 0.23, 0.41,0.06]$, $\bsym{\nu}_r = [0.01,0.51,0.95,0.61,0.45, 0.11,0.78,0.22]$, $\bsym{\tau}_c = [0.15,0.24]$, and $\bsym{\nu}_c = [0.92,0.05]$.  \\
\noindent\textbf{Closely spaced targets}: Fig.~\ref{fig:close} depicts perfect recovery in a scenario when the delay-Doppler pairs of both channels are closely located in the $[0,1]^2$ domain with $M=17$, $P=11$, and the subspace size $J=5$. Here, the minimum separation of the time-delays, and Doppler parameter were set to $\Delta_\tau = \frac{0.1}{M}$ and $\Delta_\nu = \frac{0.1}{P}$ respectively,  which is much lower than theoretical results (see \cite{heckel2016super,off_the_grid}). For this experiment, the maximum separation was set to $\frac{0.3}{M}$ and $\frac{0.3}{P}$ 
$\bsym{\tau}_r = [0.520, 0.512, 0.531]$, $\bsym{\nu}_r = [0.210, 0.225, 0.230]$ and $\bsym{\tau}_c = [0.779,0.796,  0.773]$, and $\bsym{\nu}_c = [0.233, 0.228, 0.25]$.  
\begin{figure*}[t]
    \centering
    \includegraphics[width=\linewidth]{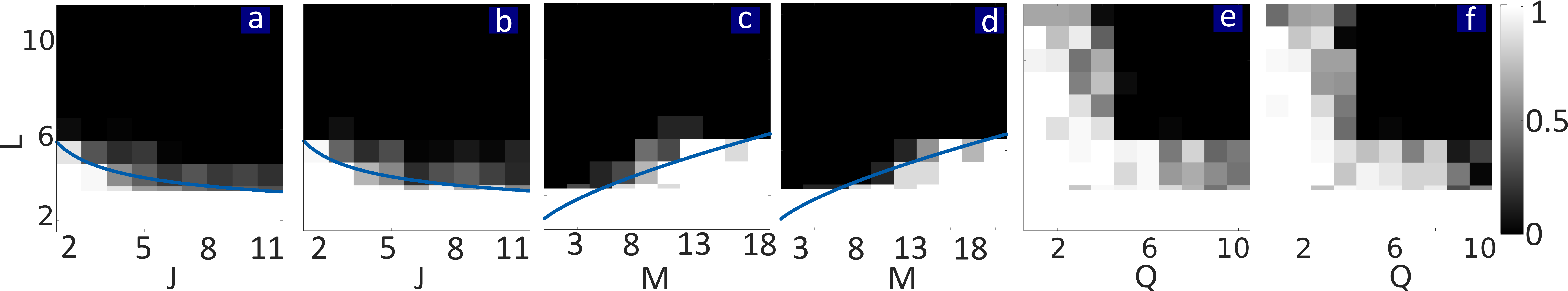}
    \caption{Probability of successfully recovering 
    (a) radar channel parameters and (b) communications messages for varying $L$ ($=Q$) and $J$ but fixed $M=13$ and $P=9$ over $100$ trials. (c)-(d) As in (a)-(b), respectively, but for fixed $P=11$ and $J=3$. (e)-(f) As in (a)-(b), respectively, but for fixed $M=13$, $P=11$, and $J=3$. {In  Figs. (a)-(d), the blue curve shows the theoretical result as predicted by Theorem 2.}}
    \label{fig:statistics}%
\end{figure*}

\subsection{Statistical performance}
As we remarked earlier, radar transmit signals and communications channels are usually known to the traditional receivers. In Figs.~\ref{fig:statistics}, we compare the statistical performance of our algorithm in recovering the conventional parameters-of-interest, i.e., radar channel $\mathbf{r}$ and communications transmit messages $\mathbf{g}$. The comparison is over several random realizations of delay and Doppler pairs with the minimum separation $\Delta_\tau = \frac{1}{M}$ and $\Delta_\nu = \frac{1}{P}$. 
We declare a successful estimation when $\Vert \mathbf{r} - \widehat{\mathbf{r}}\Vert_2<10^{-3}$ and $\Vert \mathbf{g} - \widehat{\mathbf{g}}\Vert_2<10^{-3}$. Note that the error in communications messages is obtained by concatenating all $P$ messages in a single vector. The number of targets $L$ and propagation paths $Q$ were the same. 
For fixed $M=13$ and $P=9$, Figs.~\ref{fig:statistics}a and b plot the probability of success over {$100$} trials with varying $L=Q$ and subspace size $J$ to recover radar channel and communications messages, respectively. We note an inverse relationship between $J$ and $L=Q$, as also predicted by Theorem 2, in which the theoretical curve is superposed on the figure. Figs.~\ref{fig:statistics}c and d show the recovery of the same radar and communications parameters for fixed $P=9$ but varying $L$ and time samples $M$. {Here, as suggested by Theorem~2, a logarithmic relationship is observed between $M$ and $L=Q$. This is validated by plotting the curve obtained by the right-hand side expression of Eq. \eqref{eq:th_main} in Theorem 2 on Figs~\ref{fig:statistics}a-d.} Finally, Figs.~\ref{fig:statistics}e and f show the probability of successful recovery for fixed $M=13$, $P=11$, and $J=3$ but varying $Q$ and $L$. 
\section{Summary}
\label{sec:summ}
We investigated the joint radar-communications processing from a DBD perspective. The proposed approach leverages the sparse structure of the channels and recovers the parameters of interest in this highly ill-posed problem by minimizing a SoMAN objective. We employed PhTP theories to formulate an SDP problem. Numerical experiments showed perfect recovery of radar and communications parameters in several different scenarios. 

A lot more system variables may also be considered unknown in further extensions of this problem. From a practical standpoint, some parameters such as the PRT $T$ may be accurately estimated by the common receiver over a few observations through, for instance, periodicity estimation algorithms \cite{klapuri2003multiple}. Knowing the PRT also yields an estimate of the number of pulses. In this work, we did not consider the continuous-wave transmission model for radar or even a common waveform for co-design radar-communications systems. These are interesting problems for future work. Similarly, we excluded a multiple-antenna receiver in this paper. But some of our initial results on multi-antenna DBD have recently appeared in \cite{jacome2022multid}.

Solving the linear system in (\ref{recover_gs}) provides estimates of the transmitted encoded OFDM symbols, which remain to be decoded. 
This may be achieved by the conventional blind decoding (BdC) and blind equalization (BE) algorithms, on which a large body of literature exists \cite{ghosh1998blind,via2008blind,dean2018blind}. For instance, in the case of BdC, blind selection mapping methods  \cite{joo2012new,jayalath2005slm} have been shown to recover the original symbol sequence using a maximum likelihood decoder without requiring additional side-information transmission. 
The BE techniques are used for joint channel estimation and symbol detection. For instance, \cite{de1996blind} leverages null symbols in a linear equalizer to minimize a quadratic criterion. When the channel changes on a symbol-by-symbol basis, \cite{al2012low} has proposed a BE algorithm to recover the transmitted symbols without requiring any statistical prior on the symbols.

\appendices
\section{Practical Issues}
\label{app:generalization}
We now discuss some practical impairments and adapt our DBD problem in response to those considerations.

\subsection{Lack of synchronized transmission}
In practice, the radar and communications do not transmit signals at the same time as shown in Fig.~\ref{fig:sync_fig}. It may also be difficult to synchronize the transmission if each system employs a different clock. This leads to an additional unknown time delay between radar and communications signal arrivals at the common receiver. Note that this delay is constant across all pulses/frames in the observation interval. In non-blind systems, this synchronization error is usually estimated through a variety of techniques \cite{karthik2020recent} before processing the received signal. 

\begin{figure}[t]
    \centering
    \includegraphics[width=1.0\columnwidth]{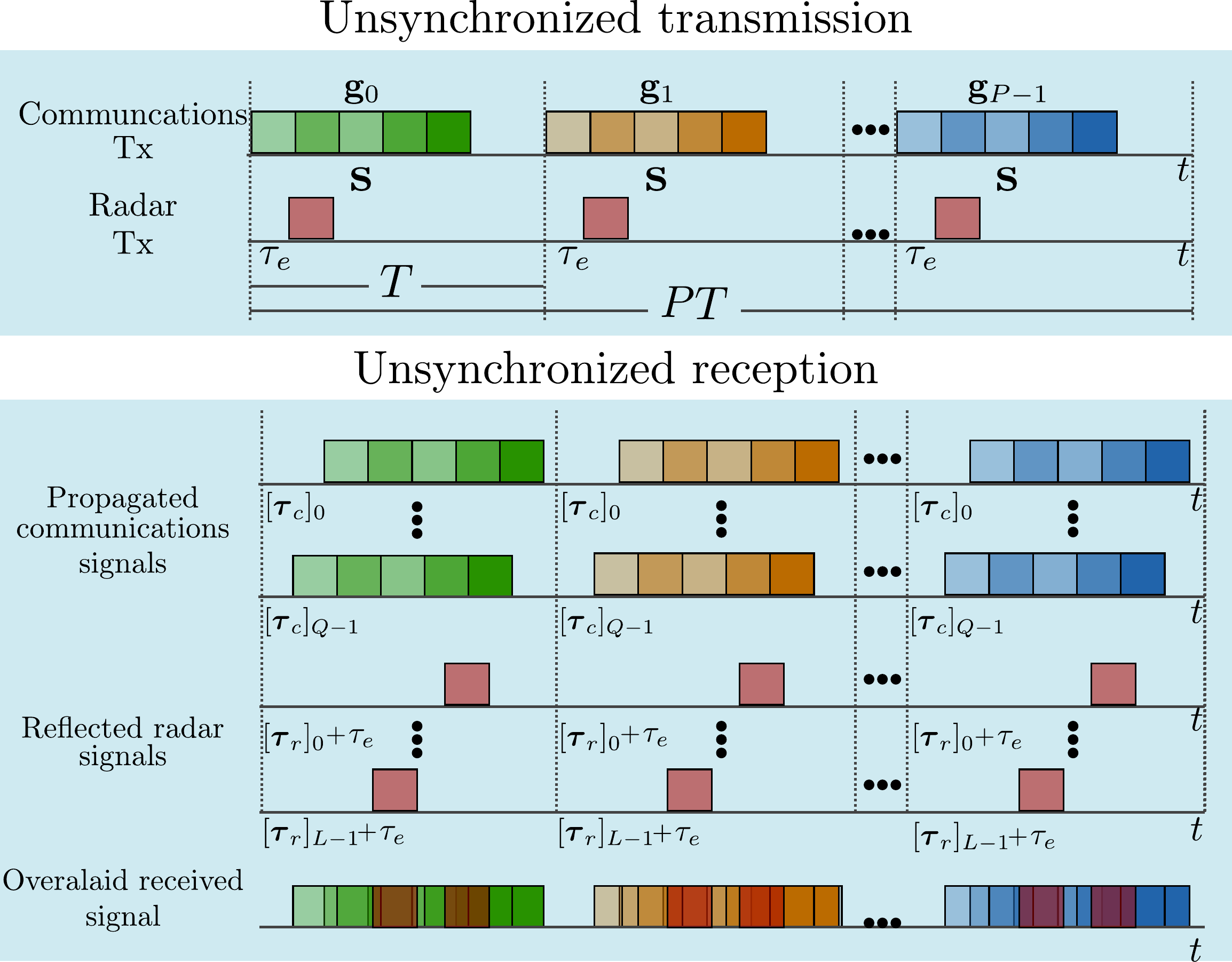}
    \caption{Sequence of transmission and reception of radar and communications signals when the transmission of the radar signal lags behind that of the communications by an unknown delay $\tau_e$.}
    \label{fig:unsync_fig}
\end{figure}

Assume that the radar transmit signal $x_r(t)$ is delayed by a continuous-valued unknown time-lag $\tau_e$ with respect to the communications signals (Fig. \ref{fig:unsync_fig}). Consequently, the receive signal in  \eqref{signal_2} becomes
\begin{align}
    [\mathbf{y}]_{\widetilde{n}} = \mathbf{h}_r^H\mathbf{E} \mathbf{e}_{\widetilde{n}} \mathbf{b}_n^H \mathbf{u} + \mathbf{h}_c^H\mathbf{e}_{\widetilde{n}}\mathbf{d}_{\widetilde{n}}^H\mathbf{v},\nonumber
\end{align} 
where $\mathbf{E} = \operatorname{diag}(\mathbf{1}_{P,1}\otimes\widetilde{\mathbf{e}}) $, $\mathbf{1}_{P,1}$ is an all-ones vector of length $P$ and $\widetilde{\mathbf{e}}(\tau_s) = [e^{\mathrm{j}2\pi\tau_s (-N)},\dots, e^{\mathrm{j}2\pi\tau_s (N)}]$. Moreover, we consider that $\Vert\mathbf{E}\Vert\leq C_e$. Define the modified radar waveform-channel matrix $\widetilde{\mathbf{Z}}_r=\mathbf{u}\mathbf{h}_r^H\mathbf{E}$ with the corresponding atomic set 
\begin{align}
    &\widetilde{\mathcal{A}}_r = \Big\{\mathbf{u}\mathbf{a}(\mathbf{r})^H\mathbf{E}: \mathbf{r}\in[0,1)^2,||\mathbf{u}||_2 = 1, \Vert\mathbf{E}\Vert\leq C_2   \Big\} \;\subset \mathbb{C}^{J\times MP},
    \label{eq:atomic_set_rad_error}
\end{align}
and its corresponding atomic norm
\begin{align}
    &||\widetilde{\mathbf{Z}}_r||_{\mathcal{A}_r} = \inf_{\stackrel{\alpharl \in \mathbb{C}, \boldsymbol{r}_\ell \in [0,1]^2}{||\mathbf{u}||_2 = 1,\Vert\mathbf{E}\Vert\leq C_2}} \Bigg\{\sum_\ell |\alpharl| \Big| \widetilde{\mathbf{Z}}_r = \sum_\ell \alpharl\mathbf{u}\mathbf{a}(\mathbf{r}_\ell)^H\mathbf{E}\Bigg\}.
    \label{eq:atomic_norm_rad_error}
\end{align}
The communications waveform-channel matrix remains the same as in \eqref{eq:comms_ch_vec}. 

Following the atomic norm minimization framework of \cite{chen2020new}, we formulate the optimization problem as
\begin{align}
    &\minimize_{\widetilde{\mathbf{Z}}_r,\mathbf{Z}_c} \Vert\mathbf{y}-\aleph(\widetilde{\mathbf{Z}}_r)  -\aleph(\mathbf{Z}_c)  \Vert_2^2 +\rho\Vert\widetilde{\mathbf{Z}}_r\Vert_{\mathcal{A}_r} +\Vert\mathbf{Z}_c\Vert_{\mathcal{A}_c},\nonumber
\end{align}
where $\rho$ is a regularization parameter that, as suggested in  \cite[Proposition 4]{chen2020new}, is chosen depending on the lag in synchronization. The corresponding Lagrangian function is
\begin{align}
    \mathcal{L}(\widetilde{\mathbf{Z}}_r,\mathbf{Z}_c,\mathbf{q},\mathbf{x}) &= \Vert\mathbf{x}-\mathbf{y}\Vert + (\rho\Vert\widetilde{\mathbf{Z}}_r\Vert_{\mathcal{A}_r}+\Vert\mathbf{Z}_c\Vert_{\mathcal{A}_c}) \nonumber\\
& \hspace{1.6cm} + \langle\mathbf{x}-\aleph(\widetilde{\mathbf{Z}}_r)_r-\aleph(\mathbf{Z}_c)_c,\mathbf{q}\rangle.\nonumber
\end{align}

This yields the dual problem as 
\begin{equation}
    \maximize_{\mathbf{q}}\min_{\widetilde{\mathbf{Z}}_r,\mathbf{Z}_c,\mathbf{x}}= \mathcal{L}(\widetilde{\mathbf{Z}}_r,\mathbf{Z}_c,\mathbf{q},\mathbf{x})=\max_\mathbf{q}\{\mathcal{L}_1(\mathbf{q})-\mathcal{L}_2(\mathbf{q})\},\nonumber
\end{equation}
where $\mathcal{L}_1 = \min_\mathbf{x}\frac{1}{2}\Vert\mathbf{x-y}\Vert_2^2+\langle\mathbf{x},\mathbf{q}\rangle$ or, equivalently, 
\begin{equation}
    \mathcal{L}_1(\mathbf{q}) = \Vert\mathbf{q}-\mathbf{y}\Vert_2^2  + \frac{1}{2}\Vert\mathbf{y}\Vert_2^2,\nonumber
\end{equation}
and $\mathcal{L}_2(\mathbf{q}) = \max_{\widetilde{\mathbf{Z}}_r,\mathbf{Z}_c}\langle\aleph(\widetilde{\mathbf{Z}}_r)_r+\aleph(\mathbf{Z}_c)_c,\mathbf{q}\rangle -(\rho\Vert\widetilde{\mathbf{Z}}_r\Vert_{\mathcal{A}_r}+\Vert\mathbf{Z}_c\Vert_{\mathcal{A}_c}) $
 or, equivalently,
\begin{equation}
    \mathcal{L}_2(\mathbf{q}) = \max_{\widetilde{\mathbf{Z}}_r,\mathbf{Z}_c}(\langle\widetilde{\mathbf{Z}}_r,\aleph^*_r(\mathbf{q})\rangle-\rho\Vert\widetilde{\mathbf{Z}}_r\Vert_{\mathcal{A}_r})+(\langle\mathbf{Z}_c,\aleph^*_c(\mathbf{q})\rangle-\Vert\mathbf{Z}_c\Vert_{\mathcal{A}_c}).\nonumber
\end{equation}
It follows from the definition of the dual norm that
\begin{equation}
    \mathcal{L}_2(\mathbf{q}) = \max_{\widetilde{\mathbf{Z}}_r,\mathbf{Z}_c}(I_r(\Vert\mathbf{q}\Vert^*_{\widetilde{\mathcal{A}}_r}\leq \rho)+I_c(\Vert\mathbf{q}\Vert^*_{\mathcal{A}_c}\leq 1)),\nonumber
\end{equation}
where 
\begin{equation}
    I_r(\Vert\mathbf{q}\Vert^*_{\widehat{\mathcal{A}}_r}\leq \rho) =  \left\{\begin{array}{cc}
        0 & \text{if }  \Vert\mathbf{q}\Vert^*_{\widetilde{\mathcal{A}}_r}\leq \rho, \\
         \infty & \text{otherwise,} \\
    \end{array}\right. \nonumber
\end{equation}
and
\begin{equation}
    I_c(\Vert\mathbf{q}\Vert^*_{\mathcal{A}_c}\leq 1) =  \left\{\begin{array}{cc}
        0 & \text{if }  \Vert\mathbf{q}\Vert^*_{\mathcal{A}_c}\leq 1, \\
         \infty & \text{otherwise,} \\
    \end{array}\right. \nonumber
\end{equation}
are the indicator functions. Then, the dual problem becomes 
\begin{equation}
    \minimize_\mathbf{q}\frac{1}{2}\Vert\mathbf{q}-\mathbf{y}\Vert_2^2 \text{  subject to } \Vert\mathbf{q}\Vert^*_{\widetilde{\mathcal{A}}_r}\leq \rho,\; \Vert\mathbf{q}\Vert^*_{\mathcal{A}_c}\leq 1.\label{eq:opt_synch}
\end{equation}
We convert the inequality constraints in \eqref{eq:opt_synch} to LMIs resulting in the following SDP 
 \begin{align}
    &\minimize_{\mathbf{q,K}}\quad \Vert\mathbf{q}-\mathbf{y}\Vert_2^2 \nonumber\\
    &\text{subject to }\mathbf{K}\succeq 0,\;\;\text{Tr}(\boldsymbol{\Theta}_\mathbf{n}\mathbf{K}) = \delta_{\mathbf{n}},\nonumber\\&\hphantom{\text{subject to }}  
    \begin{bmatrix}
        \mathbf{K} & \widehat{\mathbf{K}}_r^H \\
        \widehat{\mathbf{K}}_r &\rho^2
        \mathbf{I}_J 
        \end{bmatrix}
    \succeq0,\;\;
    \begin{bmatrix}
        \mathbf{K} & \widehat{\mathbf{K}}_c^H \\
        \widehat{\mathbf{K}}_c & \mathbf{I}_{JP} 
        \end{bmatrix}\succeq 0.
    \label{eq:dual_opt_synch}
\end{align}

\subsection{Recovery in the presence of noise}
In the presence of bounded additive noise $\mathbf{w}\in\mathbb{C}^{MP}$, the received signal in (\ref{eq:received_signal}) becomes $\mathbf{y}=\aleph_r(\mathbf{Z}_r) +\aleph_c(\mathbf{Z}_c)  + \mathbf{w}$, where $\Vert\mathbf{w}\Vert_2\leq \mu$ and $\mu$ is a constant. Then, the primal problem to recover the waveform-channel matrices is
\begin{align*}
&\minimize_{\mathbf{Z}_r,\mathbf{Z}_c} \Vert\mathbf{Z}_r\Vert_{\mathcal{A}_r}    +\Vert\mathbf{Z}_c\Vert_{\mathcal{A}_c}, \\&\text{ subject to } \Vert \mathbf{y} - \aleph_r(\mathbf{Z}_r) - \aleph_c(\mathbf{Z}_c) \Vert_2\leq \sqrt{\mu},
\end{align*}
The resulting dual problem is obtained by adding a regularization term in the noiseless case, i.e., (\ref{dual_opt}) as 
\begin{align}
    &\maximize_{\mathbf{q,K}}\quad \langle\mathbf{q,y}\rangle_{\mathbb{R}} - \mu\Vert \mathbf{q}\Vert_2^2\nonumber\\
    &\text{subject to }\mathbf{K}\succeq 0,\;\;\text{Tr}(\boldsymbol{\Theta}_n\mathbf{K}) = \boldsymbol{\delta}_{\mathbf{n}},\nonumber\\&\hphantom{\text{subject to }}  
    \begin{bmatrix}
        \mathbf{K} & \widehat{\mathbf{K}}_r^H \\
        \widehat{\mathbf{K}}_r & \mathbf{I}_J 
        \end{bmatrix}
    \succeq0,\;\;
    \begin{bmatrix}
        \mathbf{K} & \widehat{\mathbf{K}}_c^H \\
        \widehat{\mathbf{K}}_c & \mathbf{I}_{JP} 
        \end{bmatrix}\succeq 0,
\end{align} 
\subsection{Multiple emitters}
It is possible to extend the DBD problem in Section~\ref{sec:Signal_model} to multiple sources of
radar and communications signals. In this case, the received signal is 
$\mathbf{y} = \sum_{l=1}^{N_r}\aleph_{r
_l}(\mathbf{Z}_{r_l}) +\sum_{i=1}^{N_c}\aleph_{c_i}(\mathbf{Z}_{c_i}) $, where $N_r$ and $N_c$ are the number of, respectively, radar and communications signals. The received signal from each emitter has different transmit signals and channel parameters.  Therefore, the primal optimization problem in \eqref{eq:primal_problem} becomes
\begin{align*}
    &\minimize_{\stackrel{\mathbf{Z}_{r_1},\cdots, \mathbf{Z}_{r_{N_r}}}{\mathbf{Z}_{c_1}\cdots,\mathbf{Z}_{c_{N_c}}}} \sum_{l=1}^{N_r} \Vert \mathbf{Z}_{r_l}\Vert_{\mathcal{A}_r} +\sum_{i=1}^{N_c} \Vert \mathbf{Z}_{c_i}\Vert_{\mathcal{A}_c} \\&\text{ subject to } \mathbf{y} = \sum_{l=1}^{N_r}\aleph_{r
_l}(\mathbf{Z}_{r_l}) +\sum_{i=1}^{N_c}\aleph_{c_i}(\mathbf{Z}_{c_i}),
\end{align*}
for which the dual problem is
\begin{align*}
    &\maximize_{\mathbf{q}}\langle\mathbf{q,y}\rangle_{\mathbb{R}}\nonumber\\&\text{subject to } \Vert\aleph_{r_1}^\star(\mathbf{q})\Vert^\star_{\mathcal{A}_r}\leq1\;\cdots, \Vert \aleph_{r_{N_s}}^\star(\mathbf{q})\Vert^\star_{\mathcal{A}_r}\leq1\nonumber\\
    &\hphantom {\text {subject to } } \Vert \aleph_{c_1}^\star(\mathbf{q})\Vert^\star_{\mathcal{A}_c}\leq1, \cdots, \Vert \aleph_{c_{N_s}}^\star(\mathbf{q})\Vert^\star_{\mathcal{A}_c}\leq1.
\end{align*}
We identify that the $N_r+N_c$ constraints above employ, respectively, the PhTPs
$\mathbf{f}_{r_l}(\mathbf{r}) = \aleph_{r_{l}}^\star(\mathbf{q})\mathbf{a}(\mathbf{r})$, $l=1,\dots,N_r$, and $\mathbf{f}_{c_i}(\mathbf{c}) = \aleph_{c_{i}}^\star(\mathbf{q})\mathbf{a}(\mathbf{c})$, $i=1,\dots,N_c$.
We convert the constraints in the dual problem to $N_r+N_c$ LMIs. This results in the following SDP
\begin{align}
    &\maximize_{\mathbf{q,K}}\quad \langle\mathbf{q,y}\rangle_{\mathbb{R}}\nonumber\\
    &\text{subject to }\mathbf{K}\succeq 0\nonumber\\&\hphantom{\text{subject to }}  
    \begin{bmatrix}
        \mathbf{K} & \widehat{\mathbf{K}}_{r_1}^H \\
        \widehat{\mathbf{K}}_{r_1} & \mathbf{I}_J 
        \end{bmatrix}
    \succeq0,\;\cdots,\; \begin{bmatrix}
        \mathbf{K} & \widehat{\mathbf{K}}_{r_{N_s}}^H \\
        \widehat{\mathbf{K}}_{r_{N_s}} & \mathbf{I}_J 
        \end{bmatrix}
    \succeq0,\nonumber\\&\hphantom{\text{subject to }} 
    \begin{bmatrix}
        \mathbf{K} & \widehat{\mathbf{K}}_{c_1}^H \\
        \widehat{\mathbf{K}}_{c_1} & \mathbf{I}_{JP} 
        \end{bmatrix}\succeq 0,\;\cdots,\;\begin{bmatrix}
        \mathbf{K} & \widehat{\mathbf{K}}_{c_{N_s}}^H \\
        \widehat{\mathbf{K}}_{c_{N_s}} & \mathbf{I}_{JP} 
        \end{bmatrix}\succeq 0,
    \nonumber\\&\hphantom{\text{subject to }}
    \text{Tr}(\boldsymbol{\Theta}_n\mathbf{K}) = \boldsymbol{\delta}_{\mathbf{n}}.\label{m-dbd}
\end{align}

\subsection{Unequal PRI and symbol duration}
\label{subsec:generalziation_pulsewidth}
\begin{figure}[!t]
    \centering
    \includegraphics[width=1.0\columnwidth]{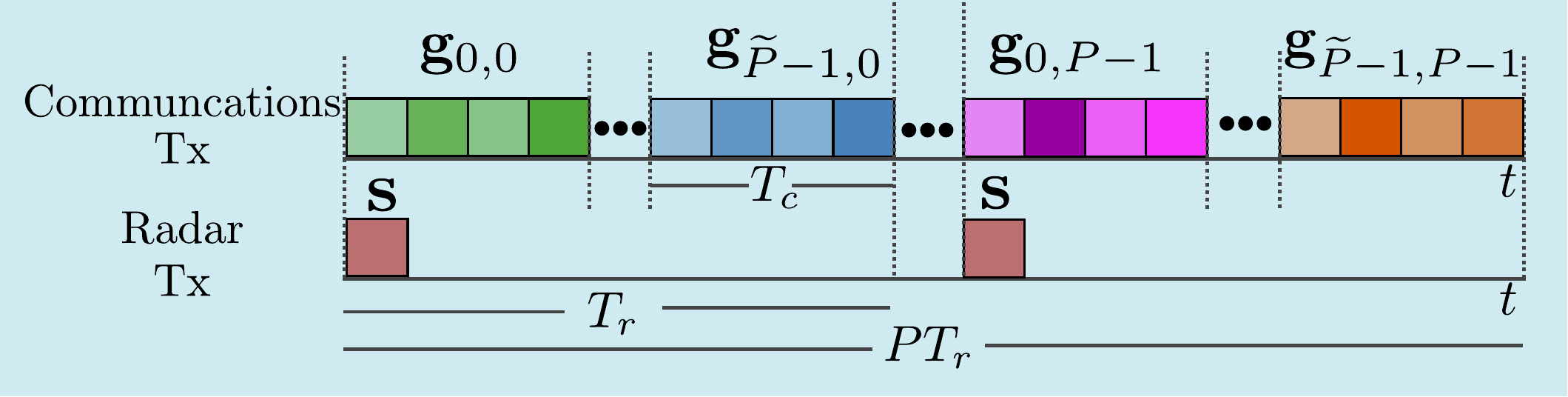}
    \caption{Transmission sequence when $T_r = \widetilde{P} T_c$. 
    As a result, multiple $\widetilde{P}$ communication messages are transmitted within a single PRI.}
    \label{fig:dp_fig_time}
\end{figure}

To simplify the model in Section~\ref{subsec:overlaid_rx}, the radar PRI $T_r$ was set equal to the communications symbol duration $T_c$. In practice, these durations may differ.  However, if a common clock is used, the PRI may be set to be an integer multiple of the symbol duration, i.e., $T_r = \widetilde{P}T_c$,  where $\widetilde{N} \in \mathbb{N}_+$. This implies that $P_c = \widetilde{P}P_r$ (Fig. \ref{fig:dp_fig_time}). Accordingly, the communications transmit signal in \eqref{comm_transmitted} becomes
\begin{align}
    x_c(t) &= \sum_{\widetilde{m}=0}^{\widetilde{P}P_r - 1} x_{\widetilde{m}} \left(t-\widetilde{m}\frac{T_r}{\widetilde{P}}\right) = \sum_{\widetilde{p}=0}^{\widetilde{P}-1}\sum_{p=0}^{P_r - 1} x_{\widetilde{p}+p\widetilde{P}} \left(t-(\widetilde{p}+p\widetilde{P})\frac{T_r}{\widetilde{P}}\right)\label{different_pulses}.
\end{align} 
Substituting \eqref{different_pulses} in \eqref{model_1} gives\par\noindent\small
\begin{align*}
y(t) &= \sum_{p=0}^{P_r-1} \left(\suml[\bsym{\alpha}_r]_\ell s(t-pT_r-[\overline{\bsym{\tau}}_r]_\ell)e^{-\mathrm{j}2\pi[\overline{\bsym{\nu}}_r]_\ell pT_r} \right. \nonumber\\
&  \left. + \sum_{\widetilde{p}=0}^{\widetilde{P}-1}\sumq [\bsym{\alpha}_c]_q x_{\widetilde{p}+\widetilde{P}p}\left(t-({\widetilde{p}+\widetilde{P}p})\frac{T_r}{\widetilde{P}}-[\overline{\bsym{\tau}}_c]_q\right)e^{-\mathrm{j}2\pi[\overline{\bsym{\nu}}_c]_qpT_r}\right)\nonumber\\
&=\sum_{p=0}^{P_r-1}\widetilde{y}_p(t),
\end{align*}\normalsize
where \par\noindent\small
\begin{align*}
    \widetilde{y}_p(t) &= \suml[\bsym{\alpha}_r]_\ell s(t-pT_r-[\overline{\bsym{\tau}}_r]_\ell)e^{-\mathrm{j}2\pi\nurl pT_r} \nonumber\\
& + \sum_{\widetilde{p}=0}^{\widetilde{P}-1}\sumq [\bsym{\alpha}_c]_q x_{\widetilde{p}+\widetilde{P}p}\left(t-({\widetilde{p}+\widetilde{P}p})\frac{T_r}{\widetilde{P}}-[\overline{\bsym{\tau}}_c]_q\right)e^{-\mathrm{j}2\pi\nucq pT_r}.
\end{align*}\normalsize
Denote the shifted signal as
\begin{align}
    y_p(t) &= \widetilde{y}_p(t+ pT) \nonumber\\
            &=\suml[\bsym{\alpha}_r]_\ell s(t-[\overline{\bsym{\tau}}_r]_\ell)e^{-\mathrm{j}2\pi\nurl pT_r} \nonumber\\
&  +\sum_{\widetilde{p}=0}^{\widetilde{P}-1}\sumq [\bsym{\alpha}_c]_q x_{\widetilde{p}+\widetilde{P}p}\left(t-\widetilde{p}\frac{T_r}{\widetilde{P}}-[\overline{\bsym{\tau}}_c]_q\right)e^{-\mathrm{j}2\pi\nucq pT_r}.\label{yp-dp}
\end{align}
Since the summation over $\widetilde{p}$ corresponds to the messages transmitted over the same radar PRI, the signal \eqref{yp-dp} is defined in terms of the index $\breve{p} = \widetilde{p}+p\widetilde{P}$ as 
\begin{align}
    y_{\breve{p}}(t) & = \suml[\bsym{\alpha}_r]_\ell s(t-[\overline{\bsym{\tau}}_r]_\ell)e^{-\mathrm{j}2\pi\nurl pT_r} \nonumber\\
&  + \sumq [\bsym{\alpha}_c]_q x_{\breve{p}}\left(t-[\overline{\bsym{\tau}}_c]_q\right)e^{-\mathrm{j}2\pi\nucq pT_r}, \label{yp_dp_f}
\end{align}
Replacing $x_{\breve{p}}(t) = \sumk [\mathbf{g}_{\breve{p}}]_k e^{\mathrm{j}2\pi k \Delta_f t} $ in \eqref{yp_dp_f} yields 
\begin{align*}
  y_{p,\widetilde{p}}(t) &= \suml[\bsym{\alpha}_r]_\ell s(t-[\overline{\bsym{\tau}}_r]_\ell)e^{-\mathrm{j}2\pi[\overline{\bsym{\nu}}_r]_\ell pT_r} \nonumber\\
& + \sumq [\bsym{\alpha}_c]_q \sumk [\mathbf{g}_{p,\widetilde{p}}]_k e^{\mathrm{j}2\pi k\Delta f (t-\taucq)}e^{-\mathrm{j}2\pi[\overline{\bsym{\nu}}_c]_qpT_c}.
\end{align*}

From here, we proceed as in Section \ref{sec:Signal_model},  compute the CTFT of $y_{p,\widetilde{p}}(t)$, and sample the frequency-domain signal to obtain the equivalent of \eqref{eq:y_p} as

\begin{align*}
    [\mathbf{y}]_{\widetilde{n},\widetilde{p}}  & = \suml[\bsym{\alpha}_r]_\ell [\mathbf{s}]_n e^{-\mathrm{j}2\pi(n[{\bsym{\tau}}_r]_\ell+ p[{\bsym{\nu}}_r]_\ell )} \nonumber\\
& + \sumq [\bsym{\alpha}_c]_q[\mathbf{g}_{p,\widetilde{p}}]_{{n}} e^{-\mathrm{j}2\pi(n[{\bsym{\tau}}_c]_q+ p[{\bsym{\nu}}_q]_\ell )}.
    \label{eq:y_np}
\end{align*}
The low-dimensional representation of radar waveform coefficients becomes $\mathbf{s} = \mathbf{B}\mathbf{u}$ where $\mathbf{B} \in \mathbb{C}^{M\times \widetilde{J}},\mathbf{u}\in\mathbb{C}^{\widetilde{J}} $ and $\widetilde{J} = \widetilde{P}J$. The similar representation for the communications messages is  $\mathbf{g}_{p,\widetilde{p}} = \mathbf{D}_{p,\widetilde{p}}\mathbf{v}_{p,\widetilde{p}}$ where $\mathbf{D}_{p,\widetilde{p}} \in \mathbb{C}^{M\times J}$ and $\mathbf{v}_{p,\widetilde{p}} \in \mathbb{C}^{J}$. Compared to the model in Section \ref{sec:Signal_model}, here we need to recover a coefficient vector for radar of size $\widetilde{J}$ and multiple coefficient vectors of size $J$ for communications. 

\begin{figure}[!t]
    \centering
    \includegraphics[width=0.90\columnwidth]{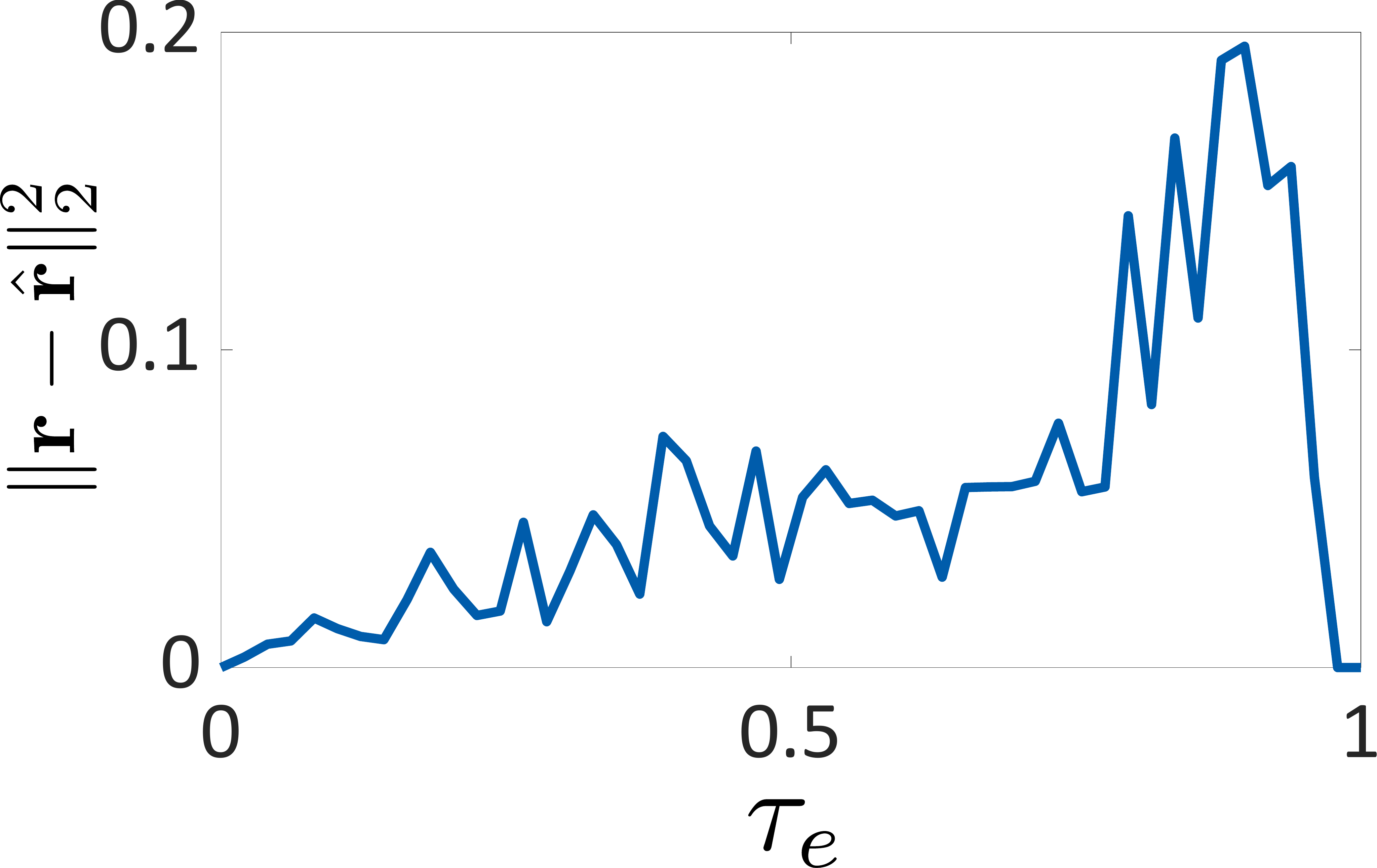}
    \caption{Estimation error (averaged over $100$ trials) for radar channel parameters with synchronization delay. The parameter values were $M=11$, $P=9$, $J=3$, and $L=Q=2$. Note that, when $\tau_e = 1$, the error vanishes because of the periodicity of the atoms.}
    \label{fig:sycnh_error}
\end{figure}
\begin{figure*}[t]
    \centering
    \includegraphics[width=0.90\linewidth]{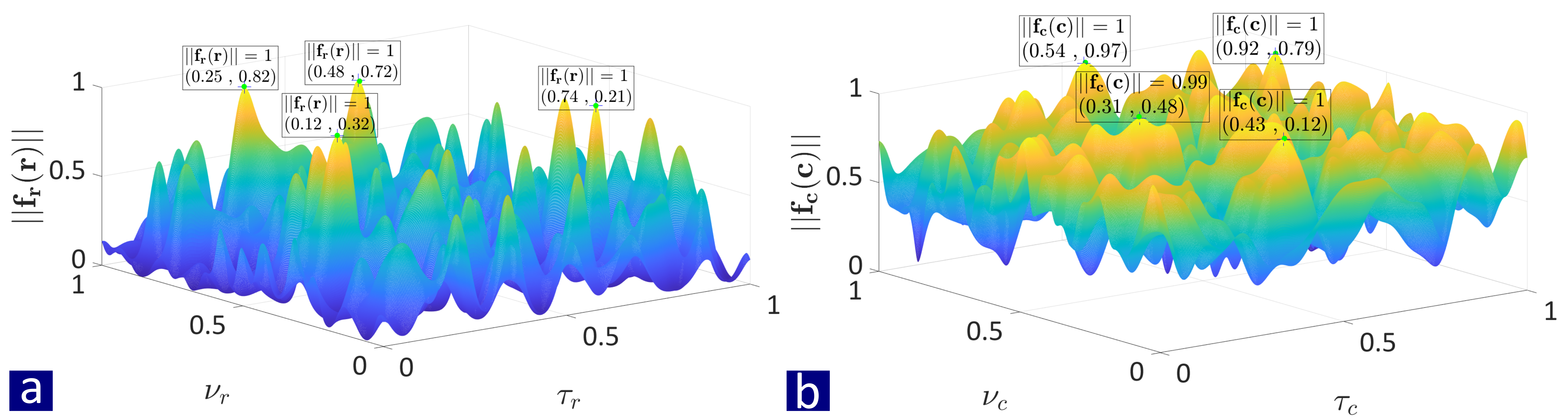}
    \caption{As in Fig.~\ref{fig:results_dual} but for $L=Q=4$, $M=17$, $P=13$, and $J=3$ in the presence of noise. The SNR was set to $5$ dB.
    }
    \label{fig:noise_pol}
\end{figure*}
\subsection{Numerical Simulations}
\begin{figure}[!t]
    \centering
    \includegraphics[width=0.90\columnwidth]{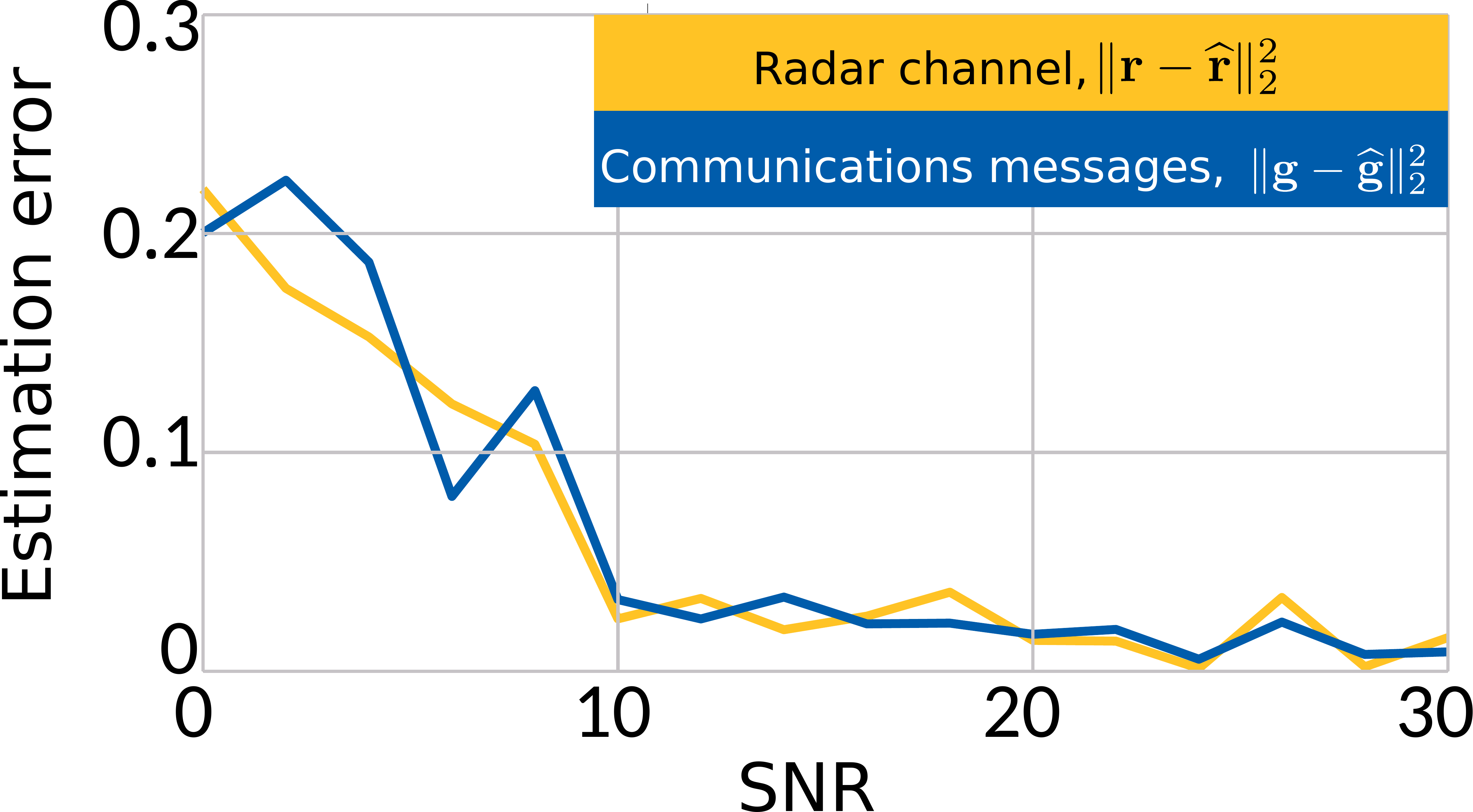}
    \caption{Estimation error for the radar channel $\mathbf{r}$ and the communications messages $\mathbf{g}$ in the presence of additive noise and averaged over $100$ trials. Here, $L=Q=4$, $M=17$, $P=13$, and $J=3$. The SNR was set to $5$ dB.}
    \label{fig:noise}
\end{figure}
\begin{figure}[!t]
    \centering
    \includegraphics[width=0.90\columnwidth]{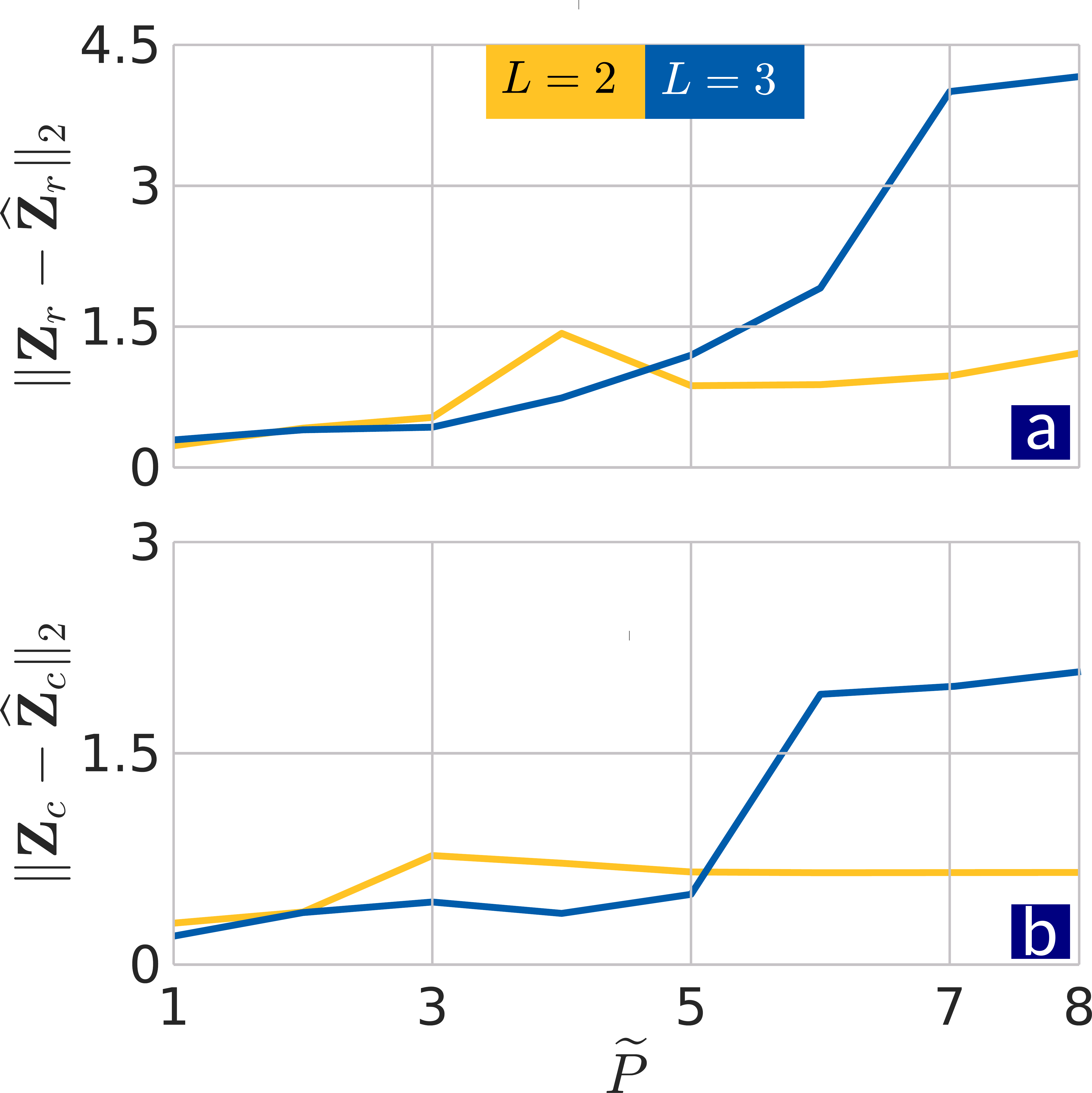}
    \caption{Estimation error (averaged over $20$ trials) in waveform-channel matrices for (a) radar and (b) communications, when $T_r = \widetilde{P}T_c$. Here, we set $M=11$, $P=11$, $J = 3$, and $L=Q$. }
    \label{fig:dif_pulse_results}
\end{figure}
\begin{figure*}
    \centering
    \includegraphics[width=1.0\textwidth]{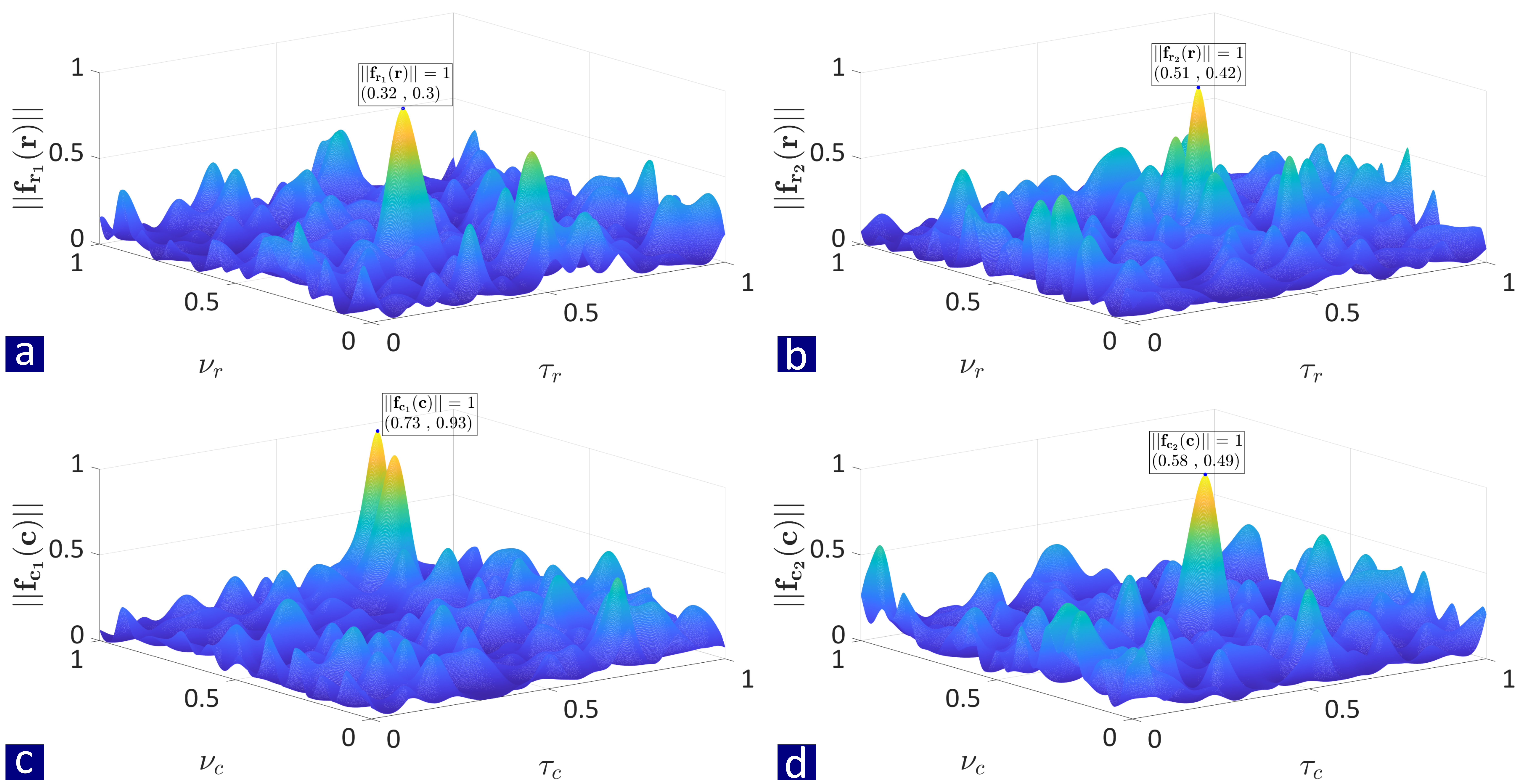}
    \caption{Noiseless channel parameter localization for (a) first radar emitter, (b) second radar emitter, (c) first communications emitter, and (d) second communications emitter. Here,  $M = 15$, $P=13$, and $J=3$. 
     }
    \label{fig:multiple_dbd}
\end{figure*}
Finally, we evaluate the practical issues discussed above through numerical experiments.\\
\textbf{Recovery with synchronization delay}: We solved the SDP in \eqref{eq:dual_opt_synch} while varying the delay $\tau_e$ from $0$ to $1$ in steps of 0.02 for $M=11$, $P=9$, $J=3$, and $L=Q=2$.  Fig.~\ref{fig:sycnh_error} plots the mean square error in radar channel estimate averaged over 100 random realizations of the channel with varying $\tau_e$. Here, the parameter $\tau_e$ represents additive noise in the target delay-time $\tau_r$. Note that the error reduces sharply when $\tau_e$ reaches a value of unity because of a $2\pi$ phase shift in the atoms. The periodicity of the complex exponential in the atoms implies that integer-valued synchronization lags do not affect delay estimation; the Doppler parameters were recovered perfectly in this experiment.     \\
\textbf{Recovery in the presence of noise}: Here, we modeled the noise $\mathbf{w}$ as a random variable drawn from a normal distribution $\sim \mathcal{N}\left(0,\frac{\Vert \mathbf{y} \Vert_2^2}{10^{\frac{\textrm{SNR}}{10}}}\right)$. We first illustrate the noisy recovery in Fig.~\ref{fig:noise_pol} for $M=17$, $P=13$, $J=3$, $\textrm{SNR} = 5$ dB, and $L=Q=4$. The channel parameters were drawn uniformly at random as before. In particular, we had $\bsym{\tau}_r = [0.82, 0.21, 0.71, 0.32]$, $\bsym{\nu}_r = [0.25, 0.74, 0.48, 0.12]$ and $\bsym{\tau}_c = [0.79, 0.12, 0.97, 0.48]$, and $\bsym{\nu}_c = [0.92, 0.43, 0.54, 0.31]$. The recovery in Fig.~\ref{fig:noise_pol} shows a slight error in the estimated parameter values. In Fig.~\ref{fig:noise}, we plot the $\ell_2$ norm error in estimating radar channel and communications messages averaged over $100$ trials while varying the SNR levels from $0$ to $30$ dB in steps of 2 dB.
\\
\textbf{Unequal PRI and symbol duration}: For $T_r \neq T_c$, we employed the formulation from Section~\ref{subsec:generalziation_pulsewidth}. We varied the integer factor $\widetilde{P}$ from $1$ (i.e., $T_r = T_c$) to $8$ (i.e., $T_r = 8T_c$) for different values of $L=Q$. Fig.~\ref{fig:dif_pulse_results} plots the $\ell-2$ norm error in the waveform-channel matrices $\widehat{\mathbf{Z}}_r$ and $\widehat{\mathbf{Z}}_c$ averaged over $20$ trials. We note that for $L=Q =2$, performance loss is not significant with increase in $\widetilde{P}$. This error, in general, increases with the number of targets/paths when $T_r$ significantly departs from $T_c$, say, for $\widetilde{P}>5$. Here, note that the communications messages do not vary only across the PRIs; the messages within the PRI are also different. \\
\textbf{Recovery with multiple emitters}: We solved the SDP in \eqref{m-dbd} with $Ns = 2$, $L = 1$, $Q=1$ and $J=3$. The set of radar and communications channel parameters were $\mathbf{r}_1 = [0.30 ,   0.32
], \mathbf{r}_2 = [0.42  ,  0.51
], \mathbf{c}_1=[0.93  ,  0.73
], \mathbf{c}_2=[ 0.49 ,  0.58
]$. All channel parameters were recovered perfectly in Fig.~\ref{fig:multiple_dbd}. This demonstrates generalization of SoMAN minimization to a more complex BdM problem.

\section{Proof of Lemma~\ref{lemma:concentration}}
\label{app:lemma_concentration}

We bound $\mathbf{H}-\overline{\mathbf{H}}$ by upper-bounding the largest entry of each block-diagonal matrix of $\mathbf{H}-\overline{\mathbf{H}}$. \newpage 
We have  
\begin{align}
&\mathbb{P}\left[\|\mathbf{H} - \mathbb{E}\left[\mathbf{H}\right] \|\geq\epsilon_1\right] \nonumber\\
&\leq { \sqrt{2}\max_{i,j} \mathbb{P} \left[\|\mathbf{H}_{i,j} -\mathbb{E}\left[\mathbf{H}_{i,j}\right]\| \geq \epsilon_1\right]  }  \nonumber \\
&\leq \mathbb{P}{ \sum_{i,j} \mathbb{P}\left[\| \mathbf{H}_{i,j} - \mathbb{E}\left[\mathbf{H}_{i,j}\right] \|\geq \epsilon_1\right]}\nonumber \\
&={\mathbb{P}{ \left[\|\mathbf{H}_1- \mathbb{E}\left[\mathbf{H}_1\right] \| \geq \epsilon_1\right]}} + 2\mathbb{P}\left[\|\mathbf{H}_2- \mathbb{E}\left[\mathbf{H}_2\right] \|\geq\epsilon_1\right]  \nonumber\\
& + {\mathbb{P}\left[{\|\mathbf{H}_3- \mathbb{E}\left[\mathbf{H}_3\right]\| \geq\epsilon_1}\right]} .
\end{align}
We make this failure probability less than $4\delta$ by bounding $\mathbb{P}{\|\mathbf{H}_i- \mathbb{E}\left[\mathbf{H}_i\right] \|\geq \epsilon_1}$ for $i=1,2,3$ by $\delta$. 

Using the expressions of $\mathbf{H}_1$, $\mathbf{H}_2$, and $\mathbf{H}_3$ from, respectively, \eqref{H1}, \eqref{H2}, and \eqref{H3} and their expected values $\overline{\mathbf{H}}_1,\overline{\mathbf{H}}_2,\overline{\mathbf{H}}_3$, define the matrices
\begin{align}
    \mathbf{S}_{1\widetilde{n}} &= \frac{1}{MP} g_{M}(n) g_{P}(p) \left(\bsym{\upsilon}_{\widetilde{n}} \bsym{\upsilon}_{\widetilde{n}}^{H}\right) \otimes\left(\mathbf{b}_{n} \mathbf{b}_{n}^{H} - \boldsymbol{I}_J\right) \in  \mathbb{C}^{3LJ \times 3LJ}, \nonumber\\
    \mathbf{S}_{2\widetilde{n}} &= \frac{1}{MP} g_{M}(n) g_{P}(p) \left(\bsym{\upsilon}_{\widetilde{n}}  \bsym{\varrho}_{\widetilde{n}}^{H}\right) \otimes\left(\mathbf{b}_{n} \mathbf{d}_{\widetilde{n}}^{H} \right) \in \mathbb{C}^{3LJ \times 3QPJ},\nonumber\\
    \mathbf{S}_{3\widetilde{n}} &= \frac{1}{MP} g_{M}(n) g_{P}(p) \left(\bsym{\upsilon}_{\widetilde{n}} \bsym{\upsilon}_{\widetilde{n}}^{H}\right) \otimes\left(\mathbf{d}_{\widetilde{n}} \mathbf{d}_{\widetilde{n}}^{H} - \boldsymbol{I}_{PJ}\right) \in \mathbb{C}^{3QPJ \times 3QPJ},
    \label{eq:S}
\end{align}
so that 
$
\mathbf{H}_i-\overline{\mathbf{H}}_i=\sum_{\widetilde{n}} \mathbf{S}_{i\widetilde{n}}
$. Note that $\mathbf{S}_{i\widetilde{n}}$, $i=1,2,3$, 
are independent random matrices with zero mean. Therefore, the non-commutative Bernstein inequality \cite{tropp2012user} is applicable for bounding $\|\mathbf{H}_i- \mathbb{E}\left[\mathbf{H}_i\right] \|$. To this end, we require the bounds on $\left\|\mathbf{S}_{i\widetilde{n}}\right\| $ and their variances. We proceed as follows.
\begin{align*}
\left\|\mathbf{S}_{1\widetilde{n}}\right\| &=\frac{1}{MP}\left|g_N(n)\right| \left|g_P(p)\right| \cdot\left\|\bsym{\upsilon}_{\widetilde{n}} \bsym{\upsilon}_{\widetilde{n}}^{H}\right\| \cdot\left\|\mathbf{b}_{n} \mathbf{b}_{n}^{H}-\mathbf{I}_{J}\right\| \\
& \leq \frac{1}{MP} \max_{n}\left|g_N(n)\right|\max_{p}\left|g_P(p)\right| \cdot\left\|\bsym{\upsilon}_{\widetilde{n}}\right\|_{2}^{2} \cdot \max \left\{\left\|\mathbf{b}_{n}\right\|_{2}^{2},\left\|\mathbf{I}_{J}\right\|\right\} \\
& \leq \frac{1}{MP} \cdot\ L \left(1 + \left(\frac{2 \pi n} {\kappa}\right)^{2}+ \left(\frac{2 \pi p}{ \kappa}\right)^{2}\right) \max \{\mu J, 1\} \\
& \leq \frac{27 L \mu J}{MP}:=R_1.
\end{align*}
where the first inequality follows from the fact that for two positive semidefinite matrices $\mathbf{A}$ and $\mathbf{B}$, $\|\mathbf{A-B}\|\leq \max\{\|\mathbf{A}\|,\|\mathbf{B}\|\}$; the second inequality follows from $g_{M}(n),g_{P}(p)\leq 1$ and the incoherence property of $F$; and the last inequality follows from the fact that $\left(\frac{2 \pi n} {\kappa}\right)^{2}, \left(\frac{2 \pi p} {\kappa}\right)^{2}\leq 13$ when $M, P \geq 4$ \cite{candes_superresolution,off_the_grid}. Similarly, 
\begin{align*}
\left\|\mathbf{S}_{2\widetilde{n}}\right\| &=\frac{1}{MP}\left|g_N(n)\right| \left|g_P(p)\right| \cdot\left\|\bsym{\upsilon}_{\widetilde{n}}  \bsym{\varrho}_{\widetilde{n}}^{H}\right\| \cdot\left\|\mathbf{b}_{n} \mathbf{d}_{\widetilde{n}}^{H}\right\| \\
& \leq \frac{1}{MP} \max_{n}\left|g_N(n)\right|\max_{p}\left|g_P(p)\right| \nonumber\\
& \hspace{1cm} \cdot \|\bsym{\upsilon}_{\widetilde{n}}\|_{2} \| \bsym{\varrho}_{\widetilde{n}}\|_{2}\cdot \left\|\mathbf{b}_{n}\right\|_{2}\left\|\mathbf{d}_{n}\right\|_{2} \\
& \leq \frac{1}{M} \cdot\sqrt{L+L\left(\frac{2 \pi n} {\kappa}\right)^{2} + L\left(\frac{2 \pi p} {\kappa}\right)^{2}} \nonumber\\
& \hspace{1cm} \times \sqrt{Q+Q\left(\frac{2 \pi n} {\kappa}\right)^{2} + Q\left(\frac{2 \pi p} {\kappa}\right)^{2}} \mu J \\
& \leq \frac{27 \sqrt{LQ} \mu J}{MP}:=R_2.
\end{align*}
\begin{align*}
\left\|\mathbf{S}_{3\widetilde{n}}\right\|&
 =\frac{1}{MP}\left|g_N(n)\right| \left|g_P(p)\right| \cdot\left\|\bsym{\upsilon}_{\widetilde{n}} \bsym{\upsilon}_{\widetilde{n}}^{H}\right\| \cdot\left\|\mathbf{d}_{\widetilde{n}} \mathbf{d}_{\widetilde{n}}^{H}-\mathbf{I}_{J}\right\| \\
& \leq \frac{1}{MP} \max_{n}\left|g_N(n)\right|\max_{p}\left|g_P(p)\right| \cdot\left\|\bsym{\varrho}_{\widetilde{n}}\right\|_{2}^{2} \cdot \max \left\{\left\|\mathbf{d}_{\widetilde{n}}\right\|_{2}^{2},\left\|\mathbf{I}_{J}\right\|\right\} \\
& \leq \frac{1}{M} \cdot\left(Q+Q\left(\frac{2 \pi n} {\kappa}\right)^{2} + Q\left(\frac{2 \pi p} {\kappa}\right)^{2}\right) \max \{\mu J, 1\} \\
& \leq \frac{27 Q \mu J}{M}:=R_3.
\end{align*}
Then, 
\begin{align*} & \left\Vert \sum_{\widetilde{n}} \mathbb{E}[{\mathbf{S}}_{1\widetilde{n}}{\mathbf{S}}_{1\widetilde{n}}^H] \right\Vert \\
& =\frac{1}{(MP)^2} \left\Vert \sum_{\widetilde{n}} \mathbb{E}\left\{ g_{M}(n)^2 g_P(p)^2  \left[(\bsym{\upsilon}_{\widetilde{n}} \bsym{\upsilon}_{\widetilde{n}}^H) \otimes (\mathbf{b}_n\mathbf{b}_n^H - {\mathbf{I}}_J) \right]  \right. \right. \nonumber\\
& \hspace{4cm} \left. \left.  \cdot\left[(\bsym{\upsilon}_{\widetilde{n}} \bsym{\upsilon}_{\widetilde{n}}^H) \otimes (\mathbf{b}_n\mathbf{b}_n^H - {\mathbf{I}}_J) \right]^H \right\} \right\Vert 
\end{align*}
\begin{align*}
& = \frac{1}{(MP)^2} \left\Vert \sum_{\widetilde{n}} g_{M}(n)^2 g_P(p)^2 \Vert \bsym{\upsilon}_{\widetilde{n}}\Vert _2^2(\bsym{\upsilon}_{\widetilde{n}} \bsym{\upsilon}_{\widetilde{n}}^H) \right.  \nonumber\\
& \hspace{2cm}  \left. \mathbb{E}\left[(\mathbf{b}_n\mathbf{b}_n^H - {\mathbf{I}}_L) (\mathbf{b}_n\mathbf{b}_n^H - {\mathbf{I}}_J) \right] \right\Vert \\
& \leq \frac{27L \mu J}{(MP)^2} \left\Vert \sum_{\widetilde{n}} g_{M}(n)^2 g_P(p)^2 (\bsym{\upsilon}_{\widetilde{n}} \bsym{\upsilon}_{\widetilde{n}}^H) \otimes {\boldsymbol{I}}_J \right\Vert \\
& \leq \frac{27L \mu J}{MP} \max_n \vert g_{M}(n)\vert \max_p \vert g_P(p)\vert \cdot\left\Vert \frac{1}{MP}\sum_{\widetilde{n}} g_{M}(n) g_{P}(p) (\bsym{\upsilon}_{\widetilde{n}} \bsym{\upsilon}_{\widetilde{n}}^H) \otimes {\mathbf{I}}_J \right\Vert \\& \leq \frac{27L \mu J}{MP} \left\Vert \overline{{\mathbf{H}}}_1\right\Vert \leq \frac{37L \mu J}{MP} := \sigma_1^2.\end{align*}
Similarly,
\begin{align*}&\quad \left\Vert \sum_{\widetilde{n}} \mathbb{E}[{\mathbf{S}}_{2\widetilde{n}}{\mathbf{S}}_{2\widetilde{n}}^H] \right\Vert \leq \frac{27 \sqrt{LQ}\mu J}{MP} \left\Vert \overline{\mathbf{H}}_2\right\Vert \leq \frac{37 \sqrt{LQ}\mu J}{MP}:=\sigma^2_2,\\
&\quad \left\Vert \sum_{\widetilde{n}} \mathbb{E}[{\mathbf{S}}_{3\widetilde{n}}{\mathbf{S}}_{3\widetilde{n}}^H] \right\Vert \leq \frac{27 {Q} \mu J}{MP} \left\Vert \overline{\mathbf{H}}_3\right\Vert \leq \frac{37Q\mu J}{MP}:=\sigma^2_3.
\end{align*}
Applying Bernstein inequality to $\mathbf{H}_i-\overline{\mathbf{H}}_i$ completes the proof.

\section{Proof of Lemma \ref{lemma:J1_last}}
\label{app:bound_J1}
\subsection{Preliminaries to the proof}
We first show through the following Lemma~\ref{lemma:J1_1} that the quantity $\Vert\bsym{\Psi}_r^{(m^\prime,n^\prime)}(\mathbf{r})-\mathbb{E} \bsym{\Psi}_r^{(m^\prime,n^\prime)}(\mathbf{r})\Vert$ is small with high probability.
\begin{lemma}\label{lemma:J1_1}
Given $\mathbf{r} \in [0,1]^2$, $\epsilon_8 \in (0,1)$ for $m^\prime=0,1,2,3$ and $n^\prime=0,1,2,3$, $\Vert\bsym{\Psi}_r^{(m^\prime,n^\prime)}(\mathbf{r})-\expec{ \bsym{\Psi}_r^{(m^\prime,n^\prime)}(\mathbf{r})}\Vert\leq 16\epsilon_8$ holds with at least $1-4\delta$ probability, where $\delta$ is specified by  
\begin{equation}
    MP\leq \frac{800\cdot 4^{2(m^\prime+n^\prime)}\mu JL}{3\epsilon_8^2}\log\left(\frac{2JL+J}{\delta}\right).
\end{equation}
\end{lemma}
\begin{IEEEproof}
We write $\boldsymbol{\Psi}_r^{m^\prime,n^\prime }(\mathbf{r})- \mathbb{E}\left[\boldsymbol{\Psi}_r^{m^\prime,n^\prime }(\mathbf{r})\right]$ as the sum of zero-mean random independent matrices as
\begin{align}
    &\boldsymbol{\Psi}_r^{(m^\prime,n^\prime )}(\mathbf{r})- \mathbb{E}\left[\boldsymbol{\Psi}_r^{(m^\prime,n^\prime)}(\mathbf{r})\right] \nonumber\\
    &= \frac{1}{MP}\sum_{\widetilde{n}} g_M(n)g_{P}(p) \left(\frac{-\mathrm{j}2\pi n}{\kappa}\right)^{m^\prime}\left(\frac{-\mathrm{j}2\pi p}{\kappa}\right)^{n^\prime} \nonumber\\
& \hspace{4cm} \times e^{-\mathrm{j}2\pi(n\tau+p\nu)}\bsym{\upsilon}_{\widetilde{n}}\otimes\left(\mathbf{b}_n\mathbf{b}_n^H-\mathbf{I}_J\right),\nonumber\\
    &= \frac{1}{MP}\sum_{\widetilde{n}}\mathbf{X}_{\widetilde{n}}^{(m^\prime,n^\prime)} (\mathbf{r}),\nonumber
\end{align}
where the last equality used the substitution 
\begin{align}
   \mathbf{X}_{\widetilde{n}}^{(m^\prime,n^\prime)} (\mathbf{r})&= g_M(n)g_{P}(p)\left(\frac{-\mathrm{j}2\pi n}{\kappa}\right)^{m^\prime}\left(\frac{-\mathrm{j}2\pi p}{\kappa}\right)^{n^\prime} \nonumber\\
& \hspace{0.5cm} \times e^{-\mathrm{j}2\pi(n\tau+p\nu)}\bsym{\upsilon}_{\widetilde{n}}\otimes\left(\mathbf{b}_n\mathbf{b}_n^H-\mathbf{I}_J\right), \in \mathbb{C}^{3JL\times J}.\nonumber
\end{align}
We now compute the bounds on $\mathbf{X}_{\widetilde{n}}^{(m^\prime,n^\prime)}(\mathbf{r})$ and its variance. We have
\begin{align*}
\left\Vert\mathbf{X}_{\widetilde{n}}^{(m^\prime,n^\prime)}\right\Vert = &\left\Vert g_M(n)g_{P}(p)\left(\frac{-\mathrm{j}2\pi n}{\kappa}\right)^{m^\prime}\left(\frac{-\mathrm{j}2\pi p}{\kappa}\right)^{n^\prime} \right. \nonumber\\
& \hspace{2cm} \left. \times e^{-\mathrm{j}2\pi(n\tau+p\nu)}\bsym{\upsilon}_{\widetilde{n}}\otimes\left(\mathbf{b}_n\mathbf{b}_n^H-\mathbf{I}_J\right)\right\Vert\\
&\leq \frac{1}{MP}4^{m^\prime}4^{n^\prime}\Vert\bsym{\upsilon}_{\widetilde{n}}\Vert_2\Vert\mathbf{b}_n\mathbf{b}_n^H-\mathbf{I}_J\Vert\\
&\leq \frac{1}{MP}4^{m^\prime+n^\prime}\triangleq R,
\end{align*}
where the first inequality follows from $\Vert g_{M}(n)\Vert_{\infty}=\Vert g_{P}(p)\Vert_{\infty}\leq 1$, and the second employs $\left\vert\frac{2\pi n}{\kappa}\right\vert\leq4$ for $M,P\geq4$.
Further, \par\noindent\small
\begin{align}
    &\left\Vert\sum_{\widetilde{n}}\expec{\mathbf{X}_{\widetilde{n}}^{(m^\prime,n^\prime)}\mathbf{X}_{\widetilde{n}}^{(m^\prime,n^\prime)H}}\right\Vert \nonumber\\
    & = \frac{1}{(MP)^2}\left\Vert\sum_{\widetilde{n}}\mathbb{E}\bigg[\vert g_{M}(n)\vert^2\vert  g_{P}(p)\vert^2\left(\left\vert\frac{-\mathrm{j}2\pi n}{\kappa}\right\vert\right)^{2m^\prime}\left(\left\vert\frac{-\mathrm{j}2\pi p}{\kappa}\right\vert\right)^{2n^\prime} \right. \nonumber\\
& \hspace{2cm} \left. \left(\bsym{\upsilon}_{\widetilde{n}}^H\otimes(\mathbf{b}_n\mathbf{b}_n^H-\mathbf{I}_J)\right)\left(\bsym{\upsilon}_{\widetilde{n}}\otimes(\mathbf{b}_n\mathbf{b}_n^H-\mathbf{I}_J)\right)\bigg]\right\Vert\nonumber\\
    &
    \leq \frac{4^{2(m^\prime+n^\prime)}}{(MP)^2}\left\Vert\sum_{\widetilde{n}}\bsym{\upsilon}_{\widetilde{n}}^H\bsym{\upsilon}_{\widetilde{n}}\expec{\left(\mathbf{b}_n\mathbf{b}_n^H-\mathbf{I}_J\right)^2}\right\Vert\nonumber\\&\nonumber
    =\frac{4^{2(m^\prime+n^\prime)}}{(MP)^2}\left\Vert\sum_{\widetilde{n}}\Vert\bsym{\upsilon}_{\widetilde{n}}\Vert_2^2\expec{\Vert\mathbf{b}_n\Vert_2^2\mathbf{b}_n\mathbf{b}_n^H-2\mathbf{b}_n\mathbf{b}_n^H }+\mathbf{I}_J\right\Vert\\&\leq \frac{4^{2(m^\prime+n^\prime)}}{(MP)^2}\left\Vert\sum_{\widetilde{n}}\Vert\bsym{\upsilon}_{\widetilde{n}}\Vert_2^2\mathbf{I}_J(\mu J-1)\right\Vert\nonumber\\&\leq \frac{100\cdot4^{2(m^\prime+n^\prime)}\mu JL}{MP}\triangleq\sigma.\label{eq:variance_x}
\end{align}\normalsize
Using the matrix Bernstein inequality on  $\boldsymbol{\Psi}_r^{(m^\prime,n^\prime )}(\mathbf{r})- \mathbb{E}\left[\boldsymbol{\Psi}_r^{(m^\prime,n^\prime)}(\mathbf{r})\right]$, we have
\begin{align*}
    \mathbb{P}\left\{\left\Vert\sum_{\widetilde{n}}\mathbf{X}_{\widetilde{n}}^{(m^\prime,n^\prime)}\right\Vert\geq\epsilon_8\right\}\leq(3JL+J)e^{\left(\frac{-3\epsilon_8^2}{8\sigma^2}\right)}.
\end{align*}
Applying the union bound for $m^\prime=0,1,2,3 $ and $n^\prime=0,1,2,3$ yields
\begin{align*}
    \mathbb{P}\left\{\left\Vert\sum_{\widetilde{n}}\mathbf{X}_{\widetilde{n}}^{(m^\prime,n^\prime)}\geq\epsilon_8\right\Vert,m^\prime,n^\prime = 0,1,2,3\right\} \leq 16\delta,
\end{align*}
where, using the bound on the variance in \eqref{eq:variance_x}, the failure probability $\delta$ is specified by
\begin{align*}
     MP\leq \frac{800\cdot 4^{2(m^\prime+n^\prime)\mu JL}}{3\epsilon_8^2}\log\left(\frac{2JL+J}{\delta}\right).
\end{align*}
\end{IEEEproof}
In the following Lemmata \ref{lemma:J1_2} and \ref{lemma:J1_3}, conditioned by the event $\mathcal{E}_{\epsilon_1}$, we employ Lemmata \ref{lemma:distance H-E[H]} and \ref{lemma:J1_1} to bound $\left\Vert\left(\bsym{\Psi}_r^{(m^\prime,n^\prime)}(\mathbf{r})^H-\expec{\bsym{\Psi}_r^{(m^\prime,n^\prime)}(\mathbf{r})}\right)^H\mathbf{L}_1\right\Vert$ and $\left\Vert\bsym{\Psi}_r^{(m^\prime,n^\prime)}(\mathbf{r})^H-\mathbb{E} \bsym{\Psi}_r^{(m^\prime,n^\prime)}(\mathbf{r})\right\Vert_F$, respectively.
\begin{lemma}\label{lemma:J1_2}
Consider $\epsilon_1\in(0,1/4]$ and a finite set of grid points $\boldsymbol{\Omega}_{\textnormal{\textrm{grid}}} =  \{\widetilde{\mathbf{r}}_d\}$ with cardinality $|\boldsymbol{\Omega}_{\textnormal{\textrm{grid}}}|$. Then, \par\noindent\small
\begin{align*}
    \mathbb{P}&\left\{\underset{\widetilde{\mathbf{r}}\in \Omega_{\textrm{grid}}}{\operatorname{sup}}\left\Vert\left(\bsym{\Psi}_r^{(m^\prime,n^\prime)}(\mathbf{\widetilde{\mathbf{r}}_d})^H-\expec{\bsym{\Psi}_r^{(m^\prime,n^\prime)}(\mathbf{\widetilde{\mathbf{r}}_d})}\right)^H\mathbf{L}_1\right\Vert \right. \nonumber\\
    & \left. \geq 16\epsilon_8, m^\prime,n^\prime = 0,1,2,3\right\}\leq \vert\Omega_{\textrm{grid}}\vert 16\delta+\mathbb{P}\left(\mathcal{E}_{1,\epsilon_1}^c\right).
\end{align*}\normalsize
\end{lemma}
\begin{IEEEproof}
Observe that
\begin{align}
    &\left\Vert\left(\bsym{\Psi}_r^{(m^\prime,n^\prime)}(\mathbf{\widetilde{\mathbf{r}}_d})^H-\expec{\bsym{\Psi}_r^{(m^\prime,n^\prime)}(\mathbf{\widetilde{\mathbf{r}}_d})}\right)^H\mathbf{L}_1\right\Vert \nonumber\\
    &\leq \left\Vert \bsym{\Psi}_r^{(m^\prime,n^\prime)}(\mathbf{\widetilde{\mathbf{r}}_d})^H-\expec{\bsym{\Psi}_r^{(m^\prime,n^\prime)}(\mathbf{\widetilde{\mathbf{r}}_d})}\right\Vert\left\Vert\mathbf{L}_1\right\Vert \nonumber\\
    &\leq \left\Vert \bsym{\Psi}_r^{(m^\prime,n^\prime)}(\mathbf{\widetilde{\mathbf{r}}_d})^H-\expec{\bsym{\Psi}_r^{(m^\prime,n^\prime)}(\mathbf{\widetilde{\mathbf{r}}_d})}\right\Vert\left\Vert\mathbf{H}^{-1}\right\Vert \nonumber\\
    &\leq\epsilon_8^2\left\Vert\expec{\mathbf{H}}^{-1}\right\Vert\nonumber\\
    &\leq 4\epsilon_8,\label{eq:result_lemma}
\end{align}
where the first inequality follows from the identity $\Vert\mathbf{AB}\Vert\leq\Vert\mathbf{A}\Vert\Vert\mathbf{B}\Vert$, the second follows from the fact that $\mathbf{L}_1$ is a sub-matrix of $\mathbf{H}^{-1}$, the third follows from $\Vert\mathbf{H}^{-1}\Vert\leq 2\Vert\expec{\mathbf{H}}^{-1}\Vert$ in \cite[Corollary IV.5]{off_the_grid} and the last inequality is obtained from using Lemma \ref{lemma:distance H-E[H]}. Applying the union bound completes the proof.
\end{IEEEproof}

\begin{lemma}
\label{lemma:J1_3}
Assume Lemma \ref{lemma:J1_2} holds. Then, conditioned on the event $\mathcal{E}_{\epsilon_8}$, we have 
\begin{align*}
\left\Vert\bsym{\Psi}_r^{(m^\prime,n^\prime)}(\mathbf{r})-\expec{\bsym{\Psi}_r^{(m^\prime,n^\prime)}(\mathbf{r})}\right\Vert_F &\leq \sqrt{J}\left\Vert\bsym{\Psi}_r^{(m^\prime,n^\prime)}(\mathbf{r})-\expec{\bsym{\Psi}_r^{(m^\prime,n^\prime)}(\mathbf{r})}\right\Vert \nonumber\\
& \leq \sqrt{J}\epsilon_8.  
\end{align*}
\end{lemma}
\begin{IEEEproof}
The first inequality follows from the identity %
$\Vert\mathbf{A}\Vert_F\leq \sqrt{J}\Vert\mathbf{A}\Vert$. Applying \eqref{eq:result_lemma} to the second inequality concludes the proof.
\end{IEEEproof}
Finally, recall the following useful result from \cite{yang2016super}.
\begin{lemma}\label{lemma:random_sampling_u}\cite{yang2016super} Assume that $\mathbf{u} \in \mathbb{C}^{J}$ are i.i.d random samples on the complex unit sphere $\mathbb{CS}^{J-1}$, consequently we obtain $\expec{\mathbf{uu}^H} = \frac{1}{J}\mathbf{I}_{J}$. 
\end{lemma}
\begin{IEEEproof}
We refer the reader to \cite[Lemma 21]{yang2016super}.
\end{IEEEproof}
\subsection{Proof of the lemma}
Given $\widetilde{\mathbf{r}}\in \Omega_{\text{grid}}$, define
\begin{align*}
    \mathbf{O}&\triangleq  \left(\bsym{\Psi}_r^{m^\prime,n^\prime}(\widetilde{\mathbf{r}})-\expec{\bsym{\Psi}_r^{m^\prime,n^\prime}(\widetilde{\mathbf{r}})}\right)^H\mathbf{L}_1 \\&=[\mathbf{O}_0,\dots,\mathbf{O}_{L-1}],
\end{align*}
where each $\mathbf{O}_\ell \in \mathbb{C}^{J\times J}$. Consider the event 
\begin{align*}
    \mathcal{E}_3 \triangleq \bigg\{\underset{\widetilde{\mathbf{r}}\in \Omega_{\textrm{grid}}}{\operatorname{sup}}&\left\Vert \left(\bsym{\Psi}_r^{m^\prime,n^\prime}(\widetilde{\mathbf{r}})-\expec{\bsym{\Psi}_r^{m^\prime,n^\prime}(\widetilde{\mathbf{r}})}\right)^H\mathbf{L}_1\right\Vert\leq 4\epsilon_2,  m^\prime,n^\prime = 0,1,2,3 \bigg\}.
\end{align*}
Recall the expression  $\mathbf{J}_1^{m^\prime,n^\prime}(\widetilde{\mathbf{r}}) =  \left(\bsym{\Psi}_r^{m^\prime,n^\prime}(\widetilde{\mathbf{r}})-\expec{\bsym{\Psi}_r^{m^\prime,n^\prime}(\widetilde{\mathbf{r}})}\right)^H\mathbf{L}_1\widetilde{\mathbf{u}},$
we re-write $\mathbf{J}_1^{(m^\prime,n^\prime)}(\widetilde{\mathbf{r}})$ as
\begin{align*}
   \mathbf{J}_1^{(m^\prime,n^\prime)}(\widetilde{\mathbf{r}}) = \sum_{\ell=0}^{L-1}\mathbf{O}_\ell\widetilde{\mathbf{u}}\triangleq \sum_{\ell=0}^{L-1}  \bsym{\Upsilon}_\ell.
\end{align*}
Since $\widetilde{\mathbf{u}}$ is a zero-mean random vector, we have $\expec{\bsym{\Upsilon}_\ell}=\mathbf{0}_{J\times 1}$. Therefore, matrix Bernstein inequality could be used to bound $\left\Vert\sum_{\ell=0}^{L-1} \bsym{\Upsilon}_\ell\right\Vert$  conditioned on the event $\mathcal{E}_3$. Note that 
\begin{align*}
    \Vert\bsym\Upsilon_\ell\Vert =& \Vert\mathbf{O}_\ell\widetilde{\mathbf{u}}\Vert \leq\Vert\mathbf{O}_\ell\Vert \leq\Vert\mathbf{O}\Vert \leq4\epsilon_2\triangleq R,
\end{align*}
where the first the first inequality follows from $\Vert \tilde{\mathbf{u}}\Vert\leq 1$ and the second inequality follows the fact that $\mathbf{O}_\ell$ is a sub-matrix of $\mathbf{O}$. Then, 
\begin{align}
    \left\Vert\sum_{\ell=0}^{L-1} \bsym{\Upsilon}_\ell^H\bsym{\Upsilon}_\ell\right\Vert&=\left\Vert\sum_{\ell=0}^{L-1}\expec{\widetilde{\mathbf{u}}^H\mathbf{O}_\ell^H\mathbf{O}_\ell\widetilde{\mathbf{u}}}\right\Vert \nonumber\\
    &=\sum_{\ell = 0}^{L-1} \operatorname{Tr}\left(\mathbf{O}_\ell^H\mathbf{O}_\ell\expec{\widetilde{\mathbf{u}}\widetilde{\mathbf{u}}^H}\right) \nonumber\\
    & =\frac{1}{J}\left\Vert\mathbf{O}\right\Vert_F^2.
\end{align}
We bound $\frac{1}{J}\left\Vert\mathbf{O}\right\Vert_F^2$ as 
\begin{align*}
    \frac{1}{J} \Vert\mathbf{O}\Vert_F^2&\leq \frac{1}{J}\left\Vert\mathbf{L}_1^H\right\Vert\left\Vert\bsym{\Psi}_r^{m^\prime,n^\prime}(\widetilde{\mathbf{r}})-\expec{\bsym{\Psi}_r^{m^\prime,n^\prime}(\widetilde{\mathbf{r}})}\right\Vert \nonumber\\
    &\leq \frac{4\cdot 1.2470^2J\epsilon_2^2}{J} \nonumber\\
    &\leq 12\epsilon_2^2 \nonumber\\
    &\triangleq \sigma^2,
\end{align*}
where the first inequality results from $\Vert\mathbf{AB}\Vert\leq \Vert\mathbf{A}\Vert_F^2\Vert\mathbf{B}\Vert_F^2$, second inequality uses the fact that $\mathbf{L}_1$ is a submatrix of $\mathbf{H}^{-1}$ i.e. $\mathbf{L_1} \leq \mathbf{H}^{-1}$, and the last inequality employed  Lemma \ref{lemma:J1_2} and Lemma 4 from \cite{yang2016super}.
Applying the matrix Bernstein inequality to $\mathbf{J}_1^{(m^\prime,n^\prime)}(\widetilde{\mathbf{r}})$ conditioned by $\mathcal{E}_3$ produces
\begin{align*}
    &\mathbb{P}\left\{\underset{\widetilde{\mathbf{r}}\in \Omega_{\text{grid}}}{\operatorname{sup}}\left\Vert\mathbf{J}_1^{(m^\prime,n^\prime)}(\widetilde{\mathbf{r}}) \right\Vert_2 \geq \epsilon_9\bigg|\mathcal{E}_3\right\} \nonumber\\
    & \leq\vert \Omega_{\text{grid}}\vert(J+1)e^{\left(\frac{-3\epsilon_9^2}{\sigma^2+\frac{R\epsilon_9}{3}}\right)} \nonumber\\
    & \leq \begin{cases}\left|\Omega_{\text{grid}}\right|(J+1) \exp \left(\frac{-3 \epsilon_{9}^{2}}{8 \sigma^{2}}\right), \; \epsilon_{9} \leq \sigma^{2} / R, \\ \left|\Omega_{\text{grid}}\right|(J+1) \exp \left(\frac{-3 \epsilon_{9}}{8 R}\right), \; \epsilon_{9} \geq \sigma^{2} / R.\end{cases}
\end{align*}
Substituting $\epsilon_{9}^{2}=\frac{800 \cdot 4^{2 (m^\prime+n^\prime)} \mu J L}{3 MP} \log \left(\frac{2 J L+J}{\delta}\right)$ and applying Lemma \ref{lemma:J1_3} yields
\begin{align*}
&\mathbb{P}\left\{\sup _{\widetilde{\mathbf{r}} \in \Omega_{\text{grid}}}\left\|\mathbf{J}_1^{(m^\prime,n^\prime)}(\widetilde{\mathbf{r}}) \right\|_{2} \geq \epsilon_{9}\right\} \nonumber\\
& \leq\left\{\begin{array}{l}
\left|\Omega_{\text{grid}}\right|(J+1) e^{ \left(\frac{-3 \epsilon_{9}^{2}}{8 \sigma^{2}}\right)}+\left|\Omega_{\text{grid}}\right| 4 \delta_{2}+\mathbb{P}\left(\mathcal{E}_{1, \epsilon_{1}}^{c}\right), \epsilon_{9} \leq\frac{\sigma^2}{R}, \\
\left|\Omega_{\text{grid}}\right|(J+1) e^{\left(\frac{-3 \epsilon_{9}}{8 R}\right)}+ \left|\Omega_{\text{grid}}\right| 4 \delta+\mathbb{P}\left(\mathcal{E}_{1, \epsilon_{1}}^{c}\right), \epsilon_9 \geq \frac{\sigma^2}{R}.\\
\end{array}\right.
\end{align*}

To ensure that the second term $\vert\Omega_{\text{grid}}\vert16\delta_2\leq \delta$, the failure probability $\delta$ needs to satisfy
\begin{equation}
M P\geq \frac{800\cdot 4^{2(m^\prime+n^\prime)} \mu JL}{3\epsilon_2^2}\log \left(\frac{16\vert\Omega_{\text{grid}}\vert(2LJ+J)}{\delta}\right).\label{eq:samp_bernsteain1}    
\end{equation}

To make the failure probability less than or equal to $\delta$ for the first term, choose
$$
\begin{cases}96 \epsilon_{2}^{2}=\frac{3 \epsilon_{9}^{2}}{\log \left(\frac{\vert\Omega_{\text{grid}}\vert(J+1)}{\delta}\right)}, & \epsilon_{9} \leq \sigma^{2} / R ,\\ 32 \epsilon_{2}=\frac{3 \epsilon_{9}}{\log \left(\frac{\vert\Omega_{\text{grid}}\vert(J+1)}{\delta}\right)}, & \epsilon_{9} \geq \sigma^{2} / R.\end{cases}
$$
Equivalently, when $\epsilon_{9} \leq \sigma^{2} / R$, we get

\begin{align}
M P\geq 
&\frac{800 \cdot 96 \cdot 4^{2 \ell} \mu J K}{9 \epsilon_{2}^{2}} \log \left(\frac{16\left|\Omega_{\text{grid}}\right|(2 LJ+J)}{\delta}\right) \nonumber\\
    & \hspace{2.2cm}
 \times \log \left(\frac{\left|\Omega_{\text{grid}}\right|(J+1)}{\delta}\right).\label{eq:samp_bernsteain2}
\end{align}

When $\epsilon_{9} \geq \sigma^{2} / R$, we obtain

\begin{align}
M P& \geq 
\frac{32^{2} \cdot 800 \cdot 4^{2 (m^\prime+n^\prime)} \mu J L}{27 \epsilon_{9}^{2}} \log \left(\frac{4\left|\Omega_{\text{grid}}\right|(2 L J+J)}{\delta}\right) \nonumber\\
    & \hspace{4.0cm} \times \log ^{2}\left(\frac{\left|\Omega_{\text{grid}}\right|(J+1)}{\delta}\right).\label{eq:samp_bernsteain3}
\end{align}
For the third term, using Lemma \ref{lemma:distance H-E[H]}, we have  $\mathbb{P}\left(\mathcal{E}_{1, \epsilon_{1}}^{c}\right) \leq \delta$, where 
\begin{align}
MP> \frac{90 \mu J}{\epsilon_1^2}\max\left(L,Q\right)\textnormal{log}\left(\max\left(\frac{6LJ}{\delta},\frac{6PQJ}{\delta}\right)\right).\label{eq:samp_bernsteain4}
\end{align}

Combining the results \eqref{eq:samp_bernsteain1}-\eqref{eq:samp_bernsteain4}  and applying the union bound, we obtain
$$
\mathbb{P}\left\{\sup _{\widetilde{\mathbf{r}} \in \Omega_{\text{grid}}}\left\|\mathbf{J}_{1}^{(m^\prime,n^\prime)}\left(\widetilde{\mathbf{r}}\right)\right\|_{2} \geq \epsilon_{9}, m^\prime,n^\prime=0,1,2,3\right\} \leq 48 \delta,
$$
the failure probability satisfies 
\begin{align*}
    MP \geq &C\mu \operatorname{max}(L,Q)J\operatorname{max}\left\{\frac{1}{\epsilon_9^2}\log\left(\frac{\vert\Omega_{\textnormal{\textrm{grid}}}\vert JL}{\delta}\right)\log^2\left(\frac{\vert\Omega_{\textnormal{\textrm{grid}}}\vert J}{\delta}\right), \right. \nonumber\\
    & \hspace{3.5cm} \left. \log\left(\frac{LJ}{\delta}\right),\log\left(\frac{PQJ}{\delta}\right)\right\}.
\end{align*}

\section{Proof of Lemma \ref{lemma:J2_last}}
\label{app:J2_last}
Recall \eqref{eq:decomposition_theta_r} that $$\mathbf{J}_{2}^{(m^\prime,n^\prime)}(\mathbf{r})=\mathbb{E}\left[ \bsym{\Psi}_r^{(m^\prime,n^\prime)}(\mathbf{r})\right]^{H} \left(\mathbf{L}_1-\mathbf{L}_1^{\prime} \otimes \mathbf{I}_{J}\right) \widetilde{\mathbf{u}}.$$
The bound on $\Vert\mathbf{J}_{2}^{(m^\prime,n^\prime)}(\mathbf{r})\Vert$ follows from the bound on $\left\Vert\expec{\bsym{\Psi}_r^{(m^\prime,n^\prime)(\mathbf{r})}}\right\Vert_F^2$. Therefore, 
\begin{align}
\left\Vert\expec{\bsym{\Psi}_r^{(m^\prime,n^\prime)}(\mathbf{r})}\right\Vert_F^2 &= \left\Vert\psi_r^{(m^\prime,n^\prime)}(\mathbf{r})\otimes \mathbf{I}_J\right\Vert_F^2 \nonumber\\
    & =J\left\Vert\psi_r^{(m^\prime,n^\prime)}(\mathbf{r})\right\Vert_2^2 \nonumber\\
    & \leq C_1J,
\label{eq:bound_E_Psi_r}
\end{align}
where $C_1$ is a numerical constant and the last inequality follows from \cite[Lemma 4.9]{off_the_grid}. Now, conditioned on the event $\mathcal{E}_{1,\epsilon_1}$, we obtain
\begin{align*}
    \bigg\Vert\left(\mathbf{L}_1-\mathbf{L}_1^{\prime} \otimes \mathbf{I}_{J}\right)^H \expec{\bsym{\Psi}_r^{(m^\prime,n^\prime)}(\mathbf{r})}\bigg\Vert_F^2 &\leq C_1J \left\Vert\left(\mathbf{L}_1-\mathbf{L}_1^{\prime} \otimes \mathbf{I}_{J}\right)\right\Vert \nonumber\\
    &\leq(2\cdot 1.2470^2\epsilon_1)^2C_1J \nonumber\\
    & \triangleq CJ\epsilon_1^2,
\end{align*}
where the first inequality follows from \eqref{eq:bound_E_Psi_r}; the second follows from  Lemma~\ref{lemma:distance H-E[H]}, \cite[Proposition 3]{suliman2022blind}, and the fact that $\mathbf{L}_1-\mathbf{L}_1^{\prime} \otimes \mathbf{I}_{J}$ is a submatrix of $\mathbf{H}^{-1}-\expec{\mathbf{H}^{-1}}$; and $C$ is a redefined constant. 

Similar to $\mathbf{J}_1$ in Appendix~\ref{app:bound_J1}, we write $\mathbf{J}_2$ as the sum of zero-mean matrices. Define 
\begin{align*}
    \widetilde{\mathbf{O}} &\triangleq \expec{\bsym{\Psi}_r^{(m^\prime,n^\prime)}(\widetilde{\mathbf{r}})}^H\left(\mathbf{L}_1 -\mathbf{L}_1^\prime \otimes \mathbf{I}_J\right)=[\widetilde{\mathbf{O}}_0,\dots,\widetilde{\mathbf{O}}_{L-1}].
\end{align*}
Consequently, 
\begin{align*}
    \mathbf{J}_2^{(m^\prime,n^\prime)} (\widetilde{\mathbf{r}}) = \sum_{\ell=0}^{L-1} \widetilde{\mathbf{O}}_\ell \widetilde{\mathbf{u}}\triangleq\sum_{\ell=0}^{L-1} \widetilde{\bsym{\Upsilon}}_\ell,
\end{align*}
where $\expec{\bsym{\Upsilon}_\ell} = \mathbf{0}_{J\times 1}$, which follows from \cite{yang2016super} (similar to the proof of Lemma~\ref{lemma:J1_last}). Conditioned on the event $\mathcal{E}_{1,\epsilon}$ and choosing $\epsilon_1 = \frac{1}{4}$, we get 
\begin{align*}
    \left\Vert\widetilde{\bsym{\Upsilon}}_\ell\right\Vert &= \left\Vert\widetilde{\mathbf{O}}\widetilde{\mathbf{u}}\right\Vert\\&\leq \left\Vert\widetilde{\mathbf{O}}\right\Vert\\&\leq \left\Vert\mathbf{L}_1 - \mathbf{L}_1^\prime \otimes\mathbf{I}_J\right\Vert\left\Vert\expec{\bsym{\Psi}_r^{(m^\prime,n^\prime)}(\widetilde{\mathbf{r}})}\right\Vert\\&= 2\cdot 1.2470^2 \epsilon_1^2\cdot CJ\\&\leq CJ\epsilon_1^2\triangleq R,
\end{align*}
where $C$ is a numerical constant. The bound on the variance term becomes
$$
\begin{aligned}
\left\|\sum_{\ell=0}^{L-1} \expec{ \widetilde{\bsym{\Upsilon}}_{\ell}^{H} \widetilde{\bsym{\Upsilon}}_{\ell}}\right\| &=\left\|\sum_{\ell=0}^{L-1} \expec{\widetilde{\mathbf{u}}^{H} \widetilde{{\mathbf{O}}}_{\ell}^{H} \widetilde{{\mathbf{O}}}_{\ell} \widetilde{\mathbf{u}}}\right\| \\&=\sum_{\ell=0}^{L-1} \operatorname{Tr}\left(\widetilde{\mathbf{O}}_{\ell}^{H} \widetilde{\mathbf{O}}_{\ell} \expec{\widetilde{\mathbf{u}} \widetilde{\mathbf{u}}^{H}}\right) \\
&=\sum_{\ell=0}^{L-1} \operatorname{Tr}\left(\widetilde{\mathbf{O}}_{\ell}^{H} \widetilde{\mathbf{O}}_{\ell} \mathbf{I}_J\right)\\
& \leq \frac{1}{J} C J \epsilon_{1}^{2} \\
&=C \epsilon_{1}^{2} \triangleq \sigma^2.
\end{aligned}
$$
Applying the matrix Bernstein inequality as in the bound of $\mathbf{J}_1$ and using the union bound concludes the proof. 

\section{Proof of Lemma \ref{lemma:theta_last} }
\label{app:bound_theta}
\subsection{Preliminaries to the proof}
Similar to the proof of Lemma \ref{lemma:J1_last}, we first show that the quantity $\Vert\bsym{\Psi}_c^{(m^\prime,n^\prime)}(\mathbf{r})\Vert$ is small with high probability.
\begin{lemma}\label{lemma:theta_1}
Given $\mathbf{r} \in [0,1]^2$, $\epsilon \in (0,1)$ for $m^\prime=0,1,2,3$ and $n^\prime=0,1,2,3$,  $\Vert\bsym{\Psi}_c^{(m^\prime,n^\prime)}(\mathbf{r})\Vert\leq 16\epsilon_{10}$  holds with at least probability $1-4\delta$, where $\delta$ is  
specified by
\begin{equation}
    MP\leq \frac{800\cdot 4^{2(m^\prime+n^\prime)}\mu JL}{3\epsilon_{10}^2}\log\left(\frac{3JPQ+J}{\delta}\right).
\end{equation}
\end{lemma}
\begin{IEEEproof}
The independence of the basis vectors $\mathbf{b}_n$ and $\mathbf{d}_{\widetilde{n}}$ implies that $\boldsymbol{\Psi}_r^{(m^\prime,n^\prime) }(\mathbf{r})$ in \eqref{eq:Psi_c} is a sum of zero-mean random independent matrices as
$$
\boldsymbol{\Psi}_c^{(m^\prime,n^\prime )}(\mathbf{r}) = \frac{1}{MP}\sum_{\widetilde{n}}\mathbf{Y}_{\widetilde{n}}^{(m^\prime,n^\prime)} (\mathbf{r}),
$$ 
where 
\begin{align}
   \mathbf{Y}_{\widetilde{n}}^{(m^\prime,n^\prime)} (\mathbf{r})&= g_M(n)g_{P}(p)\left(\frac{-\mathrm{j}2\pi n}{\kappa}\right)^{m^\prime}\left(\frac{-\mathrm{j}2\pi p}{\kappa}\right)^{n^\prime}  \nonumber\\
& \hspace{2cm} \cdot e^{-\mathrm{j}2\pi(n\tau+p\nu)} \bsym{\varrho}_{\widetilde{n}}\otimes\mathbf{b}_n\mathbf{d}_{\widetilde{n}}^H.\nonumber
\end{align}
To bound  $\|\boldsymbol{\Psi}_c^{(m^\prime,n^\prime) }(\mathbf{r})\|$, we employ the matrix Bernstein inequality. To this end, we compute the following bounds on $\|\mathbf{Y}_{\widetilde{n}}^{(m',n')}(\mathbf{r})\|$ and the variance term.   
\begin{align*}
\left\Vert\mathbf{Y}_{\widetilde{n}}^{(m^\prime,n^\prime)}\right\Vert = &\left\Vert g_M(n)g_{P}(p)\left(\frac{-\mathrm{j}2\pi n}{\kappa}\right)^{m^\prime}\left(\frac{-\mathrm{j}2\pi p}{\kappa}\right)^{n^\prime} \right. \nonumber\\
& \hspace{2cm} \left. \times e^{-\mathrm{j}2\pi(n\tau+p\nu)} \bsym{\varrho}_{\widetilde{n}}\otimes\mathbf{b}_n\mathbf{d}_{\widetilde{n}}^H\right\Vert\\
&\leq \frac{1}{MP}4^{m^\prime}4^{n^\prime}\Vert \bsym{\varrho}_{\widetilde{n}}\Vert_2\Vert\mathbf{b}_n\Vert \Vert\mathbf{d}_{\widetilde{n}}\Vert\\
&\leq \frac{1}{MP}4^{m^\prime+n^\prime}\triangleq R,
\end{align*}
where the first inequality is the consequence of $\Vert g_{M}(n)\Vert_{\infty}=\Vert g_{P}(p)\Vert_{\infty}\leq 1 $ for  $M,P\geq4$,  and the second uses $\Vert  \bsym{\varrho}_{\widetilde{n}}\Vert\leq 27Q$ for $\left\vert\frac{2\pi n}{\kappa}\right\vert\leq4$. Further, \par\noindent\small
\begin{align*}
    &\left\Vert\sum_{\widetilde{n}}\expec{\mathbf{Y}_{\widetilde{n}}^{(m^\prime,n^\prime)}\mathbf{Y}_{\widetilde{n}}^{(m^\prime,n^\prime)H}}\right\Vert \\
    & = \frac{1}{(MP)^2}\left\Vert\sum_{\widetilde{n}}\mathbb{E}\bigg[\vert g_{M}(n)\vert^2\vert g_{P}(p)\vert^2\left(\left\vert\frac{-\mathrm{j}2\pi n}{\kappa}\right\vert\right)^{2m^\prime}\left(\left\vert\frac{-\mathrm{j}2\pi p}{\kappa}\right\vert\right)^{2n^\prime} \right. \nonumber\\
& \hspace{2cm} \left. \cdot\left(\bsym{\upsilon}_{\widetilde{n}}^H\otimes\mathbf{d}_{\widetilde{n}}\mathbf{b}_n^H\right)\left(\bsym{\upsilon}_{\widetilde{n}}\otimes(\mathbf{b}_n\mathbf{d}_{\widetilde{n}}^H)\right)\bigg]\right\Vert\\    
    &\leq\frac{4^{2(m^\prime+n^\prime)}}{(MP)^2}\left\Vert\sum_{\widetilde{n}}\Vert \bsym{\varrho}_{\widetilde{n}}\Vert_2^2\expec{\Vert\mathbf{b}_n\Vert_2^2\mathbf{d}_{\widetilde{n}}\mathbf{d}_{\widetilde{n}}^H}\right\Vert\\
    &\leq \frac{4^{2(m^\prime+n^\prime)}}{(MP)^2}\left\Vert\ \sum_{\widetilde{n}}\Vert \bsym{\varrho}_{\widetilde{n}}\Vert_2^2 \mu J \mathbf{I}_J\right\Vert\\
    &\leq \frac{100\cdot4^{2(m^\prime+n^\prime)}\mu JQ}{MP}\triangleq \sigma.
\end{align*}\normalsize
We use the Bernstein inequality on $\bsym{\Psi}_c^{(m^\prime,n^\prime)}(\mathbf{r})$ yields 
\begin{align*}
    \mathbb{P}\left\{\left\Vert\sum_{\widetilde{n}}\mathbf{Y}_{\widetilde{n}}^{(m^\prime,n^\prime)}\geq\epsilon_{10}\right\Vert\right\}\leq(3JPQ+J)e^{\left(\frac{-3\epsilon_{10}^2}{8\sigma^2}\right)},
\end{align*}
Applying the union bound for $m^\prime=0,1,2,3 $ and $n^\prime=0,1,2,3$ yields
\begin{align*}
    \mathbb{P}\left\{\left\Vert\sum_{\widetilde{n}}\mathbf{Y}_{\widetilde{n}}^{(m^\prime,n^\prime)}\geq\epsilon_{10}\right\Vert,m^\prime,n^\prime = 0,1,2,3\right\} \leq 16\delta.
\end{align*}
where the failure probability $\delta_5$ satisfies
\begin{align*}
     MP\leq \frac{800\cdot 4^{2 (m^\prime+n^\prime)\mu JL}}{3\epsilon_{10}^2}\log\left(\frac{3JPQ+J}{\delta}\right).
\end{align*}
\end{IEEEproof}
As demonstrated in the following Lemma \ref{lemma:theta_2}, conditioned on the event $\mathcal{E}_{\mathcal{E}_1}$, the quantities $\left\Vert\bsym{\Psi}_c^{(m^\prime,n^\prime)}(\mathbf{r})^H(\mathbf{r})^H\widehat{\mathbf{L}}_1\right\Vert_2$ and $\left\Vert\bsym{\Psi}_c^{(m^\prime,n^\prime)}(\mathbf{r})\right\Vert_F$ are small.
\begin{lemma}\label{lemma:theta_2}
Considering that $\epsilon_1\in(0,1/4]$ and a finite set of grid points $\boldsymbol{\Omega}_{\textnormal{\textrm{grid}}} =  \{\widetilde{\mathbf{r}}_d\}$ with cardinality $|\boldsymbol{\Omega}_{\textnormal{\textrm{grid}}}|$. Then, we have that 
\begin{align*}
    \mathbb{P}&\left\{\underset{\widetilde{\mathbf{r}}\in \Omega_{\textnormal{\textrm{grid}}}}{\operatorname{sup}}\left\Vert\bsym{\Psi}_c^{(m^\prime,n^\prime)}(\mathbf{r})^H\widehat{\mathbf{L}}_1\right\Vert\geq 4\epsilon_5, m^\prime,n^\prime = 0,1,2,3\right\} \nonumber\\ 
    & \leq \vert\Omega_{\textnormal{\textrm{grid}}}\vert 16\delta_5+\mathbb{P}\left(\mathcal{E}_{1,\epsilon_1}^c\right)
\end{align*}
\end{lemma}
\begin{IEEEproof}
The proof follows the same procedure as in Lemma \ref{lemma:J1_last} using the bound of $\bsym{\Psi}_c^{(m^\prime,n^\prime)}(\mathbf{r})$ is a zero-mean matrix, hence, using the matrix Bernstein inequality and conditioned by the event $\mathcal{E}_{\varepsilon_5}$ concludes the proof. 
\end{IEEEproof}
\subsection{Proof of the lemma}
The proof is similar to that of Lemma~\ref{lemma:J1_last} in Appendix~\ref{app:bound_J1}. Briefly, using the bounds in Lemmata \ref{lemma:theta_1} and \ref{lemma:theta_2}, we compute the bound on  $\bsym{\Theta}_c^{(m^\prime,n^\prime)}$ using the matrix Bernstein inequality. Then, applying the union bound completes the proof. 

\section{Proof of Lemma \ref{lemma:certificate_less_1_everywhere}}
\label{app:certificate_less_1_everywhere}
Denote the $k$-th column of $\boldsymbol{\Psi}_{r}^{(m^\prime,n^\prime)}(\mathbf{r})$ by $\boldsymbol{\Psi}_{r}^{(m^\prime,n^\prime)}(\mathbf{r};k)$. Then,
$$
\begin{aligned}
\left\|\boldsymbol{\Psi}_{r}^{(m^\prime,n^\prime)}(\tau ; k)\right\|_{2}&= \bigg\|  \frac{1}{MP} \sum_{\widetilde{n}} g_{M}(n)g_{P}(p)\left(\frac{-\mathrm{j} 2 \pi n}{\kappa}\right)^{m'}  \nonumber\\
&  \times \left(\frac{-\mathrm{j} 2 \pi p}{\kappa}\right)^{n'} e^{-\mathrm{j} 2 \pi (n \tau+p\nu)} \bsym{\upsilon}_{\widetilde{n}} \otimes \mathbf{b}_{n} \mathbf{b}_{n}^{H}(k)\bigg\| \\
& \leq \frac{(MP) 4^{2(m^\prime+n^\prime)}}{MP} \mu \sqrt{J}\|\bsym{\upsilon}_{\widetilde{n}}\|_{2} \\
& \leq \frac{\sqrt{27} \cdot(MP) 4^{\ell}}{M} \mu \sqrt{J L} \\
&= C \mu \sqrt{JL},
\end{aligned}
$$
for some constant $C$, where the first inequality results from $\left|[\boldsymbol{b}_{n}]_k\right| \leq \sqrt{\mu}$ (using the incoherence property) and the second uses the fact that  $\|\bsym{\upsilon}_{\widetilde{n}}\|_{2}^{2} \leq 27 L$ for $M, P\geq 4$. Similarly, $$\left\|\boldsymbol{\Psi}_{c}^{(m^\prime,n^\prime)}(\tau ; k)\right\|_2 \leq C\mu\sqrt{PJQ}.$$

Denote the $k$-th entry of the vector-valued polynomial $\mathbf{f}_r^{(m^\prime,n^\prime)}(\mathbf{r})$ by $\mathbf{f}_r^{(m^\prime,n^\prime)}(\mathbf{r};k)$. Conditioned on the event $\mathcal{E}_{1, \epsilon_{1}}$ with $\epsilon_1 \in\left(0, \frac{1}{4}\right]$, we obtain
\par\noindent\small
$$
\begin{aligned}
\left|\frac{1}{\kappa} \mathbf{f}_{r}^{\left(m^{\prime}, n^{\prime}\right)}(\mathbf{r} ; k)\right| &= \bigg\| \boldsymbol{\Psi}_{r}^{\left(m^{\prime}, n^{\prime}\right)}(\mathbf{r} ; k)^{H}\left(\mathbf{L}_{1} \widetilde{\mathbf{u}} \right. \nonumber\\
& \hspace{1cm} \left. + \mathbf{L}_{2} \widetilde{\mathbf{v}}\right)+ \bsym{\Psi}_{c}^{\left(m^{\prime}, n^{\prime}\right)}(\mathbf{r} ; k)^{H}\left(\widehat{\mathbf{L}}_{1} \widetilde{\mathbf{u}}+\widehat{\mathbf{L}}_{2} \widetilde{\mathbf{v}}\right) \bigg\| \\
&\leq \left\|\bsym{\Psi}_{r}^{\left(m^{\prime}, n^{\prime}\right)}(\mathbf{r} ; k)\right\|\left(\left\|\mathbf{L}_{1}\right\|\|\widetilde{\mathbf{u}}\|+\left\|\mathbf{L}_{2}\right\|\|\widetilde{\mathbf{v}}\|\right) \nonumber\\ 
& \hspace{1cm} +\left\|\bsym{\Psi}_{c}^{\left(m^{\prime}, n^{\prime}\right)}(\mathbf{r} ; k)\right\|\left(\left\|\widetilde{\mathbf{L}}_{1}\right\|\|\widetilde{\mathbf{u}}\|+\left\|\widetilde{\mathbf{L}}_{2}\right\|\|\widetilde{\mathbf{v}}\|\right)\\ & \leq C\mu(L\sqrt{J}+Q\sqrt{PJ}),
\end{aligned}
$$
\normalsize
for some constant $C$. Then,
applying the Bernstein's polynomial inequality \cite{off_the_grid}, \cite{harris1996bernstein} yields
$$
\begin{aligned}
&\left|\frac{1}{\kappa^{(m^\prime,n^\prime)}} \mathbf{f}_r^{(m^\prime,n^\prime)}\left(\mathbf{r}_{a} ; k\right)-\frac{1}{\kappa^{(m^\prime,n^\prime)}} \mathbf{f}_r^{(m^\prime,n^\prime)}\left(\mathbf{r}_{b} ; k\right)\right| \\
&\leq\left|e^{\mathrm{j} 2 \pi \tau_{a}}-e^{\mathrm{j} 2 \pi \tau_{b}}\right|\cdot\left|e^{\mathrm{j} 2 \pi \nu_{a}}-e^{\mathrm{j} 2 \pi \nu_{b}}\right| \sup _{z=e^{\mathrm{j} 2 \pi \mathbf{r}}} \bigg| \frac{d \frac{1}{\kappa}\mathbf{f}_r^{(m^\prime,n^\prime)}(z ; k)}{dz}   \bigg|\\
&\leq 16 \pi^2 MP\left|\tau_{a}-\tau_{b}\right|\cdot\left|\nu_{a}-\nu_{b}\right|  \sup _{\tau}\left|\frac{1}{\sqrt{\left|\varphi_{M}^{\prime \prime}(0)\right|^{\ell}}} \frac{1}{\kappa}\mathbf{f}_r^{(m^\prime,n^\prime)}(\mathbf{r} ; k)\right| \\
&\leq C MP \mu 
(L\sqrt{J}+
Q\sqrt{PJ})\left|\tau_{a}-\tau_{b}\right|\left|\cdot|\nu_{a}-\nu_{b}\right|,\end{aligned}
$$
for some numerical constant $C$. Hence, 
$$
\begin{aligned}
&\left\|\frac{1}{\kappa}\mathbf{f}_r^{(m^\prime,n^\prime)}(\mathbf{r}_a)-\frac{1}{\kappa}\mathbf{f}_r^{(m^\prime,n^\prime)}(\mathbf{r}_b)\right\|_{2} \nonumber\\ 
&\leq C MP \mu 
(LJ+
QPJ)\left|\tau_{a}-\tau_{b}\right|\left|\cdot|\nu_{a}-\nu_{b}\right|\\&\leq C(MP)^2\left|\tau_{a}-\tau_{b}\right|\left|\cdot|\nu_{a}-\nu_{b}\right|,
\end{aligned}$$
where the second inequality follows from $MP \geq  \mu 
(LJ+
QPJ)$. We then choose $\Omega_{\text{grid}}$  such that for any  $\mathbf{r} \in[0,1)^2$, there exists a point $\widetilde{\mathbf{r}} \in \Omega_{\text{grid}}$ with $\left\|\mathbf{r}-\widetilde{\mathbf{r}}\right\| \leq \frac{\epsilon}{3 C (MP)^{2}}$, where $\left|\Omega_{\text{grid}}\right| \leq \frac{3 C (MP)^{2}}{\epsilon}$. Using this $\Omega_{\text{grid}}$ and conditioned on the events $\mathcal{E}$ and $\mathcal{E}_{1, \epsilon_{1}}$ with $\epsilon_{1} \in\left(0, \frac{1}{4}\right]$, we have
$$
\begin{aligned}
&\left\|\frac{1}{\kappa^{m'+n'}}\mathbf{f}_r^{(m^\prime,n^\prime)}(\mathbf{r})-\frac{1}{\kappa^{m'+n'}}   f^{(m^\prime, n^\prime)}(\mathbf{r})\mathbf{u}\right\|_{2} \\
&\leq\left\|\frac{1}{\kappa^{m'+n'}}\mathbf{f}_r^{(m^\prime,n^\prime)}(\mathbf{r})-\frac{1}{\kappa^{m'+n'}}\mathbf{f}_r^{(m^\prime,n^\prime)}(\widetilde{\mathbf{r}})\right\|_{2} \nonumber\\
& \hspace{1cm} \times \left\|\frac{1}{\kappa^{m'+n'}}\mathbf{f}_r^{(m^\prime,n^\prime)}(\widetilde{\mathbf{r}})-\frac{1}{\kappa^{m'+n'}}   f^{(m^\prime, n^\prime)}(\mathbf{r})\mathbf{u}\right\|_{2} \\
&+\left\|\frac{1}{\kappa^{m'+n'}}   f^{(m^\prime, n^\prime)}(\widetilde{\mathbf{r}})\mathbf{u}-\frac{1}{\kappa^{m'+n'}}   f^{(m^\prime, n^\prime)}(\mathbf{r})\mathbf{u}\right\|_{2} \\
&\leq C(MP)^2\left|\tau-\widetilde{\tau}\right|\left|\cdot|\nu-\widetilde{\nu}\right|+\frac{\epsilon}{3}+C(MP)^2\left|\tau-\widetilde{\tau}\right|\left|\cdot|\nu-\widetilde{\nu}\right| \\
&\leq \epsilon, \quad \forall \mathbf{r} \in[0,1)^2.
\end{aligned}
$$
Using the chosen $\Omega_{\text{grid}}$, it follows from Lemma \ref{lemma:bound_f_f_tilde_discrete} that the bound on $MP$ is \par\noindent\small
\begin{align*}
    &MP \geq  C\mu \operatorname{max}(L,Q)J\operatorname{max}\left\{\frac{1}{\epsilon^2}\log\left(\frac{MP QJ}{\delta}\right), \frac{1}{\epsilon^2}\log\left(\frac{MP LJ}{\delta}\right), \right. \nonumber\\
& \hspace{4.2cm} \left. \log^2\left(\frac{MP J}{\delta}\right), \log\left(\frac{LJ}{\delta}\right), \right. \nonumber\\
& \hspace{4.2cm} \left. \log\left(\frac{PQJ}{\delta}\right)\right\},
\end{align*}\normalsize
with a redefined numerical constant $C$. This completes the proof. 

\bibliographystyle{IEEEtran}
\bibliography{main}

\end{document}